\documentclass[showpacs,aps,prd,reprint]{revtex4-1}
\usepackage{graphicx,amsmath,amssymb,epsfig,color,lineno}
\usepackage{verbatim} 
\usepackage{xspace,afterpage,natbib,psfrag,multirow}
\input{babarsym}

\newcommand{\BABARPubYear}    {13}
\newcommand{\BABARPubNumber}  {001}
\newcommand{\SLACPubNumber} {15381}

\newcommand{\dss}{\ensuremath{D^{**}}\xspace}
\newcommand{\ds}{\ensuremath{D^{(*)}}\xspace}
\newcommand{\dspiz}{\ensuremath{D^{(*)}\piz}\xspace}
\newcommand{\dspizl}{\ensuremath{D^{(*)}\piz\ell}\xspace}
\newcommand{\Btag}{\ensuremath{B_{\rm tag}}\xspace}

\newcommand{\dsl}{\ensuremath{\ds\ell}\xspace}

\newcommand{\BDtaunu}{\ensuremath{\Bb \rightarrow D\tau^-\nutb}\xspace}
\newcommand{\BDstaunu}{\ensuremath{\Bb \rightarrow D^*\tau^-\nutb}\xspace}
\newcommand{\BDxtaunu}{\ensuremath{\Bb \rightarrow D^{(*)}\tau^-\nutb}\xspace}
\newcommand{\BDsstaunu}{\ensuremath{\Bb \rightarrow D^{**}\taum\nub_{\tau}}\xspace}

\newcommand{\BDlnu}{\ensuremath{\Bb \rightarrow D\ell^-\nub_{\ell}}\xspace}
\newcommand{\BDslnu}{\ensuremath{\Bb \rightarrow D^{*}\ell^-\nub_{\ell}}\xspace}
\newcommand{\BDxlnu}{\ensuremath{\Bb \rightarrow D^{(*)}\ell^-\nub_{\ell}}\xspace}
\newcommand{\BDsslnu}{\ensuremath{\Bb \rightarrow D^{**}\ellm\nub_{\ell}}\xspace}

\newcommand{\BDxltnu}{\ensuremath{\Bb \rightarrow \ds(\tau^-/\ellm)\nub}\xspace}
\newcommand{\BDssltnu}{\ensuremath{\Bb \rightarrow D^{**}(\tau^-/\ellm)\nub}\xspace}

\newcommand{\Dxltnu}{\ensuremath{D^{(*)}(\ell/\tau)\nu}\xspace}
\newcommand{\Dsslnu}{\ensuremath{D^{**}(\ell/\tau)\nu}\xspace}

\newcommand{\Dssltnu}{\ensuremath{D^{**}(\ell/\tau)\nu}\xspace}
\newcommand{\Dzl}{\ensuremath{D^{0}\ell}\xspace}
\newcommand{\Dpl}{\ensuremath{D^{+}\ell}\xspace}
\newcommand{\Dszl}{\ensuremath{D^{*0}\ell}\xspace}
\newcommand{\Dspl}{\ensuremath{D^{*+}\ell}\xspace}

\newcommand{\RDss}{\ensuremath{{\cal R}(D^{**})}}
\newcommand{\RDx}{\ensuremath{{\cal R}(D^{(*)})}}
\newcommand{\RDs}{\ensuremath{{\cal R}(D^*)}}
\newcommand{\RD}{\ensuremath{{\cal R}(D)}}
\newcommand{\RDz}{\ensuremath{{\cal R}(\Dz)} }
\newcommand{\RDp}{\ensuremath{{\cal R}(\Dp)} }
\newcommand{\RDstarz}{\ensuremath{{\cal R}(\Dstarz)}}
\newcommand{\RDstarp}{\ensuremath{{\cal R}(\Dstarp)}}

\newcommand{\mmiss}{\ensuremath{m_{\rm miss}^2}\xspace}

\newcommand{\eextra}{\ensuremath{E_{\rm extra}}\xspace}
\newcommand{\pstarl}{\ensuremath{|\boldsymbol{p}^*_\ell|}\xspace}
\newcommand{\fDss}{\ensuremath{f_{D^{**}}}\xspace}
\newcommand{\tanB}{\ensuremath{{\rm tan}\beta}\xspace}
\newcommand{\mH}{\ensuremath{m_{H^+}}\xspace}
\newcommand{\tBmH}{\ensuremath{\tanB/\mH}\xspace}
\newcommand{\pD}{\ensuremath{|\boldsymbol{p}^*_{D}|}\xspace}
\newcommand{\pDs}{\ensuremath{|\boldsymbol{p}^*_{\Dstar}|}\xspace}
\newcommand{\pDx}{\ensuremath{|\boldsymbol{p}^*_{\ds}|}\xspace}
\newcommand{\mB}{\ensuremath{m_B}\xspace}

\newcommand{\mDs}{\ensuremath{m_{\Dstar}}\xspace}

\newcommand{\qt}{\ensuremath{q^{2}}\xspace}

\begin{document}

\preprint{\babar-PUB-\BABARPubYear/\BABARPubNumber} 
\preprint{SLAC-PUB-\SLACPubNumber} 
  
\begin{flushleft}
SLAC-PUB-\SLACPubNumber\\
\babar-PUB-\BABARPubYear/\BABARPubNumber \\
\end{flushleft}
  
\title{Measurement of an Excess of {\boldmath \BDxtaunu} Decays and\\ Implications
for Charged Higgs Bosons} \vspace{0.3in}
%
\author{J.~P.~Lees}
\author{V.~Poireau}
\author{V.~Tisserand}
\affiliation{Laboratoire d'Annecy-le-Vieux de Physique des Particules (LAPP), Universit\'e de Savoie, CNRS/IN2P3,  F-74941 Annecy-Le-Vieux, France}
\author{E.~Grauges}
\affiliation{Universitat de Barcelona, Facultat de Fisica, Departament ECM, E-08028 Barcelona, Spain }
\author{A.~Palano$^{ab}$ }
\affiliation{INFN Sezione di Bari$^{a}$; Dipartimento di Fisica, Universit\`a di Bari$^{b}$, I-70126 Bari, Italy }
\author{G.~Eigen}
\author{B.~Stugu}
\affiliation{University of Bergen, Institute of Physics, N-5007 Bergen, Norway }
\author{D.~N.~Brown}
\author{L.~T.~Kerth}
\author{Yu.~G.~Kolomensky}
\author{M.~Lee
}
\author{G.~Lynch}
\affiliation{Lawrence Berkeley National Laboratory and University of California, Berkeley, California 94720, USA }
% ...
\collaboration{The \babar\ Collaboration}
\noaffiliation

\date{March 2, 2013}

\begin{abstract}
Based on the full \babar\ data sample, we report improved measurements of the ratios
$ \RD = {\cal B}(\BDtaunu)/{\cal B}(\BDlnu)$ and $ \RDs = {\cal B}(\BDstaunu)/{\cal B}(\BDslnu)$,  
where $\ell$ refers to either an electron or muon.
These ratios are sensitive to new physics contributions in the form of a charged Higgs boson. 
We measure $\RD  = 0.440\pm 0.058\pm 0.042$ and $\RDs = 0.332\pm 0.024\pm 0.018$,
which exceed the Standard Model expectations by $2.0\sigma$ and $2.7\sigma$, respectively. 
Taken together, the results disagree with these expectations at 
the $3.4\sigma$ level. This excess cannot be explained by a charged Higgs boson in the
type II two-Higgs-doublet model. 
Kinematic distributions presented here exclude large portions of the more general 
type III two-Higgs-doublet model, but there are solutions within this model compatible
with the results.
\end{abstract}

\maketitle

\section{Introduction} \label{sed:Introduction}

In the Standard Model (SM), semileptonic decays of $B$ mesons proceed via first-order 
electroweak interactions and are mediated  by the $W$ boson
\cite{Heiliger:1989yp,Korner:1989qb,Hwang:2000xe}. 
Decays involving  electrons and muons are expected to be insensitive to non-SM contributions 
and therefore have been the bases of the determination of the Cabibbo-Kobayashi-Maskawa (CKM) 
matrix elements \Vcb\ and \Vub~\cite{Amhis:2012bh}. 
Decays involving the higher-mass $\tau$ lepton provide additional information on SM processes and 
are sensitive to additional amplitudes, such as those involving an intermediate charged Higgs boson~\cite{Tanaka:1994ay,Itoh:2004ye,Nierste:2008qe,Tanaka:2010se,Fajfer:2012vx}.
Thus, they offer an excellent opportunity to search for this and other non-SM contributions.

Over the past two decades,
the development of heavy-quark effective theory (HQET) and precise measurements of 
\BDxlnu decays~\cite{convention} at the $B$ factories~\cite{Antonelli:2009ws,Nakamura:2010zzi}
have greatly improved our understanding of exclusive semileptonic decays.  
The relative rates 
\begin{equation}
\RD = \frac{{\cal B}(\BDtaunu)}{{\cal B}(\BDlnu)},\hspace{2mm} 
 \RDs = \frac{{\cal B}(\BDstaunu)}{{\cal B}(\BDslnu)} 
\end{equation}
are independent of the CKM element 
$|V_{cb}|$ and also, to a large extent, of the parameterization of the hadronic matrix elements.
SM expectations~\cite{Fajfer:2012vx} for the ratios 
\RD\ and \RDs\ have uncertainties of less than 6\% and 2\%, respectively.
Calculations~\cite{Tanaka:1994ay,Itoh:2004ye,Nierste:2008qe,Tanaka:2010se,Fajfer:2012vx} 
based on two-Higgs-doublet models predict a
substantial impact on the ratio \RD, and a smaller effect on \RDs\ due to the spin of the \Dstar meson.

The decay \BDstaunu was first observed in 2007 by the Belle Collaboration~\cite{Matyja:2007kt}.
Since then, both \babar\ and Belle have published improved measurements, 
and have found evidence for \BDtaunu decays~\cite{Aubert:2007dsa,Adachi:2009qg,Bozek:2010xy}. 
Up to now, the measured values for \RD\ and \RDs\ have consistently exceeded the 
SM expectations, though the significance of the excess is low due to the large statistical uncertainties. 

We recently presented an update of the earlier measurement~\cite{Aubert:2007dsa} based on the full 
\babar\ data sample~\cite{Lees:2012xj}. This update included improvements to the event reconstruction 
that increased the signal efficiency by more than a factor of 3.  In the following, we describe
the analysis in greater detail, present the distributions of some important 
kinematic variables, and expand the interpretation of the results.

We choose to reconstruct only the purely leptonic decays of the $\tau$ lepton,
$\tau^- \to e^- \nueb \nut$ and $\tau^- \to \mu^- \numb\nut$, so that \BDxtaunu 
and \BDxlnu decays are identified by the same particles in the final
state. This leads to the cancellation of various detection efficiencies and the reduction of
related uncertainties on the ratios \RDx. 

Candidate events originating from $\FourS\to\BB$ decays
are selected by fully reconstructing the hadronic decay  of one of the $B$ mesons (\Btag), and 
identifying the semileptonic decay of the other $B$ by a charm meson 
(charged or neutral $D$ or $D^*$  meson),
a charged lepton (either $e$ or $\mu$) and the missing momentum and energy in the whole event.

Yields for the signal decays \BDxtaunu\ and the normalization decays \BDxlnu\ are extracted 
by an unbinned maximum-likelihood fit to the two-dimensional distributions of the invariant mass 
of the undetected particles 
$\mmiss=p^2_{\rm miss}=(p_{\epem} - p_{B_{\rm tag}} - p_{D^{(*)}} - p_{\ell})^2$ 
(where $p_{\epem}$, $p_{B_{\rm tag}}$, $p_{D^{(*)}}$, and $p_{\ell}$ refer to the four-momenta of the colliding beams, the \Btag, the \ds, and the charged lepton, respectively) 
versus the lepton three-momentum in the $B$ rest frame, \pstarl.
The \mmiss distribution for decays with a single missing neutrino peaks at zero, whereas signal 
events, which have three missing neutrinos,
have a broad \mmiss distribution that extends to about $9 \gev ^2$.
The observed lepton in signal events is a secondary particle from the $\tau$ decay, so its 
\pstarl spectrum is softer than for primary leptons in normalization decays.

The principal sources of background originate from \BB\ decays and
from continuum events, {\it i.e.,}  $\epem\to f\overline{f}(\gamma)$ pair production, 
where $f=u,d,s,c,\tau$. The yields and distributions of these two background sources 
are derived from selected data control samples. The background decays that are most difficult 
to separate from signal decays come from semileptonic decays to higher-mass, excited charm mesons, 
since they can produce similar \mmiss and \pstarl values to signal decays and their branching fractions 
and decay properties are not well known.  
Thus, their impact on the signal yield is examined in detail.

The choice of the selection criteria and fit configuration are based on samples of simulated and data 
events. To avoid bias in the determination of the signal yield, the signal region  was blinded
for data until the analysis procedure was settled.

\section{Theory of $\boldsymbol{\BDxtaunu}$ Decays } 
\label{sec:Theory}

\subsection{Standard Model}
\label{sec:SM}
\begin{figure}\begin{center}
\includegraphics[width=2.2in]{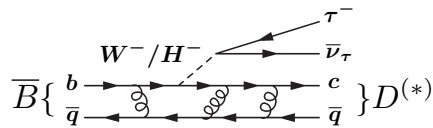}
\caption{Parton level diagram for \BDxtaunu decays. The gluon lines illustrate the QCD interactions that 
affect the hadronic part of the amplitude. }
\label{fig:Parton}
\end{center}\end{figure}

Given that leptons are not affected by 
quantum chromodynamic (QCD) interactions (see Fig.~\ref{fig:Parton}), 
the matrix element of \BDxtaunu decays can be factorized in the form \cite{Tanaka:1994ay}
\begin{equation}\label{eq:mat_w}
\mathcal M^{\lambda_\tau}_{\lambda_{\ds}}(q^2,\theta_\tau) =
  \frac{G_F V_{cb}}{\sqrt{2}}  \sum_{\lambda_W}
  \eta_{\lambda_W}L^{\lambda_\tau}_{\lambda_W}(q^2,\theta_\tau)H^{\lambda_{\ds}}_{\lambda_W}(q^2),
\end{equation}
where $L^{\lambda_\tau}_{\lambda_W}$ and $H^{\lambda_{\ds}}_{\lambda_W}$ are the leptonic and 
hadronic currents defined as
\begin{linenomath} 
\begin{align}
L^{\lambda_\tau}_{\lambda_W}(q^2,\theta_\tau) \; \equiv  & \;\;\epsilon_\mu(\lambda_W) \left< 
\tau\; \nutb| \overline{\tau}\; \gamma^\mu(1-\gamma_5)\; \nu_\tau|0 \right>, \\
H^{\lambda_{\ds}}_{\lambda_W}(q^2) \; \equiv & \;\;\epsilon^*_\mu(\lambda_W) \left< \ds\;
| \overline{c}\; \gamma^\mu(1-\gamma_5)\; b|\overline{B} \right>. \label{eq:HAmplitudes}
\end{align}
\end{linenomath} 
Here, the indices $\lambda$ refer to the helicities of the $W$, \ds, and $\tau$, 
$q=p_B-p_{\ds}$ is the four-momentum of the virtual $W$, and $\theta_\tau$ is the angle between the $\tau$
and the \ds three-momenta measured in the rest frame of the virtual $W$.
The metric factor $\eta$ in Eq.~\ref{eq:mat_w} is $\eta_{\{\pm,0,s\}}= \{1,1,-1\}$,
where $\lambda_W=\pm$, 0, and $s$ refer to the four helicity states of the virtual $W$ boson ($s$ is the 
scalar state which, of course, has helicity 0).

The  leptonic currents can be calculated analytically with the standard framework of electroweak interactions. 
In the rest frame of the virtual $W$ ($W^*$), they take the form  \cite{Hagiwara:1989cu}:
\begin{linenomath} 
\begin{align}
L^-_\pm &= -2\sqrt{q^2}vd_\pm,
&L^+_\pm &= \mp\sqrt{2}m_\tau vd_0,  \label{eq:Lpm}\\
L^-_0 &= -2\sqrt{q^2}vd_0,
&L^+_0 &= \sqrt{2}m_\tau v(d_+-d_-),   \label{eq:Lz}\\
L^-_s &= 0 , 
&L^+_s &= -2m_\tau v, \label{eq:Ls}
\end{align}
\end{linenomath} 
with
\begin{equation}
v = \sqrt{1-\frac{m_\tau^2}{q^2}}, \hspace{4mm}
d_\pm = \frac{1\pm\cos\theta_\tau}{\sqrt{2}},\hspace{4mm}
d_0 =\sin\theta_\tau.
\end{equation}
Given that the average $q^2$ in \BDxtaunu decays is about 8 $\gev^2$, the fraction of  \taum 
leptons with positive helicity is about 30\% in the SM.

Due to the nonperturbative nature of the QCD interaction at this energy scale,
the hadronic currents cannot be calculated analytically.  They are 
expressed in terms of form factors (FF) as functions of $q^2$ (see Secs.~\ref{sec:Theory:FFDs}
and \ref{sec:Theory:FFD}).

The differential decay rate, integrated over angles,
is derived from Eq.~\ref{eq:mat_w} and  Eqs..\ref{eq:Lpm}--\ref{eq:Ls}~ \cite{Korner:1989qb}:
\begin{linenomath} 
\begin{align}
 \frac{{\rm d}\Gamma_\tau}{{\rm d}q^2} = &
\frac{G_F^2\; |V_{cb}|^2\; \pDx \; q^2}{96\pi^3\mB^2}  \left(1-\frac{m_\tau^2}{q^2} \right)^2 
\Bigl[(|H_{+}|^2+|H_{-}|^2 \nonumber \\
&+|H_{0}|^2)  \left(1+\frac{m_\tau^2}{2q^2}  \right)
 + \frac{3 m_\tau^2}{2q^2}|H_{s}|^2  \Bigr],
 \label{eq:Gamma_q2}
\end{align}
\end{linenomath} 
where \pDx is the three-momentum of the \ds meson in the $B$ rest frame. For simplicity,  the helicities of 
the \ds meson and the $q^2$ dependence of the hadron helicity amplitudes $H_{\pm,0,s}$  have been 
omitted. The assignment is unambiguous because in \BDstaunu decays, 
$H_\pm$ only receive contributions from 
$\lambda_{\Dstar}=\pm$, while $H_{0,s}$ require $\lambda_{\Dstar}=0$. In
\BDtaunu decays, only $\lambda_{D}=s$ is possible, which implies $H_\pm=0$.

\subsubsection{Form factor parameterization of \BDstaunu decays} 
\label{sec:Theory:FFDs}
Four independent FFs, $V$, $A_0$, $A_1$, and $A_2$, describe the non-perturbative 
QCD interactions in \BDstaunu decays. Based on the FF convention of Ref.~\cite{Fajfer:2012vx}, 
the hadronic currents take the following form:
\begin{linenomath} 
\begin{align}
H_{\pm}(q^2)&=(m_B+m_{D^*})A_1(q^2)\mp\frac{2m_B}{m_B+m_{D^*}}\pDs V(q^2),\nonumber\\
H_{0}(q^2)&=\frac{-1}{2m_{D^*}\sqrt{q^2}}\Biggl[ \frac{4m_B^2 \pDs^2}{m_B+m_{D^*}}A_2 (q^2)  \nonumber\\
&-(m_B^2-m_{D^*}^2-q^2)(m_B+m_{D^*})A_1(q^2)\Biggr], \nonumber\\
H_{s}(q^2)&=\frac{2m_B \pDs}{\sqrt{q^2}}A_0(q^2)\,.
\end{align}
\end{linenomath} 
In this analysis, we use an HQET-based parameterizations for the FFs that is expressed in 
terms of the scalar product  of the $B$ and \Dstar four-velocities
\begin{equation} \label{eq:wDs}
w \equiv v_B\cdot v_{\Dstar} = \frac{\mB^2+\mDs^2-q^2}{2\mDs\mB} .
\end{equation}
Its minimum value $w_{\rm min}=1$ corresponds to $q^2_{\rm max} = (m_B-\mDs)^2$.
The maximum value is obtained for the lowest possible value of $q^2$, which is the square of the
mass of the lepton. 
Thus, $w_{\rm max}=1.35$ for \BDstaunu decays and $w_{\rm max}=1.51$ for \BDslnu decays.

In this framework, the FFs are usually expressed in terms of a
universal form factor $h_{A_1}(w)$ and ratios $R_i(w)$:
\begin{linenomath} 
\begin{align}
A_1(w) &= \frac{w+1}{2} r_{\Dstar} h_{A_1}(w), &
A_0(w) &= \frac{R_0(w)}{r_{\Dstar}} h_{A_1}(w), \nonumber \\
A_2(w) &= \frac{R_2(w)}{r_{\Dstar}} h_{A_1}(w), &
V(w) &= \frac{R_1(w)}{r_{\Dstar}} h_{A_1}(w), \nonumber 
\end{align}
\end{linenomath} 
where $r_{\Dstar}=2\sqrt{\mB\mDs}/(\mB+\mDs)$. Using dispersion relations and analyticity constraints
\cite{Caprini:1997mu,Fajfer:2012vx}, the universal FF and the ratios can be expressed in terms
of just five parameters:
\begin{linenomath} 
\begin{align}
h_{A_1}(w) &= h_{A_1}(1) \; [1-8\rho^2_{\Dstar} z(w)+(53\rho^2_{\Dstar}-15)z(w)^2 \nonumber \\
& \hspace{17mm}- (231\rho^2_{\Dstar}-91) 
z(w)^3] , \nonumber \\
R_1(w) &= R_1(1) - 0.12(w-1)+0.05(w-1)^2, \nonumber \\
R_2(w) &= R_2(1) + 0.11(w-1)-0.06(w-1)^2, \nonumber \\
R_0(w) &= R_0(1) - 0.11(w-1)+0.01(w-1)^2. \nonumber 
\end{align}
\end{linenomath} 
Here, $z(w) = (\sqrt{w+1}-\sqrt{2})/(\sqrt{w+1}+\sqrt{2})$.
The factor $h_{A_1}(1)$ only affects the overall
normalization, so it cancels in the ratio \RDs.

Three of the remaining four FF parameters, $R_1(1)$, $R_2(1)$,  and 
$\rho_{\Dstar}^2$, have been measured in analyses of \BDslnu decays. 
The most recent averages by the Heavy Quark Averaging Group (HFAG)~\cite{Amhis:2012bh} 
and their correlations $C$ are:
\begin{linenomath} 
\begin{align}
\rho_{\Dstar}^2 = &1.207\pm 0.028, & C(\rho_{\Dstar}^2,R_1(1)) = &0.566, \nonumber \\
R_1(1) =& 1.401\pm0.033, & C(\rho_{\Dstar}^2,R_2(1)) = &-0.807,\nonumber  \\
R_2(1) = &0.854\pm0.020,     &C(R_1(1),R_2(1)) = &-0.758.\nonumber 
\end{align}
\end{linenomath} 

$R_0(w)$ affects the decay rate only  via the scalar hadronic amplitude $H_s(q^2)$. The corresponding leptonic amplitude $L_s(q^2,\theta_\tau)$ is helicity suppressed, 
{\it i.e.,} its rate is proportional to the mass of the lepton (Eq.~\ref{eq:Lz}).
As a result, \BDslnu decays are not sensitive to this FF, and $R_0(w)$ has not been measured.
We therefore rely on a theoretical estimate, $R_0(1)=1.14\pm0.07$,  based on HQET \cite{Fajfer:2012vx}.

\subsubsection{Form factor parameterization of \BDtaunu decays} 
\label{sec:Theory:FFD}
The non-perturbative QCD interactions in \BDtaunu decays are described by two independent
FFs, referred to as $V_1$ and $S_1$~\cite{Tanaka:2010se}.  
The helicity amplitudes take the form:
\begin{linenomath} 
\begin{align}
H_0(w) = & \sqrt{m_B m_D} \, \frac{m_B+m_D}{\sqrt{q^2(w)}}\sqrt{w^2-1}\,V_1(w),\\
H_s(w) = & \sqrt{m_B m_D} \, \frac{m_B-m_D}{\sqrt{q^2(w)}} (w+1)\,S_1(w).
\end{align}
\end{linenomath} 
The amplitudes corresponding to the helicities $\lambda_W=\pm$
vanish because the $D$ meson has spin 0. For this decay mode, the variable $w$ is defined as in 
Eq.~\ref{eq:wDs}, except that the \Dstar meson mass is replaced by the $D$ meson
mass $m_D$.

Taking into account dispersion relations \cite{Caprini:1997mu}, $V_1$ can be expressed as
\begin{linenomath} 
\begin{align}
V_1(w) =    V_1(1) \times [&1-8\rho_D^2 z(w)+(51\rho_D^2-10)z(w)^2 \nonumber \\
& - (252\rho_D^2-84)z(w)^3] ,
\end{align}
\end{linenomath} 
where $V_1(1)$ and $\rho_D^2$ are FF parameters. 
The normalization $V_1(1)$ cancels in the ratio \RD. Based on
\BDlnu decays, the average value of the shape parameter is $\rho_D^2 = 1.186 \pm 0.055$~\cite{Amhis:2012bh}.
As for \BDstaunu decays, the scalar hadronic amplitude is helicity suppressed and as a result,
$S_1(w)$ cannot be measured with \BDlnu decays.
We use instead the following estimate based on HQET~\cite{Tanaka:2010se}:
\begin{linenomath} 
\begin{align}
S_1(w)= V_1(w)\bigl\{1+\Delta[&-0.019+0.041(w-1) \nonumber \\
&-0.015(w-1)^2] \bigr\},
\end{align}
\end{linenomath} 
with $\Delta = 1\pm1$. 

We have employed this FF parameterization to generate \BDtaunu
and \BDlnu decays, as described in Sec.~\ref{sec:mc-ffparameters}.  Though we used the same 
FF definitions and parameters, we found a difference of 1\% between the value of
\RD\ that we obtained by integrating Eq.~\ref{eq:Gamma_q2} and the value quoted in Ref.~\cite{Tanaka:2010se}. 

On the other hand, if we adopt the FF parameters of Ref.~\cite{Kamenik:2008tj}, we perfectly 
reproduce the \RD\ predictions presented there. 
The translation of the FF parameterization of Ref.~\cite{Kamenik:2008tj} into standard
hadronic amplitudes is not straightforward, so we do not use 
these FFs in the Monte Carlo simulation. Since both parameterizations 
yield essentially identical $q^2$ spectra, they are equivalent with respect to Monte Carlo generation,
which is not sensitive to differences in normalization.

\subsubsection{SM calculation of \RDx\ and $q^2$ spectrum} 
\label{sec:Theory:SMRDx}

We determine the SM predictions for the 
ratios \RDx\ integrating the expression for the differential decay rate (Eq.~\ref{eq:Gamma_q2}) as follows:
\begin{equation}
\RDx 
\equiv \frac{{\cal B}(B\to \ds\tau\nu)}{{\cal B}(B\to \ds \ell\nu)}
= \frac{\int_{m^2_\tau}^{q^2_{\rm max}} \frac{{\rm d}\Gamma_\tau}{{\rm d}q^2}\; {\rm d} q^2}
{\int_{m^2_\ell}^{q^2_{\rm max}}  \frac{{\rm d}\Gamma_\ell}{{\rm d}q^2} \; {\rm d}q^2},
\label{eq:RDxRate}\end{equation}
with $q^2_{\rm max}=(\mB-m_{\ds})^2$.

The uncertainty of this calculation is determined by generating one million random sets of
values for all the FF parameters assuming Gaussian distributions for the uncertainties and including 
their correlations. We calculate \RDx\ with each set of values, and assign the root mean square (RMS) of 
its distribution as the uncertainty. We apply this procedure for \Bz and \Bm decays, and for $\ell=e$ 
and $\mu$, and average the four results to arrive at the following predictions,
\begin{linenomath} 
\begin{align}
\RD_{\rm SM} &= 0.297 \pm 0.017,\\
\RDs_{\rm SM} &= 0.252 \pm 0.003. 
\end{align}
\end{linenomath} 
Additional uncertainties that have not been taken into account could contribute at 
the percent level. For instance, some electromagnetic corrections could affect \BDxlnu and \BDxtaunu 
decays differently \cite{Fajfer:2012vx}. 
The experimental uncertainty on \RDx\ is expected to be considerably larger.

\begin{figure}[t]
\psfrag{De}[Bl]{$\ds e\nu$}
\psfrag{Dmu}[Bl]{$\ds \mu\nu$}
\psfrag{Dtau}[Bl]{$\ds \tau\nu$}
\psfrag{x}[Br]{\footnotesize{${\rm d}\Gamma/{\rm d}q^2 (10^{16}\gev^{-1})$}}
\psfrag{q}[Bc]{\footnotesize{$q^2$ (GeV$^2$)}}
\includegraphics[width=3.4in]{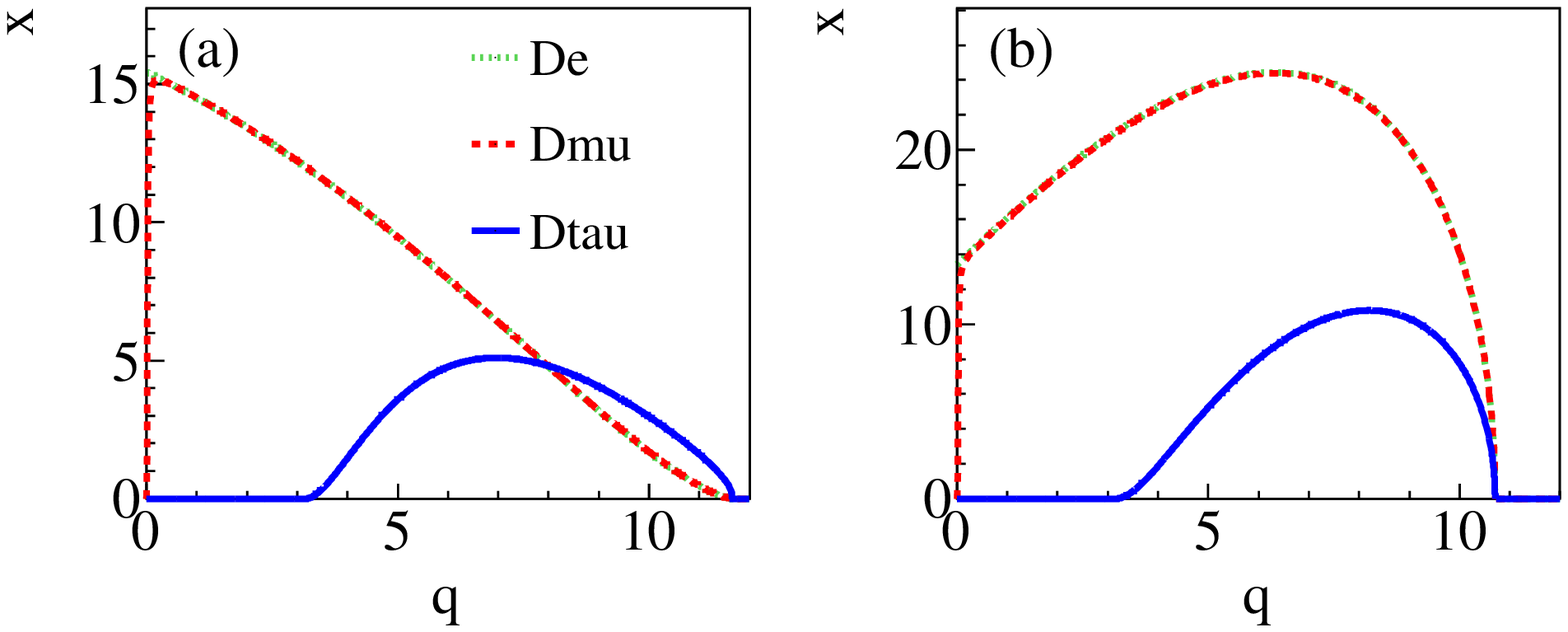} 
\caption{(Color online). \label{fig:SMq2} Predicted $q^2$ spectra for (a) \BDtaunu\ and \BDlnu\ decays 
for $V_1(1)V_{cb} = 0.0427$ and (b) \BDstaunu\ and \BDslnu\ decays for 
$h_{A_1}(1)V_{cb} = 0.0359$~\cite{Amhis:2012bh}.}
\end{figure}

The $q^2$ spectra for \BDxtaunu decays in Fig.~\ref{fig:SMq2} clearly show the threshold
 at $q^2_{\rm min}=m^2_\tau$, while for \BDxlnu decays  $q^2_{\rm min}\sim 0$. 
We take advantage of this difference in the signal selection by imposing $q^2>4\gev^2$. 
The spectra for $\ell=e$ and $\mu$ are almost identical, except for $q^2<m^2_\mu=0.011\gev^2$.

\subsection{Two-Higgs-Doublet Model Type II} \label{sec:theory:2hdm}
As we noted in the introduction, \BDxtaunu decays are potentially sensitive to 
new physics (NP) processes. 
Of particular interest is the two-Higgs-doublet model (2HDM) of type II, which describes the 
Higgs sector of the Minimal Supersymmetric model at tree level. 
In this model, one of the two Higgs doublets couples to up-type quarks, 
while the other doublet couples to down-type quarks and leptons.

The contributions of the charged Higgs  to \BDxtaunu\ decays can be encapsulated in the scalar helicity 
amplitude in the following way \cite{Tanaka:1994ay, Kamenik:2008tj}:
\begin{equation}\label{eq:Hsapprox}
H^{\rm 2HDM}_{s} \approx H^{\rm SM}_{s} \times \left( 1-\frac{{\rm tan}^2\beta}{m_{H^\pm}^2}\frac{q^2}{1\mp 
m_c/m_b} \right).
\end{equation}
Here, $\tanB$ is the ratio of the vacuum expectation values of the two Higgs doublets, 
$m_{H^\pm}$ is the
mass of the charged Higgs, and $m_c/m_b = 0.215\pm0.027$ \cite{Xing:2007fb} is the ratio of
the  $c$- and $b$-quark masses at a common mass scale. 
The negative sign in Eq.~\ref{eq:Hsapprox} applies to \BDtaunu decays and the positive sign 
applies to \BDstaunu decays.
This expression is accurate to 1\% for $\mH$ larger than $15\gev$.
The region for $\mH\leq 15\gev$ has already been excluded by $B\to X_s\gamma$ 
measurements~\cite{Misiak:2006zs}.

The \tBmH dependence of the ratios \RDx\ in the type II 2HDM can be studied by substituting
$H^{\rm 2HDM}_{s}$ for $H^{\rm SM}_{s}$ in Eq.~\ref{eq:Gamma_q2}.
Given that charged Higgs bosons are not expected to contribute significantly to \BDxlnu\  decays,
$\RDx_{\rm 2HDM}$ can be described by a parabola
in the variable ${\rm tan}^2\beta/m_{H^+}^2$,
\begin{equation}
\RDx_{\rm 2HDM}=\RDx_{\rm SM}+A_{\ds} \frac{{\rm tan}^2\beta}{m^2_{H^+}}+
B_{\ds} \frac{{\rm tan}^4\beta}{m^4_{H^+}}, \label{eq:RDNP}
\end{equation}
Table \ref{tab:2HDMCurves} lists the values of $A_{\ds}$ and $B_{\ds}$, which are determined 
by  averaging over \Bz and \Bm decays.
The uncertainty estimation includes the uncertainties on the mass ratio
 $m_c/m_b$ and the FF parameters, as well as their correlations.

\begin{table}
\caption{Dependence of  \RDx\ on  \tBmH in the 2HDM according to Eq.~\ref{eq:RDNP} for \BDtaunu and \BDstaunu decays: the values of \RDx, the parameters A and B  with their uncertainties, and correlations $C$.} 
\label{tab:2HDMCurves} 
\begin{tabular}{l  r @{ $\pm$ } l  r @{ $\pm$ } l}\hline\hline
&  \multicolumn{2}{c}{\BDtaunu} & \multicolumn{2}{c}{\BDstaunu} \\ \hline
$\RDx_{\rm SM}$			& 0.297 		& 0.017 	& 0.252 		& 0.003 \\
$A_{\ds}$ (GeV$^2$)		& $-3.25$		& 0.32	& $-0.230$	& 0.029 \\
$B_{\ds}$ (GeV$^4$)		& 16.9		& 2.0		&0.643		& 0.085 \\ \hline
$C(\RDx_{\rm SM}, A_{\ds})$	& \multicolumn{2}{c}{$-0.928$}		& \multicolumn{2}{c}{$-0.946$} \\
$C(\RDx_{\rm SM},B_{\ds})$	& \multicolumn{2}{c}{$0.789$}		& \multicolumn{2}{c}{$0.904$} \\
$C(A_{\ds},B_{\ds})$	& \multicolumn{2}{c}{$-0.957$}		& \multicolumn{2}{c}{$-0.985$} \\
\hline\hline
\end{tabular}
\end{table}

Due to the destructive interference between the SM and 2HDM amplitudes in Eq.~\ref{eq:Hsapprox},
charged Higgs contributions depress the ratios \RDx\ for low values of \tBmH.
For larger values of \tBmH, the Higgs contributions dominate and \RD\ and \RDs\ increase rapidly.
As the coefficients of Table \ref{tab:2HDMCurves} show, the 2HDM impact is expected to be 
larger for \RD\ than 
for \RDs. This is because charged Higgs contributions only affect the scalar amplitude $H_s^{\rm 2HDM}$, but
\BDstaunu decays also receive contributions from $H_\pm$, diluting the effect on the total rate.

\begin{figure}[t]
\psfrag{x}[Br]{\footnotesize{${\rm d}\Gamma/{\rm d}q^2\,\, (\gev^{-1})$}}
\psfrag{q}[Bc]{\footnotesize{$q^2$ (GeV$^2$)}}
\includegraphics[width=3.4in]{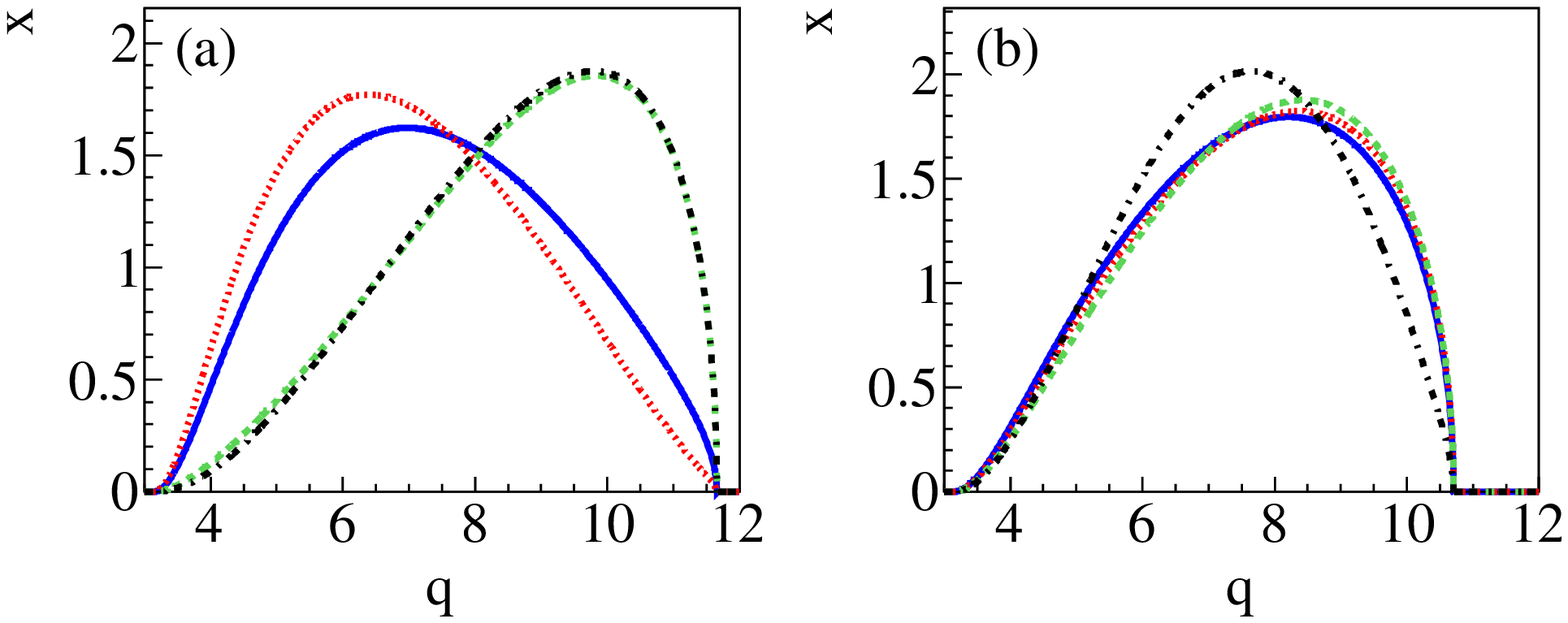}\\ \vspace{1mm}
\psfrag{a}[bl]{\footnotesize{ SM}}
\psfrag{b}[bl]{\footnotesize{ $\tBmH=0.3\gev^{-1}$}}
\psfrag{c}[bl]{\footnotesize{ $\tBmH=0.5\gev^{-1}$}}
\psfrag{d}[bl]{\footnotesize{ $\tBmH=1\gev^{-1}$}}
\includegraphics[width=3.4in]{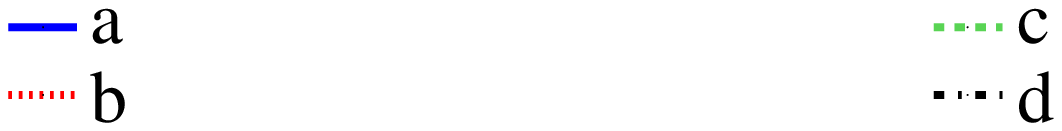} 
\caption{(Color online).  Predicted $q^2$ distributions for (a) \BDtaunu and (b) \BDstaunu decays for different values 
of \tBmH. All curves are normalized to unit area.}
\label{fig:2HDMq2}
\end{figure}

Figure \ref{fig:2HDMq2} shows the impact of the 2HDM on the $q^2$ spectrum. Given that
the $B$ and $D$ mesons have spin $J=0$, the SM decays $B\to D W^*\to D\tau\nu$ proceed via $P$-wave for
$J_{W^*}=1$, and via $S$-wave for $J_{W^*}=0$. For the $P$-wave decay, which accounts for about
96\% of the total amplitude, the decay rate receives an additional factor $\pD^2$, which suppresses the 
$q^2$ spectrum at high values.
Since charged Higgs bosons have $J_H=0$, their contributions proceed via $S$-wave, and, thus, 
have a larger average $q^2$ than the SM contributions.
As a result, for low values of $\tBmH$ where the negative interference depresses $H^{\rm 2HDM}_s$, 
the $q^2$ spectrum shifts to lower values.
For large values of \tBmH, the Higgs contributions dominate the decay rate and 
the average $q^2$ significantly exceeds that of the SM.

The situation is different for \BDstaunu\ decays because the \Dstar
meson has spin $J_{D^*}=1$. The SM decays can proceed via $S$, $P$, or $D$-waves, while
the decay via an intermediate Higgs  boson must proceed via $P$-wave,
suppressing the rate at high $q^2$. 

When searching for charged Higgs contributions, it is important to account for the 
changes in the $q^2$ spectrum. This distribution has a significant impact on the analysis due to the
close relation between $q^2$ and \mmiss, one of the fit variables.

Charged Higgs contributions also affect the \pstarl\ distribution.  
Given the spin 0 of the Higgs boson and the positive helicity (right-handedness) of the anti-neutrino,
the decays $H^-\to\taum\nutb$ always produce \taum\ leptons with positive helicities ($\lambda_{\tau}=+$). 
As a result, the fraction of right-handed \taum leptons produced in \BDxtaunu decays changes from 
30\% in the SM, to close to 100\% when the 2HDM contributions dominate.

The lepton spectrum of polarized $\tau^\pm\to\ell^\pm\nul\nut$ decays is well known~\cite{Tsai:1971vv}.
For $\taum$ leptons with $\lambda_{\tau^-}=-$, the \ellm is emitted preferentially in the 
$\taum$ direction, while the opposite is true for positive helicities. 
In the $B$ rest frame, leptons of a certain momentum in the $\taum$ rest frame have larger momentum 
if they are emitted in the direction of the \taum\ momentum than in the opposite direction. 
As a result, the \pstarl\ spectrum for SM decays is harder than
for Higgs dominated decays.
For low values of \tBmH for which the destructive interference depresses the \BDxtaunu rate,
the proportion of left-handed \taum leptons increases, and therefore, the 
\pstarl spectrum is harder than in the SM.

,\section{Data Sample, Detector and Simulation } \label{sec:detector}
\subsection{Data Sample} \label{sec:data}
This analysis is based on the full data sample recorded with the \babar\ detector~\cite{Aubert:2001tu} 
at the \pep2 energy-asymmetric $e^+e^-$ storage rings~\cite{Seeman:2008zz}.
It operated at a center-of-mass (c.m.) energy of 10.58 \gev, equal to the mass of the $\FourS$ resonance.  
This resonance decays almost exclusively to \BB\ pairs.
The collected data sample of 471 million $\FourS \to \BB$ events (on-peak data),
corresponds to an integrated luminosity of 426~$\mathrm{fb}^{-1}$ \cite{Lees:2013aa}.  
To study continuum background, 
an additional sample of $40\invfb$ (off-peak data) was recorded approximately 40 \mev below the 
\FourS resonance, {\it i.e.,} below the threshold for \BB\ production.

\subsection{The \babar\ Detector and Single Particle Reconstruction} \label{sec:babar}
The \babar\ detector and event reconstruction have been described in
detail elsewhere~\cite{Aubert:2001tu}.  The momentum and angles of charged particles 
were measured in a tracking system consisting of a 5-layer, double-sided silicon-strip 
detector (SVT) and a 40-layer, small-cell drift chamber (DCH) filled with a helium-isobutane 
gas mixture. Charged particles of different masses were distinguished by their
ionization energy loss in the tracking devices and by a ring-imaging Cerenkov
detector (DIRC). A finely segmented CsI(Tl) calorimeter (EMC) measured the energy and position 
of electromagnetic showers generated by electrons and photons.  The EMC was surrounded by 
a superconducting solenoid providing a 1.5-T magnetic field and by a segmented flux return with 
a hexagonal barrel section and two endcaps.
The steel of the flux return was instrumented (IFR) with resistive plate chambers and limited
streamer tubes to detect particles penetrating the magnet coil and steel.

Within the polar angle acceptance of the SVT and DCH ($0.4<\theta_{\mathrm lab}<2.6$)
the efficiency for the reconstruction of charged particles exceeds 99\% for momenta above 1 \gev. 
For low momentum pions, especially from  $\Dstarp \to D^0 \pi^+$ decays, the efficiency drops to 
about 90\% at 0.4 \gev and to 50\% at 0.1 \gev.  

The electron and muon identification efficiencies and the probabilities to misidentify a pion, a kaon, 
or a proton as an electron or a muon are measured as a function of the laboratory momentum and
angles using high-purity data samples.

Electrons are separated from charged hadrons primarily on the basis of the ratio of the energy 
deposited in the EMC to the track momentum.  A special algorithm has been developed to identify 
photons from bremsstrahlung in the inner detector, and to correct the electron momentum for the 
energy loss.  Within the polar angle acceptance, the average electron efficiency for laboratory 
momenta above 0.5 \gev is 97\%, largely independent of momentum.  The average pion 
misidentification rate is less than 0.5\%. 

Muon identification relies on a new multivariate algorithm that significantly increases the
reconstruction efficiency at low muon momenta, $|\boldsymbol{p}_\mu|<1\gev$. 
This algorithm combines information on the 
measured DCH track, the track segments in the IFR, and the energy deposited in the EMC.
The average muon efficiency is close to 90\% independent of momentum, except in the forward 
endcap, where it decreases for laboratory momenta below 1 \gev.  The average pion 
misidentification rate is about 2\%
above 1.2 \gev, rising at lower momenta and reaching a maximum of 9\% at 0.8 \gev. 

By choosing a fairly loose selection of charged leptons and taking advantage of improved PID algorithms, 
we increased the lepton efficiencies by 6\% for electrons and 50\% for muons compared to the previous \babar\ 
analysis~\cite{Aubert:2007dsa}. 

Charged kaons are identified up to 4 \gev\ on the basis of information from the DIRC, SVT, and DCH. 
The efficiency exceeds 80\% over most of the momentum range and varies with polar angle. 
The probability that a pion is misidentified as a kaon is close to 2\%, varying by about 1\% as 
a function of momentum and polar angle.

The decays $\KS \to \pi^+\pi^-$ are reconstructed as pairs of tracks of opposite charge originating 
from a displaced vertex.  The invariant mass of the pair $m_{\pi \pi}$ is required to be in the range 
$m_{\pi \pi} \in[0.491,0.506]\gev$. 
No attempt is made to identify interactions of \KL\ in the EMC or IFR.

To remove beam-generated background in the EMC and electronic  noise, photon candidates are 
required to have a minimum  energy of 30 \mev\ and a shower shape that is consistent with 
that of an electromagnetic shower. 
Neutral pions are reconstructed from pairs of photon candidates 
with an invariant mass in the range $m_{\gamma \gamma} \in[120, 150] \mev$.

\subsection{Monte Carlo Simulation}\label{sec:montecarlo}

\subsubsection{Simulated Samples} \label{sec:mc-sample}
This analysis relies on Monte Carlo (MC) techniques to simulate the production 
and decay of continuum and \BB events. The simulation is based on the EvtGen generator \cite{Lange:2001uf}.
The \qqbar fragmentation is performed by Jetset \cite{Sjostrand:1993yb}, and the detector response 
by Geant4 \cite{Agostinelli:2002hh}. Radiative effects such as bremsstrahlung in the detector material 
and initial-state and final-state radiation \cite{Barberio:1993qi} are included. 

We derive predictions for the distributions and efficiencies of the signal and backgrounds from the simulation. 
The size of the simulated sample of generic \BB\ events exceeds that of the \BB\ data sample by
about a factor of ten, while the sample for \qqbar\ events corresponds to twice the size of the
off-peak data sample.
We assume that the \FourS\ resonance decays exclusively to \BB\ pairs and use recent 
measurements of branching fractions~\cite{Nakamura:2010zzi} for all produced particles.  The
impact of their uncertainties on the final results is assessed as a systematic uncertainty. 

Information extracted from studies of selected data control samples is used to improve the accuracy of 
the simulation.  Specifically, we reweight simulated events to account for small differences 
observed in comparisons of data and simulation (Sec.~\ref{sec:Control_samples}).

\subsubsection{Implementation of the Form Factor Parameterizations}\label{sec:mc-ffparameters}
For reasons of simplicity, the simulation of \BDlnu and \BDxtaunu\ decays is based on the
ISGW2 model~\cite{Scora:1995ty}, and \BDslnu\ decays are generated using an HQET-based 
parameterization~\cite{Isgur:1989ed}. 
A change to a different FF parameterization is implemented  by reweighting the generated events 
with the weights
\begin{equation} 
\label{eq:FFWeightD}
    w_{\rm HQET}(q^2,\theta_i)= 
    \left( \frac{{\cal M}(q^2,\theta_i)_{\rm HQET}}
                {{\cal M}(q^2,\theta_i)_{\rm MC}} \right)^2 
         \times \frac{{\cal B}_{\rm MC}}{{\cal B}_{\rm HQET}}. 
\end{equation}
\noindent
Here, ${\cal M}(q^2,\theta_i)_{HQET}$ refers to the matrix element for the FF parameterizations
described in Secs. \ref{sec:Theory:FFDs} and \ref{sec:Theory:FFD}, and ${\cal M}(q^2,\theta_i)_{MC}$ 
is the matrix element employed in the MC generation. The matrix element of decays involving the 
scalar $D$ meson depends on one angular variable,  the lepton helicity angle 
$\theta_{\ell}$, with ${\ell = e,\mu,\tau}$. In addition to $\theta_\ell$, the matrix element of decays 
involving the vector meson $D^*$ is sensitive to two additional angular variables describing the \Dstar
decay.  
The ratio of the branching fractions ${\cal B}_{\rm MC} /{\cal B}_{\rm HQET}$ ensures that the sum of 
all weights  equals the number of generated events. 

In the SM, this reweighting results in a small shift of the $q^2$ distribution to higher values, while the 
changes in the helicity angle $\theta_{\tau}$  and the $\tau$ polarization are negligible.
Therefore, the distributions of the secondary charged lepton are not affected. 

In the presence of a charged Higgs boson, however, the $\tau$ polarization can change substantially, 
affecting the momentum of the secondary lepton $\ell$ originating from the $\tau\to\ell\nul\nut$ decays.
We account for the potential presence of a charged Higgs of 2HDM type II by reweighting 
the simulation with the following weights,
\begin{linenomath} 
\begin{align} 
\label{eq:FFWeightH}
w_{\rm 2HDM}(q^2,\theta_i,\pstarl) =& \left( \frac{{\cal M}(q^2,\theta_i)_{\rm 2HDM}} {{\cal M}(q^2,\theta_i)_{\rm MC}} \right)^2 \times \nonumber \\
& \frac{\Gamma(\pstarl)_{\rm 2HDM}}{\Gamma(\pstarl)_{\rm MC}} \times \frac{{\cal B}_{\rm MC}}{{\cal B}_{\rm 2HDM}}. 
\end{align}
\end{linenomath} 
where $\theta_i $ refers again to the angular variables. 
The second factor represents the ratio of the $\pstarl$ distributions $\Gamma(\pstarl)$ in the 2HDM 
parameterization and in the MC simulation. This factorization is necessary because in the MC 
generation the polarization is handled in a probabilistic manner, so it cannot be corrected on an
event-per-event basis. It is only applicable if  $\pstarl$  
is uncorrelated with $q^2$ and the angular variables, which is largely the case.
In some regions of phase space, the 2HDM weights have a much larger dispersion than the weights 
applied in the SM reweighting, leading to larger statistical uncertainties for the simulation of the 
Higgs boson contributions.

\subsubsection{Simulation of \BDssltnu  decays}\label{sec:mc-bf}
By \dss we refer to excited charm resonances heavier than the \Dstar meson.
We include in the simulation the \BDsstaunu and \BDsslnu decays that involve the four \dss states 
with $L=1$ that have been measured \cite{Amhis:2012bh}.
This simulation takes into account their helicities \cite{Leibovich:1997em} and the following decay modes:
$D^*_0, D^*_2 \to D \pi$ and  $D'_1, D_1, D^*_2 \to D^* \pi$.  
Three-body decays $D^{**} \to D^{(*)} \pi \pi$ are not included in the nominal fit
for lack of reliable measurements.

To estimate the rate of $B \to D^{**} \tau \nu_{\tau}$ decays, we rely on ratios of the available 
phase space $\Phi$, 
\begin{equation}
\RDss \equiv \frac{{\cal B} (\BDsstaunu)} {{\cal B} (\BDsslnu)} 
\approx \frac {\Phi (\BDsstaunu)} {\Phi (\BDsslnu)}. 
\end{equation}
The value of this ratio depends on the mass of the \dss state involved in the \BDssltnu decay.  
We use the largest of  the four possible choices, $\RDss=0.18$.

Possible contributions from non-resonant $\Bb \to D^{(*)} \pi (\pi) \ellm \nulb$ decays 
and semileptonic  decays involving higher-mass excited charm mesons are not included in the nominal fit, 
and will be treated as a systematic uncertainty.

\section{Event selection } 
\label{sed:Event_selection}
The event selection proceeds in two steps. First, we select \BB
events in which one of the $B$ mesons, the \Btag, is fully reconstructed in a hadronic  decay,
while the other $B$ meson decays semileptonically.  
To  increase the event selection efficiency compared to earlier analyses, we have added 
more decay chains to the $B_{\rm tag}$ selection and have chosen a looser charged lepton selection.  
This leads to significantly higher backgrounds,  primarily combinatorial background from \BB\ and 
continuum events, and charge-crossfeed events. Charge-crossfeed events are  
\BDxltnu\ decays in which the charge of the reconstructed $B_{\rm tag}$ 
and $D^{(*)}$ mesons are wrong, primarily because of an 
incorrectly assigned low-momentum $\pi^{\pm}$.

Semileptonic decays to higher mass charm mesons have a signature similar to that of signal events and 
their composition is not well measured.
This background is fitted in selected control samples that are enriched with these decays.

As the second step in the event selection, we introduce kinematic criteria that increase the fraction
of selected signal events with respect to normalization and background decays.
We also apply a multivariate algorithm to further improve the signal-to-background ratio.  

\subsection{Selection of Events with a ${\boldsymbol B_{\text{tag}}}$ and a 
Semileptonic $\boldsymbol{B}$ Decay}

$\Upsilon(4S)\to \BB$ events  are tagged by the hadronic decay of 
one of the $B$ mesons.
We use a semi-exclusive algorithm which includes additional \Btag decay chains and enhances 
the efficiency by a factor of 2 compared to  the earlier version employed by \babar~\cite{Aubert:2007dsa}.
We look for decays of the type $B_{\rm tag} \to S X^{\pm}$, where $S$ refers to a {\it seed} meson 
and $X^{\pm}$ is a charged state comprising of up to five hadrons, pions or kaons, among them up 
to two neutral mesons, $\pi^0$ or \KS. 
The seed mesons, $D$, $D^*$, $D_s$, $D_s^*$, and $\jpsi$, are reconstructed in 56 decay modes.
As a result, the \Btag is reconstructed in 1,680 different decay chains, which are further subdivided into
2,968 kinematic modes.

To isolate the true tag decays from combinatorial background, 
we use two kinematic variables: the energy substituted mass 
$m_{ES}=\sqrt{E^{2}_{\rm beam} - \mathbf{p}^{2}_{\rm tag}}$ 
and the energy difference $\Delta E = E_{\rm tag} - E_{\rm beam}$.  Here $\mathbf{p}_{\rm tag}$ 
and $E_{\rm tag}$ refer to the c.m.~momentum and energy of the $B_{\rm tag}$, and  $E_{\rm beam}$ 
is the c.m.~energy of a single beam particle.  
These variables make optimum use of the precisely known energies of the colliding beams. 
For correctly reconstructed $B$ decays, the $m_{ES}$ distribution is centered at the $B$-meson 
mass with a resolution of 2.5 \mev, while $\Delta E$ is centered at zero with a resolution of 
18\mev which is dominated by the detector resolution. We require $m_{ES} > 5.27 \gev$ and 
$|\Delta E| <0.072 \gev$. 

For each \Btag candidate in a selected event, we look for the signature of the semileptonic 
decay of the second $B$ meson, a $D$ or \Dstar meson and a charged lepton $\ell$. 
We combine charged \Btag candidates with $D^{(*)0}\ellm$ systems and neutral \Btag candidates 
with both $D^{(*)+}\ellm$ and $D^{(*)-}\ellp$ systems, where the inclusion of both charge combinations 
allows for neutral $B$ mixing.
We require all charged particles to be associated with the $\Btag\ds\ell$ candidate,
but we allow for any number of additional photons in the event.

The laboratory momentum of the electron 
or muon is required to exceed 300 \mev or 200 \mev, respectively.
For $D$ mesons, we reconstruct the following decay modes:
$\Dz \to \Km \pip$, $\Km \Kp$, $\Km \pip \piz$, $\Km \pip \pim \pip$, $\KS \pip \pim$, 
and $\Dp \to \Km \pip \pip$, 
$\Km \pip \pip \piz$, $\KS \pip$, $\KS \pip \pip \pim$, $\KS \pip \piz,$ $\KS \Kp,$ with $\KS \to \pi^+ \pi^-$.  
The reconstructed invariant mass of $D$ candidates is required to be consistent with the 
nominal $D$ mass to within four standard deviations ($\sigma$).
The combined reconstructed branching fractions are 35.8\% and 27.3\% for $\Dz$ and $\Dp$, respectively.
We identify $D^*$ mesons by their decays $D^{*+}\to D^0\pi^+,D^+\pi^0$, 
and $D^{*0}\to D^0 \pi^0,D^0\gamma$.  For these decays, the c.m. momentum of the pion or 
the c.m. energy of the photon are required to be less than 
400 \mev.  Furthermore, the mass difference $\Delta m = m(D^*)-m(D)$ is required to differ by 
less than 4$\sigma$ from the expected value \cite{Nakamura:2010zzi}.

To further reduce the combinatorial background, we perform a kinematic fit to the event, 
constraining  tracks of secondary charged particles to the appropriate $B$, \ds, or \KS\ decay vertices.   
The  fit also constrains the reconstructed masses of the $D$, $D^*$, and \KS\ mesons to their nominal values.
The vertex of the $\FourS \to \BB$ decay has to be compatible with a beam-beam interaction. 
Candidates for which this fit does not converge are rejected. 
The \mmiss resolution improves by about 25\% and becomes more symmetric for the remaining candidates.

To select a single \BB candidate, we determine $\eextra= \sum_{i} E_i^{\gamma}$, 
the sum of the energies  of all photons that are not associated with the reconstructed \BB\ pair.  
We only include photons of more than 50 \mev, thereby eliminating about $99 \%$  of the 
beam-generated background. We retain the candidate with the lowest value of \eextra, 
and if more than one candidate survives, we select the one with the smallest $|\Delta E|$.  
This procedure preferentially selects  $\Dstar\ell$ candidates over
$D\ell$ candidates. Thus, we reduce the fraction of misreconstructed
events with a $\Dstar\to D(\pi/\gamma)$ decay for which the pion or
photon is not properly assigned to the \Dstar meson.

As a consequence of the rather loose lepton selection criteria and the addition of decay modes with 
multiple neutral pions and $\KS$  for the \Btag selection, the number of 
$\Btag\ds\ell$ candidates per event is very large. To address this problem, we identify
the \Btag decay modes that contribute primarily to the combinatorial background.
Specifically, we determine for each of the 2,968 kinematic modes $R_{\rm tc}$, 
the fraction of events for which all charged particles in the \Btag final state are correctly reconstructed 
and associated with the tag decay.
This assessment is based on a large sample of simulated \BB\ events equivalent to 700 \invfb.  
We observe that for decay chains with low multiplicity final states and no neutral hadrons the 
signal-to-background ratio ($S/B$) is very high. 
For instance,  for the $B^-_{\rm tag} \to \jpsi (\to \mu^+ \mu^-) \Km$ decay, we obtain 
$S/B = 316/79$, whereas for the decay 
$\Bz_{\rm tag} \to D^-(\to \KS \pi^-) \pi^+\pi^+ \pi^+ \pi^-\pi^-$ this 
ratio is $S/B=20/145$.  For this decay mode, typically 3.5  of the 8 \Btag final state particles 
are incorrectly associated with the second $B$ decay in the event or otherwise misidentified.  
Based on this study, we only retain $B_{\rm tag}$ decay chains with $R_{\rm tc}> 0.3$.  With this criterion, 
we remove 2100 $B_{\rm tag}$ kinematic modes, eliminate 2/3 of the combinatorial background, 
and retain 85\% of the signal $\BDxtaunu$ decays.
Thanks to this procedure, the average number of candidates per event before single candidate selection 
is reduced to 1.8 for the \Dzl and \Dpl samples,  and 3.1 and 4.8 for the \Dszl and \Dspl samples, respectively.

\subsection{Selection of the $\boldsymbol{\dspizl}$ Control Samples}

To constrain the \BDssltnu background, we select four \dspizl control samples, identical to the \dsl samples
except for an additional reconstructed \piz. The \piz is selected in the mass range 
$m_{\gamma \gamma} \in[120, 150] \mev$. 
Decays of the form $B\to\ds\pi\ell\nu$ peak at $\mmiss=0$ in these samples.
As a result, we can extract their yields together with the signal and normalization yields by
fitting the \dsl and \dspizl samples simultaneously.

More than half of the events in these control samples originate from continuum
$\epem \to \qqbar (\gamma)$ events.  Since the fragmentation of light quarks leads to a 
two-jet event topology, this background is very effectively suppressed by the requirement
$|\cos \Delta \theta_{\rm thrust}|<0.8$, where $\Delta \theta_{\rm thrust}$ is the angle between the 
thrust axes of the \Btag and of the rest of the event.  
Since $B$ mesons originating from \FourS\ decays are produced just above threshold, 
their final state particles are emitted almost isotropically, and, therefore, the 
$\cos \Delta \theta_{\rm thrust}$ distribution is uniform.  As a result,  the loss of \BDssltnu decays 
due to this restriction is significantly smaller than the amount of continuum events rejected.

\begin{figure}
\psfrag{a}[Bl]{\footnotesize{\Dxltnu}}
\psfrag{b}[Bl]{\footnotesize{\Dssltnu}}
\psfrag{c}[Bl]{\footnotesize{\BB bkg.}}
\psfrag{d}[Bl]{\footnotesize{Cont.~bkg.}}
\psfrag{t0}[Bc]{\footnotesize{\eextra (GeV)}}
\psfrag{t1}[Bc]{\footnotesize{\DeltaE (GeV)}}
\psfrag{t2}[Bc]{\footnotesize{$m(D^{*0}_{\rm sig})$ (GeV)}}
\psfrag{t3}[Bc]{\footnotesize{$\Delta m_{\rm sig}$ (GeV)}}
\psfrag{t4}[Bc]{\footnotesize{$m(D_{\rm tag})$ (GeV)}}
\psfrag{t5}[Bc]{\footnotesize{$\Delta m_{\rm tag}$ (GeV)}}
\psfrag{t6}[Bc]{\footnotesize{\Btag charged multiplicity}}
\psfrag{t7}[Bc]{\footnotesize{$\cos \Delta \theta_{\rm thrust}$}}
\includegraphics[width=3.4in]{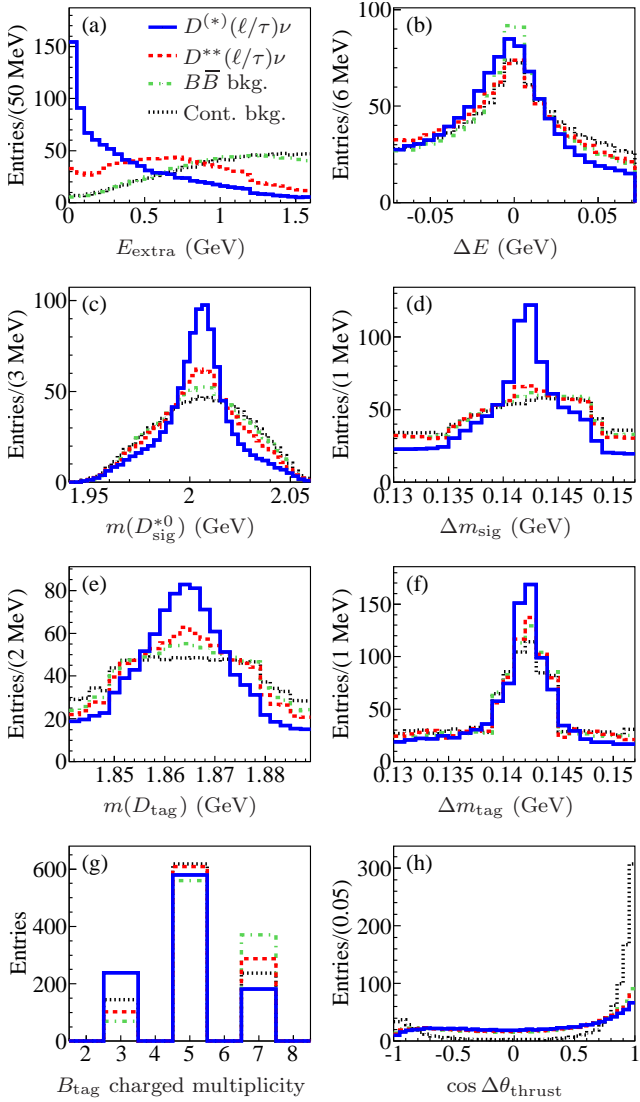}
\caption{(Color online). Input variables for the BDT selector trained on the $\Dstarz\ell$ sample.
Histograms are normalized to 1000 entries.} \label{fig:EvtSel_BDTVar}
\end{figure}

\subsection{Optimization of the Signal Selection} 
We introduce criteria that discriminate signal from background, and also differentiate between 
signal \BDxtaunu\ and \BDxlnu\ decays.
For semileptonic decays the minimum momentum transfer is largely determined by the mass 
of the charged lepton.  For decays involving $\tau$ leptons, 
$q^2_{\rm min}= m^2_{\tau}\simeq 3.16 \gev^2$.  Thus the 
selection $q^2 > 4 \gev^2$ retains 98\% of the \BDxtaunu\ decays and rejects more that 30\% 
of the \BDxlnu\ decays. The event sample with 
$q^2 < 4 \gev^2$ is dominated by \BDxlnu\ and serves as a very clean data sample for 
comparisons with the MC simulation.
To reject background from hadronic $B$ decays in which a pion is misidentified as muon, 
we require $|\boldsymbol{p}_{\rm miss}|>200\mev$, where
 $|\boldsymbol{p}_{\rm miss}|$ is the missing momentum in the c.m. frame.

\begin{figure*} 
\psfrag{Dl}[Bl]{\footnotesize{$D(\ell/\tau)\nu$}}
\psfrag{Dsl}[Bl]{\footnotesize{$\Dstar(\ell/\tau)\nu$}}
\psfrag{Dssl}[Bl]{\footnotesize{$D^{**}(\ell/\tau)\nu$}}
\psfrag{BB}[Bl]{\footnotesize{\BB bkg.}}
\psfrag{uds}[Bl]{\footnotesize{Cont.~bkg.}}
\psfrag{x0}[bc]{\small{\pstarl (GeV)}}
\psfrag{x1}[bc]{\small{\pstarl (GeV)}}
\psfrag{x2}[bc]{\small{\pstarl (GeV)}}
\psfrag{x3}[bc]{\small{\eextra (GeV)}}
\psfrag{x4}[bc]{\small{\mes (GeV)}}
\psfrag{x5}[bc]{\small{\mes (GeV)}}
\includegraphics[width=7in]{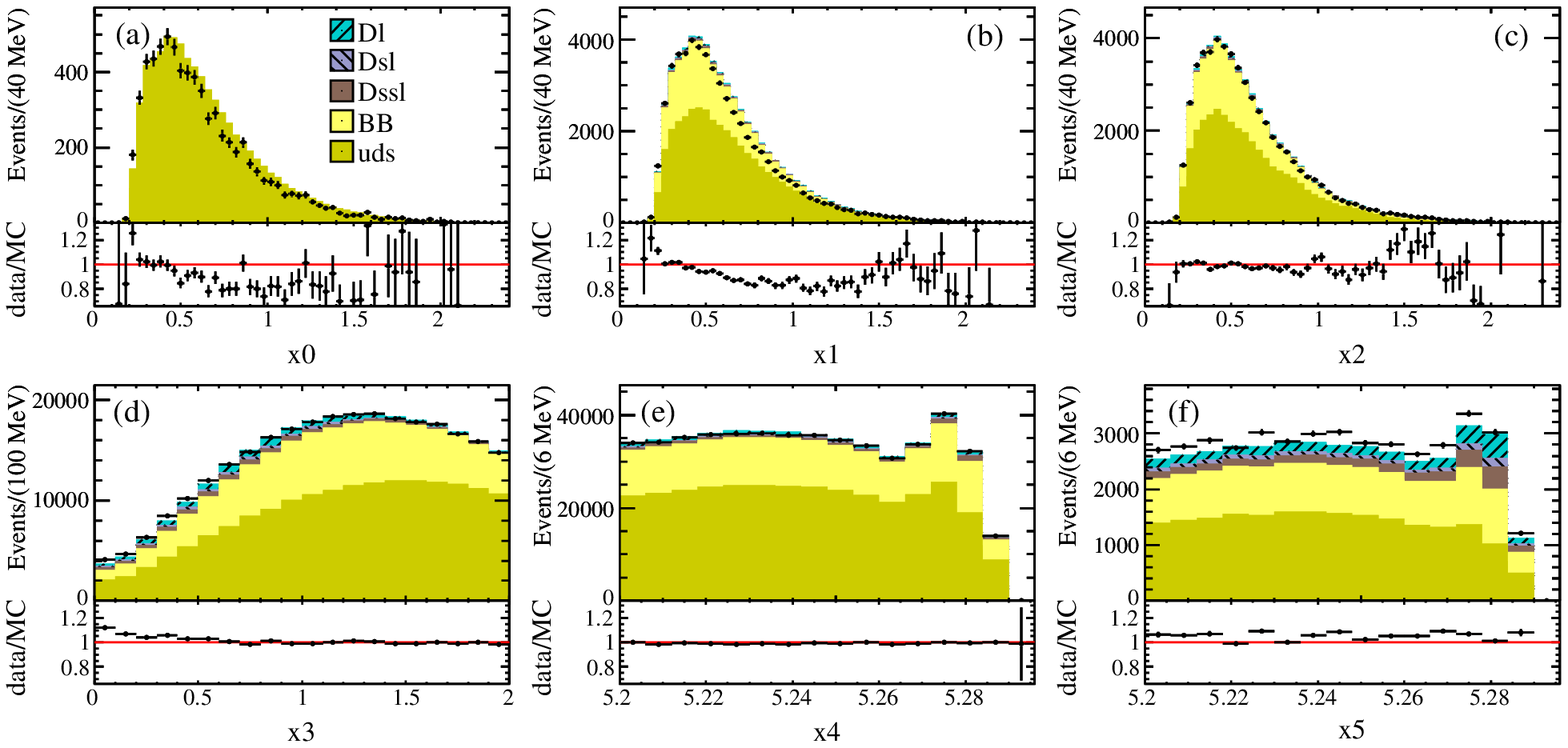}
\caption{(Color online). Comparison of data control samples (data points) with MC simulated 
samples (histograms) of
the \pstarl\ distributions for
(a) off-peak data prior to \pstarl reweighting,
(b) the intermediate \eextra  sample prior to \pstarl reweighting, and  
(c) the intermediate \eextra\ sample after \pstarl reweighting; 
(d) the \eextra\ distribution  for the combinatorial background;  
and the \mes distributions for
(e) the intermediate \eextra\ sample, and
(f) the low \eextra\ sample.
The results are shown for the four \dsl samples combined.}
\label{fig:Control_Samples}
\end{figure*}

To further improve the separation of well-reconstructed signal and normalization 
decays from various backgrounds, we employ a boosted 
decision tree (BDT) multivariate method~\cite{BDT}.  
This method relies on simple classifiers which determine signal and background regions 
by using binary selections on various input distributions. 
For each of the four $\ds\ell$ samples, we train a BDT to select 
signal and normalization events and reject \BDssltnu and charge cross-feed backgrounds.  
Each BDT selector relies on the simulated distributions of the following variables:
(a) $\eextra$; 
(b) $\Delta E$; 
(c) the reconstructed mass of the signal \ds meson; 
(d) the mass difference for the reconstructed signal $D^*$: $\Delta m= m(D\pi) - m(D)$; 
(e) the reconstructed mass of the seed meson of the \Btag; 
(f) the mass difference for a $D^*$ originating from the \Btag, $\Delta m_{\rm tag}= m(D_{\rm tag}\pi) - 
m(D_{\rm tag})$;
(g) the charged particle multiplicity of the \Btag candidate; and
(h) $\cos \Delta \theta_{\rm thrust}$. 
The input distributions for one of the BDT selectors are shown in Fig.~\ref{fig:EvtSel_BDTVar}.
For the \dspizl samples, we use similar BDT selectors that are trained to reject continuum, 
$\ds(\ell/\tau)\nu$, and other \BB background.
After the BDT requirements are applied, the fraction of events attributed to signal in the 
$\mmiss>1.5\gev^2$ region, which excludes most of the
normalization decays, increases from 2\% to 39\%.
The background remaining in that region is composed of normalization events (10\%), 
continuum (19\%), \Dsslnu events (13\%), and other 
\BB events (19\%), primarily from $B\to\ds D^{(*)+}_s$ decays with $\Ds \to \taup\nut$.

\section{Correction and Validation of the MC Simulation} 
\label{sec:Control_samples}

The simulation of the full reconstruction of high-multiplicity events, including 
the veto of events with extra tracks or higher values of \eextra\ is a rather challenging task.
To validate the simulation, we compare simulated distributions with data control samples,
and, when necessary, correct the MC simulations for the observed differences. 
The figures shown in this section combine events from all four channels (\Dzl, \Dszl, \Dpl, and \Dspl);
the observed differences are similar in the individual samples.

The control samples are selected to have little or no contamination from signal decays.
Specifically we select,
\begin{itemize}
\item{\it Continuum events:} off-peak data.
\item{\it Normalization decays:} $q^2\leq 4\gev^2$.
\item{\it Combinatorial \BB\ and continuum backgrounds:} $5.20 < \mes < 5.26 \gev$.
\item{\it Incorrectly reconstructed events:}  events in
three \eextra\ intervals, {\it high} ($1.2<\eextra<2.4\gev$), {\it intermediate} 
($0.5<\eextra<1.2\gev$), and {\it low} ($\eextra < 0.5\gev$ for events that fail the 
BDT selection). 
N.B.~the BDT selection results in the elimination of all events with $\eextra>0.4\gev$.
\end{itemize}

The off-peak data sample confirms the \mmiss distribution of simulated continuum events, but 
shows discrepancies in the \pstarl\ spectrum and overall normalization of the simulation
[Fig.~\ref{fig:Control_Samples}(a)]. These features  are also observed in other control samples,
such as on-peak data with high \eextra [Fig.~\ref{fig:Control_Samples}(b)]. 
We correct the simulated \pstarl\ spectrum and yield of the continuum contribution by reweighting it to match 
off-peak data, on an event-by-event basis. 
After this correction, the \pstarl\ distributions of the expected backgrounds agree well 
in independent control samples down to low lepton momenta where the misidentification
rates are significant [Fig.~\ref{fig:Control_Samples}(c)].
We observe that in the high \eextra region, the simulation exceeds data yield by $( 1.3\pm 0.5)$\%.
This small excess is corrected by decreasing the expected \BB\ background yield by $(4.3\pm1.9)\%$. 
After this correction, the simulation provides accurate yield predictions for the backgrounds
at intermediate and high \eextra. For instance, the ratio of the expected to observed yield 
of events with $\mmiss>1.5\gev^2$ is $0.998\pm0.006$.
The \mmiss distributions of the continuum and \BB backgrounds are described well in all control samples.

The region of low \eextra, which includes the signal region, is more difficult to model, 
primarily due to low energy photons and \KL mesons interacting in the EMC. 
Figure \ref{fig:Control_Samples}(d) shows that the data in the \mes\ sideband agree well 
with the combinatorial background predictions for $\eextra>0.5\gev$,
but are underestimated for low \eextra.
This, and small differences in the other BDT input distributions,
result in a underestimation of the combinatorial background when the BDT requirements are applied.
Based on the  $5.20<\mes<5.26\gev$ sideband, we find scale factors of
$1.099\pm0.019$ and $1.047\pm0.034$ for the combinatorial background in the $D\ell$ and 
$\Dstar\ell$ samples, respectively. The uncertainties are given by the statistics of the data
and simulated samples.
The ratio between the observed and expected \mes distribution is independent of \eextra
[Figs.~\ref{fig:Control_Samples}(e,f)], so we apply these corrections to the continuum and
\BB backgrounds in the signal region.
The same correction is applied to \BDssltnu events, which cannot be easily isolated, because 
their simulated \eextra\ distributions are very similar to those of combinatorial background.  
These corrections affect the fixed \BB\ and continuum yields in the fit, as well as the relative
efficiency of \BDssltnu events in the \dsl and \dspizl samples. As a result, these corrections are the 
source of the dominant systematic uncertainties.

Relying on the $q^2\leq4\gev^2$ control sample, where \BDxlnu decays account for 96\% of the events,
we correct the \eextra\ distribution and an 8.5\% overestimation of the simulated  
normalization events. 
We apply the same correction to simulated signal events 
which are expected to have a similar \eextra\ distribution. 
This procedure does not affect the relative efficiency of signal to normalization events,
so it has a very small impact on the \RDx\ measurements.

We use the same $q^2\leq4\gev^2$ control sample to compare and validate the \pstarl distributions 
of \BDxlnu events. We observe that the \mmiss\ resolution of the narrow peaks at 
$\mmiss=0$ is slightly underestimated by the simulation. 
This effect is corrected by convolving the simulated distributions with a Gaussian 
resolution function, for which the width is adjusted by iteration.

\begin{table*}
\caption{Contributions to the four \dsl samples. The expected relative abundance of events 
in each data sample is represented by $f_\mathrm{exp}$. The columns labeled \emph{Yield} 
indicate whether the contribution is free in the fit, fixed, or linked to another component 
through a cross-feed
constraint. The charged cross-feed components, marked with \emph{Fix./It.}, are fixed in the
fit, but updated in the iterative process.} \label{tab:ContributionsDsl} 
\begin{tabular}{l |rrl |rrl |rrl |rrl}\hline\hline
& \multicolumn{3}{c|}{\Dzl} 	& \multicolumn{3}{c|}{\Dszl} 	& \multicolumn{3}{c|}{\Dpl} 	
& \multicolumn{3}{c}{\Dspl} 	\\ 
Source	\hspace{32mm} & \hspace{1mm}$f_\mathrm{exp}$ (\%) && Yield \hspace{3mm}	& \hspace{1mm}$f_\mathrm{exp}$ (\%) && Yield \hspace{3mm}	
& \hspace{1mm}$f_\mathrm{exp}$ (\%) && Yield \hspace{3mm}	& \hspace{1mm}$f_\mathrm{exp}$ (\%) && Yield \hspace{3mm}	\\ \hline
$\ds\tau\nu$  signal				& 2.6 	&& Free			& 4.9 	&& Free			& 4.3 	&& Free  	        		& 5.0 	&& Free \\		
$\ds\tau\nu$ signal feed-down/up	& 2.8 	&& $\Dstarz\ell$ 	& 0.4		&& $\Dz\ell$ 		& 1.8		&& $\Dstarp\ell$ 	& 0.1 	&& $\Dp\ell$ \\	
$\ds\ell\nu$  normalization		& 24.5 	&& Free 		  	& 80.7 	&& Free 	    	  	& 37.3 	&& Free	 		& 88.0	&& Free \\		             	  
$\ds\ell\nu$ norm.~feed-down/up	& 53.5 	&& Free 			& 2.7 	&& $\Dz\ell$ 		& 35.0	&& Free 	       		& 0.3 	&& $\Dp\ell$ \\		
\Dsslnu background				& 4.3 	&& $\Dz\piz\ell$ 	& 3.6 	&& $\Dstarz\piz\ell$	& 6.6		&& $\Dp\piz\ell$ 	& 3.0 	&& $\Dstarp\piz\ell$ \\	
Cross-feed background			& 3.8		&& Fix./It. 			& 1.3		&& Fix./It.  			& 2.1		&& Fix./It.  			& 0.4 	&& Fix./It.  \\		
\BB background				& 4.1		&& Fixed 		  	& 3.7 	&& Fixed  	    		& 7.1 	&& Fixed       		& 2.8 	&& Fixed \\ 			      	  
Continuum background			& 4.4		&& Fixed 			& 2.6		&& Fixed        	  	& 5.9		&& Fixed	   		& 0.5		&& Fixed \\ \hline\hline                    
\end{tabular}
\end{table*}

\begin{table*}
\caption{Contributions to the four \dspizl samples. The expected relative abundance of events 
in each data sample is represented by $f_\mathrm{exp}$. The columns labeled \emph{Yield} 
indicate whether the contribution is free in the fit, fixed, or linked to another component 
through a cross-feed constraint. The $D(\ell/\tau)\nu$ components are linked to the $\Dstar(\ell/\tau)\nu$ 
components, and the cross-feed constraint is updated in the iteration. 
The charged cross-feed components, marked with \emph{Fix./It.}, are fixed in the
fit, but updated in the iterative process.} \label{tab:ContributionsDspizl}
\begin{tabular}{l |rrl |rrl |rrl |rrl}\hline\hline
& \multicolumn{3}{c|}{$\Dz\piz\ell$} 	& \multicolumn{3}{c|}{$\Dstarz\piz\ell$} 	
& \multicolumn{3}{c|}{$\Dp\piz\ell$} 	& \multicolumn{3}{c}{$\Dstarp\piz\ell$} 	\\ 
Source	\hspace{28mm} &  \hspace{1mm}$f_\mathrm{exp}$ (\%) && Yield \hspace{3mm}	& \hspace{1mm}$f_\mathrm{exp}$ (\%) && Yield \hspace{3mm}	
& \hspace{1mm}$f_\mathrm{exp}$ (\%) && Yield \hspace{3mm}	& \hspace{1mm}$f_\mathrm{exp}$ (\%) && Yield \hspace{3mm}	\\ \hline
\Dsslnu background 			& 20.1 	&& Free   			& 16.4 	&& Free 			& 19.9 	&& Free 			& 22.1 	&& Free \\							
$\Dstar(\ell/\tau)\nu$  feed-up 	& 19.1 	&& Free			& 20.6 	&& Free			& 10.0 	&& Free			& 25.2 	&& Free \\							
$D(\ell/\tau)\nu$  feed-up 		& 6.4 	&& $\Dz\piz\ell$ 	& 2.3 	&& $\Dstarz\piz\ell$ 	& 4.7		&& $\Dp\piz\ell$ 	& 0.8		&& $\Dstarp\piz\ell$ \\		
Cross-feed background 		& 4.9		&& Fix./It. 			& 3.6		&& Fix./It.  			& 4.4		&& Fix./It. 			& 2.5		&& Fix./It.  \\						
\BB background 			& 28.4 	&& Free 			& 36.4 	&& Free  			& 38.7 	&& Free  			& 37.4 	&& Free \\							
Continuum background 		& 21.0 	&& Fixed  			& 20.8 	&& Fixed			& 22.2 	&& Fixed			& 12.0	&& Fixed \\  \hline\hline			
\end{tabular}
\end{table*}

\section{Fit procedure and results } \label{sec:Fit}

\subsection{Overview} \label{sec:fit:overview}
We extract the signal and normalization yields from an extended, unbinned maximum-likelihood fit 
to two-dimensional \mmiss--\pstarl distributions.
The fit is performed simultaneously to the four \dsl samples and the four \dspizl samples. 
The distribution of each \dsl and \dspizl sample is fit to the sum of eight or six 
contributions, respectively. Each of the $4\times8+4\times6=56$ contributions is described by 
a probability density function (PDF). Their relative scale factor determines the number 
of events from each source.
Tables \ref{tab:ContributionsDsl} and \ref{tab:ContributionsDspizl} summarize 
the contributions to the fit for the four \dsl sample and the four \dspizl  samples. 
These tables also list the relative yield  for each contribution
as estimated from MC simulation (for SM signal), and specify whether the yield is free, fixed, or constrained 
in the fit.

We introduce the following notation to uniquely identify each contribution to the fit:
\emph{source} $\Rightarrow$ \emph{sample}. For instance, $\Dstarz\tau\nu\Rightarrow\Dszl$
refers to signal $\Dstarz\tau\nu$ decays that are correctly reconstructed in the \Dszl sample, 
while $\Dstarz\tau\nu\Rightarrow\Dzl$
refers to the same decays, but incorrectly reconstructed in the \Dzl sample. 
We refer to the latter as feed-down. 
Contributions of the form $D(\tau/\ell)\nu\Rightarrow\Dstar(\tau/\ell)$ and
$\ds(\tau/\ell)\nu\Rightarrow\dss(\tau/\ell)$ are referred to as feed-up.

The contributions from the continuum, \BB, and cross-feed backgrounds, with the 
exception of \BB\ background in the \dspizl samples, are fixed to the 
yields determined by MC simulation after small adjustments based on data control regions. 
The yields of the remaining  36 contributions are determined in the fit.
Some of these contributions share the same source and therefore the
ratio of their yields is constrained to the expected value,
\emph{e.g.},~$\Dstarz\tau\nu\Rightarrow\Dszl$
and $\Dstarz\tau\nu\Rightarrow\Dzl$. 
Of special importance are the constraints linking the \Dsslnu yields in the \dsl samples
($N_{D^{**}\Rightarrow \ds}$) to the yields in the
\dspizl samples ($N_{D^{**}\Rightarrow \dspiz}$),
\begin{equation}\label{eq:fDss}
\fDss = \frac{N_{D^{**}\Rightarrow \ds}}{N_{D^{**}\Rightarrow \dspiz}} =
\frac{\eps_{D^{**}\Rightarrow \ds}}{\eps_{D^{**}\Rightarrow \dspiz}},
\end{equation}
Given that these constraints share the same source, $\fDss$ is equivalent 
to the  ratio of the \Dsslnu reconstruction efficiencies for the two samples.

\begin{table}[hbt] \begin{center}
\caption{Number of free parameters in the isospin-unconstrained ($N_{\rm un}$) and
constrained ($N_{\rm cons}$)  fits.} 
\label{tab:FreePar} 
\begin{tabular}{l l  cc} \hline\hline
Sample \hspace{2mm}	& Contribution 		& $N_{\rm un}$   & $N_{\rm cons}$   \\ \hline
\dsl		& $\ds\tau\nu$ signal			& 4 & 2\\
\dsl		& $\ds\ell\nu$ normalization		& 4 & 2 \\
\dsl		& $\Dstar\ell\nu$ norm.~feed-down 	& 2 & 1 \\
\dspizl	& \Dsslnu background 			& 4& 4 \\
\dspizl	& $\ds\ell\nu$ norm.~feed-up		& 4 & 4 \\
\dspizl	& \BB background 				& 4 & 4 \\
\hline\hline
\end{tabular}
\end{center}\end{table}

Taking into account the constraints imposed on event yields from a common source,
there are 22 free parameters in the standard fit, as listed in Table \ref{tab:FreePar}.
In addition, we perform a fit in which we 
impose the isospin relations 
$\RDz=\RDp\equiv\RD$ and $\RDstarz=\RDstarp\equiv\RDs$.
We choose not to impose isospin relations for the \dspizl samples. 
Consequently, this fit has a total of 17  free parameters.

The following inputs are updated by iterating the fit:
\begin{itemize}
\item The eight $\ds(\ell/\tau)\nu\Rightarrow\dspizl$ PDFs are recalculated taking into account
the fitted $\ds\ell\nu$ and $\ds\tau\nu$ contributions to the \dsl samples. 
\item The fixed charge cross-feed yields are updated 
based on the deviation of the fitted $\ds\ell\nu$ yields from the expected values.
\item The continuum, \BB, and \Dsslnu background corrections are recalculated. 
They have a slight dependence on the fitted $\ds\ell\nu$ events because some of these events 
extend into the \mes sideband.
\item The correction to the \mmiss\ resolution of the normalization contributions is readjusted.
\item The two feed-down constraints for $\Dstar\tau\nu$ are updated using the fitted
feed-down constraints for the normalization contributions in the following way:
\begin{align}\label{eq:Feeddown_Iter}
\left.\frac{N_{\Dstar\tau\nu\Rightarrow D\ell}}{N_{\Dstar\tau\nu\Rightarrow \Dstar\ell}}\right|_{\rm Iter.} =&
\left.\frac{N_{\Dstar\tau\nu\Rightarrow D\ell}}{N_{\Dstar\tau\nu\Rightarrow \Dstar\ell}}\right|_{\rm MC} \times
\left.\frac{N_{\Dstar\ell\nu\Rightarrow D\ell}}{N_{\Dstar\ell\nu\Rightarrow \Dstar\ell}}\right|_{\rm Fit}
\nonumber\\ & \times
\left.\frac{N_{\Dstar\ell\nu\Rightarrow \Dstar\ell}}{N_{\Dstar\ell\nu\Rightarrow D\ell}}\right|_{\rm MC}..
\end{align}
\end{itemize}
The iterations continue until the change on the values of \RDx\ is less than 0.01\%.
The update of the feed-down rates has a significant impact on the fits to the \Dz and \Dp
samples because of the large signal feed-down.
The other iterative updates have only a marginal impact.

\subsection{Probability Density Functions and Validation} 
\label{sec:fit:pdfs}

\begin{figure} 
\psfrag{m}[Bc]{\small{\mmiss (GeV$^2$)}}
\psfrag{p}[Bc]{\small{\pstarl (GeV)}}
\includegraphics[width=3.4in]{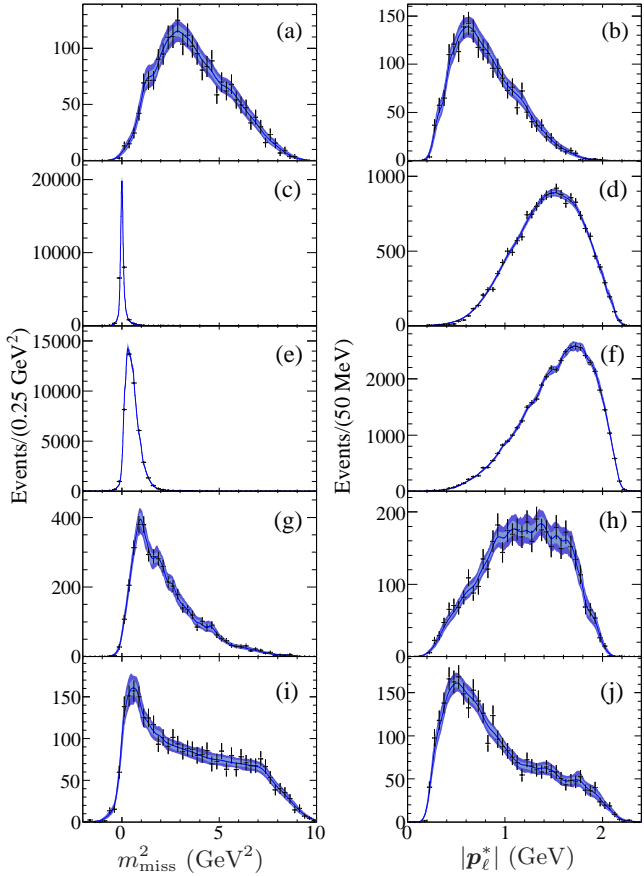}
\caption{(Color online) Projections of the simulated \mmiss and \pstarl distributions and the PDFs 
for the following contributions to the \Dzl\  sample:
(a), (b) $\Dz\tau\nu$; (c), (d) $\Dz\ell\nu$; (e), (f) $\Dstarz\ell\nu$;  (g), (h) $\Dsslnu$, and (i), (j) \BB\ background. 
The light and dark blue (gray) bands mark the $1\sigma$ and $2\sigma$ envelopes of the 
variations of the PDF projections due to their statistical uncertainty.} 
\label{fig:PDFs}
\end{figure}

The fit relies on 56 PDFs, which are derived
from MC samples of continuum and \BB\ events equivalent to 2 and 9 times the size of the 
data sample, respectively. The two-dimensional \mmiss--\pstarl
distributions for each of the 56 contributions to the fit are described by smooth 
non-parametric kernel estimators \cite{Cranmer:2000du}.
These estimators enter a two-dimensional Gaussian function centered at the \mmiss and
\pstarl values of each simulated event. The width of the Gaussian function determines
the smoothness of the PDF.  We find the optimum level of global smoothing with a
cross-validation algorithm \cite{bootstrap}. For PDFs that have variations in shape that require more 
than one level of smoothing, we combine estimators with different Gaussian widths in up to 
four areas in the \mmiss--\pstarl space.  For instance, 
we use different levels of smoothing in the $\Dstarz\ell\nu\Rightarrow\Dstarz\ell$ contribution 
for the narrow peak at $\mmiss=0$ and the smooth \mmiss 
tail that extends up to $7\gev^2$.
Figure \ref{fig:PDFs} shows one-dimensional projections of five two-dimensional PDFs. 
The bands indicate the statistical uncertainty on the PDFs estimated with a bootstrap 
algorithm~\cite{bootstrap}.

The \mmiss distributions of signal and normalization are very distinct due to the
different number of neutrinos in the final state. The \mmiss\ distributions of the backgrounds
resemble those of the signal, and therefore these contributions to the fit are either fixed or
constrained by the \dspizl samples.

To validate the PDFs and the fit procedure, we divide the large sample of simulated \BB events
into two: sample A with about $3.3\times 10^9$ \BB events, and sample B with $9.4\times 10^8$ 
\BB events.
We determine the PDFs with sample A, and create  histograms by integrating the PDFs in bins of 
their \mmiss and \pstarl projections. We compare the resulting histograms with the events in sample A, 
and derive a $\chi^2$ based on the statistical significance of the difference for each bin. The distribution 
of the corresponding $p$ values for these PDFs is uniform, as expected for an unbiased estimation.
As another test, we extract the signal and normalization yields
from  fits to the events of sample B, using the PDFs obtained from sample A. 
Again, the results are compatible with an unbiased fit.
Furthermore, we  validate the fit procedure based on a large number of pseudo experiments 
generated from these PDFs.
Fits to these samples also show no bias in the extracted signal and normalization yields.

\subsection{Fit Results} 
\label{sec:fit:results}

\begin{figure*}
\psfrag{m}[Bc]{\mmiss (GeV$^2$)}
\psfrag{p}[Bc]{\pstarl (GeV)}
\psfrag{Dtau}[Bl]{\footnotesize{$D\tau\nu$}}
\psfrag{Dstau}[Bl]{\footnotesize{$\Dstar\tau\nu$}}
\psfrag{Dl}[Bl]{\footnotesize{$D\ell\nu$}}
\psfrag{Dsl}[Bl]{\footnotesize{$\Dstar\ell\nu$}}
\psfrag{Dssl}[Bl]{\footnotesize{$D^{**}(\ell/\tau)\nu$}}
\psfrag{Bkg}[Bl]{\footnotesize{Bkg.}}
\psfrag{D0}[Bl]{$\Dz\ell$}
\psfrag{D\*0}[Bl]{$\Dstarz\ell$}
\psfrag{D\+}[Bl]{$\Dp\ell$}
\psfrag{D\*\+}[Bl]{$\Dstarp\ell$}
\includegraphics[width=3.4in]{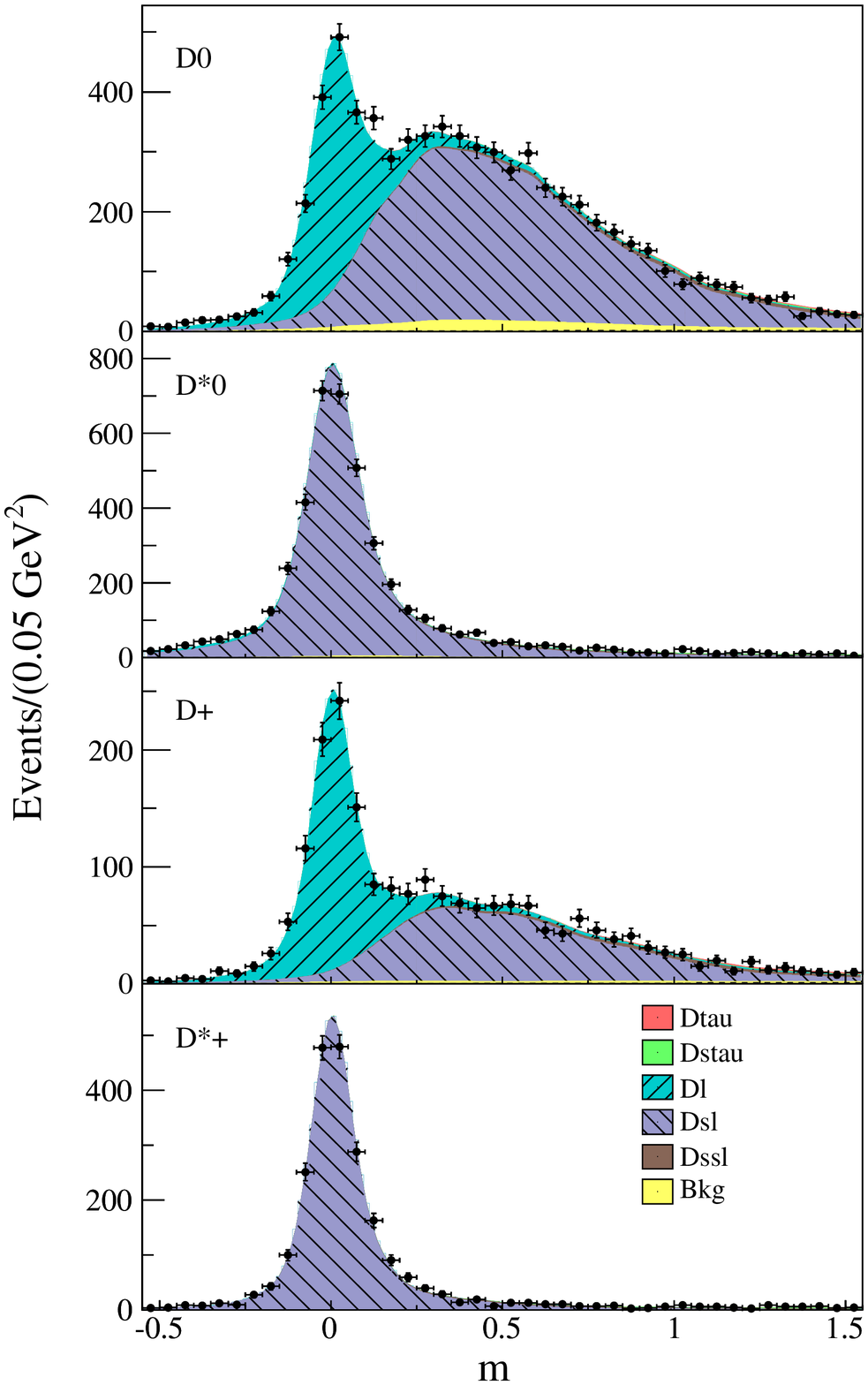}
\includegraphics[width=3.4in]{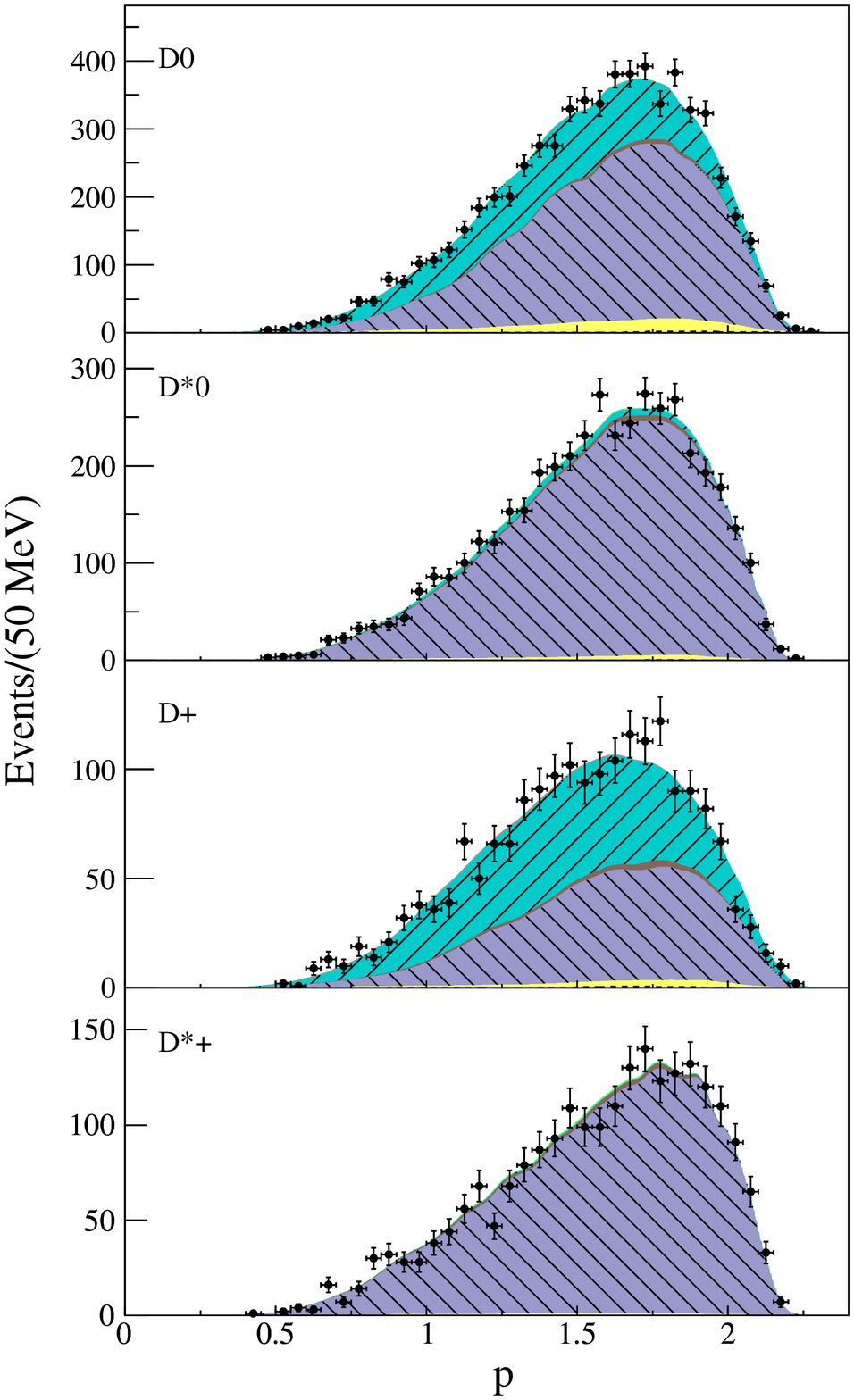}
\caption{(Color online). Comparison of the \mmiss and \pstarl distributions of the \dsl samples (data points) 
with the projections of the results of the isospin-unconstrained fit (stacked colored distributions). 
The \pstarl distributions show the normalization-enriched region with $\mmiss<1\gev^2$, 
thus excluding most of the signal events in these samples.}
\label{fig:Fit_Norm}
\end{figure*}

\begin{figure*}
\psfrag{m}[Bc]{\mmiss (GeV$^2$)}
\psfrag{p}[Bc]{\pstarl (GeV)}
\psfrag{Dtau}[Bl]{\footnotesize{$D\tau\nu$}}
\psfrag{Dstau}[Bl]{\footnotesize{$\Dstar\tau\nu$}}
\psfrag{Dl}[Bl]{\footnotesize{$D\ell\nu$}}
\psfrag{Dsl}[Bl]{\footnotesize{$\Dstar\ell\nu$}}
\psfrag{Dssl}[Bl]{\footnotesize{$D^{**}(\ell/\tau)\nu$}}
\psfrag{Bkg}[Bl]{\footnotesize{Bkg.}}
\psfrag{D0}[Bl]{$\Dz\ell$}
\psfrag{D\*0}[Bl]{$\Dstarz\ell$}
\psfrag{D\+}[Bl]{$\Dp\ell$}
\psfrag{D\*\+}[Bl]{$\Dstarp\ell$}
\includegraphics[width=3.4in]{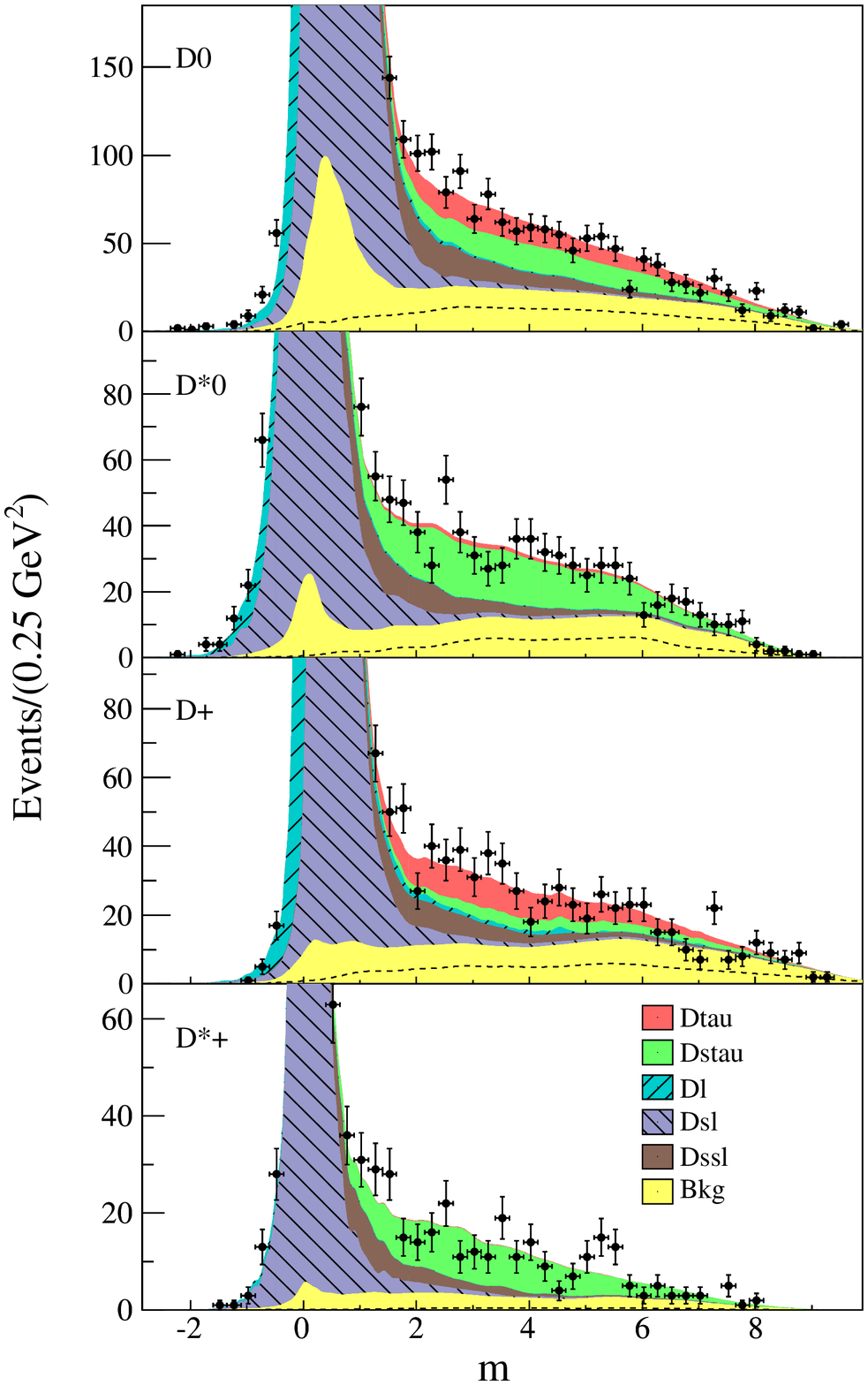}
\includegraphics[width=3.4in]{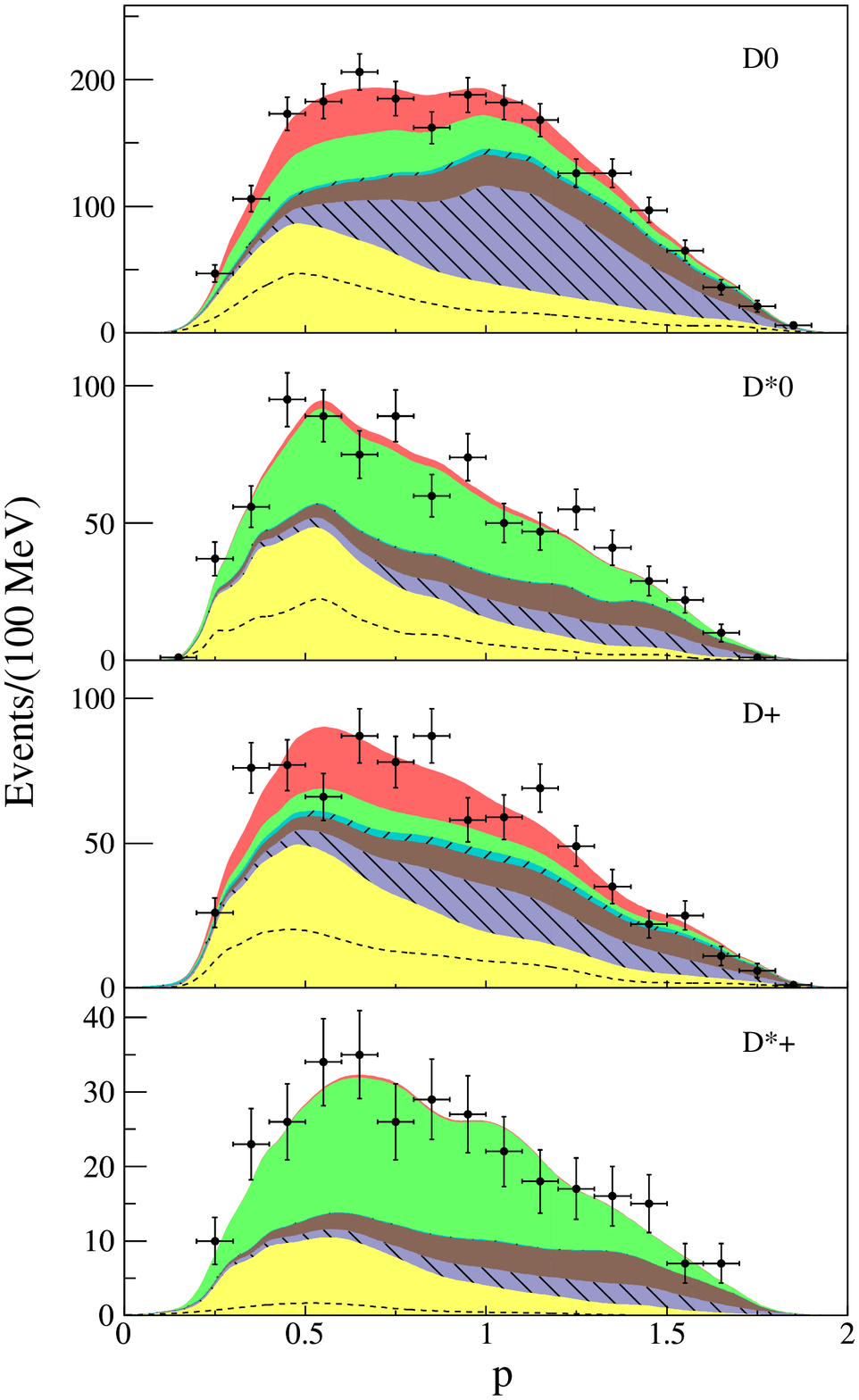}
\caption{(Color online). Comparison of the \mmiss and \pstarl distributions of the \dsl samples (data points) 
with the projections of the results of the isospin-unconstrained fit (stacked colored distributions). 
The region above the dashed line of the background component corresponds to \BB background
and the region below corresponds to continuum. The peak at $\mmiss=0$ in the
background component is due to charge cross-feed events.
The \pstarl distributions show the signal-enriched region with $\mmiss\geq1\gev^2$, 
thus excluding most of the normalization events in these samples.}
\label{fig:Fit_Sig}
\end{figure*}

Figures \ref{fig:Fit_Norm} and \ref{fig:Fit_Sig} show the \mmiss and \pstarl projections of the
fits to the \dsl samples. 
In Fig.~\ref{fig:Fit_Norm}, the \pstarl projections do not include events with $\mmiss>1\gev^2$, {\it i.e.,} 
most of the signal events.  In Fig.~\ref{fig:Fit_Sig}, the vertical scale is enlarged and the horizontal 
axis is extended for the \mmiss\ projection to reveal the signal and background contributions. 
The \pstarl projections emphasize the signal events by excluding events with $\mmiss<1\gev^2$.  
Both figures demonstrate that
the fit describes the data well and the observed differences are consistent with the statistical 
and systematic uncertainties on the PDFs and the background contributions. 

\begin{figure*}
\psfrag{m}[Bc]{\mmiss (GeV$^2$)}
\psfrag{p}[Bc]{\pstarl (GeV)}
\psfrag{Dl}[Bl]{\footnotesize{$D(\ell/\tau)\nu$}}
\psfrag{Dsl}[Bl]{\footnotesize{$\Dstar(\ell/\tau)\nu$}}
\psfrag{Dssl}[Bl]{\footnotesize{$D^{**}(\ell/\tau)\nu$}}
\psfrag{Bkg}[Bl]{\footnotesize{Bkg.}}
\psfrag{D0}[Bl]{$\Dz\piz\ell$}
\psfrag{D\*0}[Bl]{$\Dstarz\piz\ell$}
\psfrag{D\+}[Bl]{$\Dp\piz\ell$}
\psfrag{D\*\+}[Bl]{$\Dstarp\piz\ell$}
\includegraphics[width=3.4in]{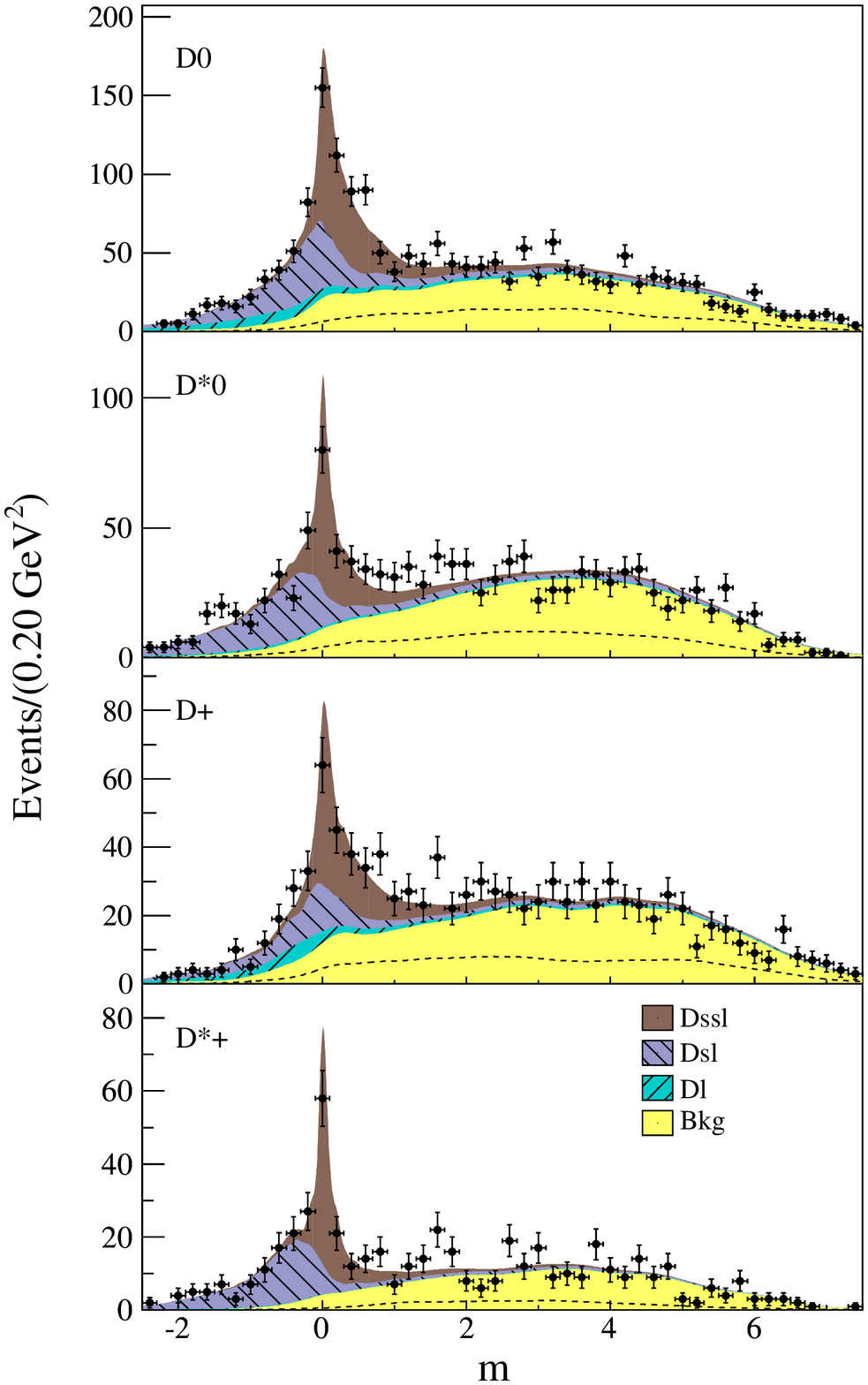}
\includegraphics[width=3.4in]{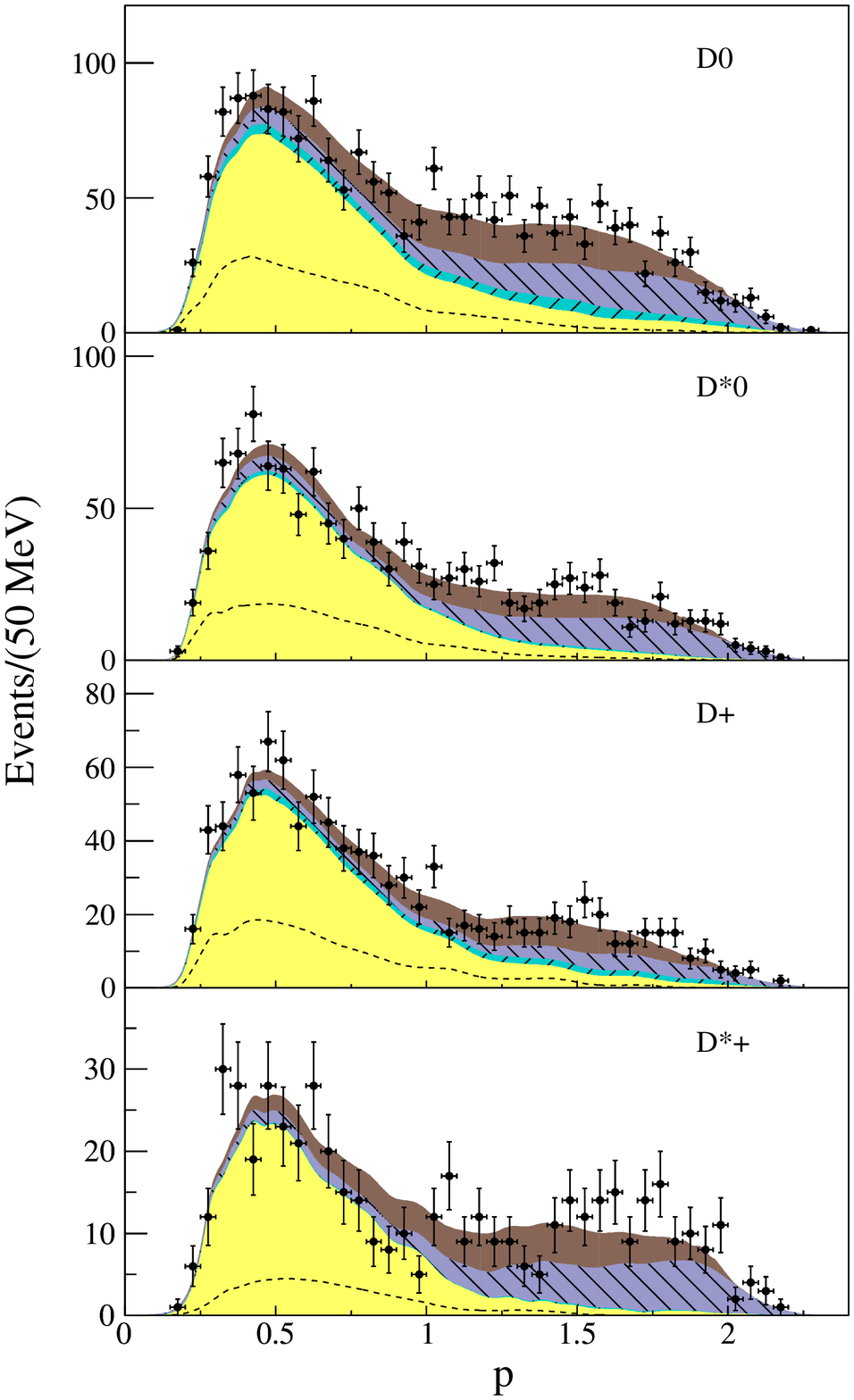}
\caption{(Color online). Comparison of the \mmiss and \pstarl distributions of the \dspizl samples 
(data points) with the projections of the results of the isospin-unconstrained fit (stacked colored distributions). 
The region above the dashed line of the background component corresponds to \BB background
and the region below corresponds to continuum.}
\label{fig:Fit_Dss}
\end{figure*}

Figure \ref{fig:Fit_Dss} shows the  \mmiss and \pstarl projections of the fit
to the four \dspizl samples. 
The narrow \mmiss peak is described well by the fit. It tightly constrains contributions from
$B\to\ds\pi\ell\nu$ decays, including the nonresonant $\ds\pi$ states as well as decays of
\dss\ states, narrow or wide. 
There appears to be  a small excess of  events in the data for $1<\mmiss<2\gev^2$.
This might be  an indication for an underestimation of the \Dsslnu background. 
The impact of this effect is assessed as a systematic uncertainty.

The fit determines, for each signal decay mode, the number of signal events in the data sample, 
$N_{\text{sig}}$, and the corresponding number of normalization events, $N_{\text{norm}}$. 
We derive the ratios 
of branching fractions as
\begin{equation}\label{eq:Rds}
\RDx = \frac{N_{\text{sig}}}{N_{\text{norm}}} \frac{\eps_{\rm norm}}{\eps_{\rm sig}} ,
\end{equation}
where $\eps_{\text{sig}}/\eps_{\text{norm}}$ is the ratio of
efficiencies (including the $\tau^{\pm}$ branching fractions) taken from MC simulation. 
These relative efficiencies  are larger for \RD\ than for \RDs, because 
the $q^2>4\gev^2$ requirement rejects a larger fraction of \BDlnu\ decays than of
\BDslnu decays, while keeping almost 100\% of \BDxtaunu decays..

The results of the fits in terms of the number of events, the efficiency ratios, 
and \RDx\ are listed in Table \ref{tab:FinalResults}, for both the standard  and the 
isospin-constrained fits.
Due to the large signal feed-down, there are significant negative correlations between
the fits to the $D\ell$ and $\Dstar\ell$ samples.
The statistical correlations are $-0.59$ for \RDz and \RDstarz,
$-0.23$ for \RDp and \RDstarp, and $-0.45$ for \RD\ and \RDs.

\section{Systematic uncertainties } 
\label{sec:Systematics}

Table \ref{tab:Syst} lists the systematic uncertainties considered in this analysis, as well
as their correlations in the measurements of \RD\ and \RDs.
We distinguish two kinds of uncertainties that affect the measurement of \RDx:
\emph{additive} uncertainties which impact the signal and background yields 
and thereby the significance of the results, and \emph{multiplicative} uncertainties that  affect  
the $\eps_{\rm sig}/\eps_{\rm norm}$ ratios and, thus, do not change the significance. 
The limited size of the simulated signal and background samples impact both 
additive and multiplicative uncertainties.

\begin{table*}
\caption{Systematic uncertainties and correlations on \RDx\
for the isospin-unconstrained (columns 1--4 and 7--8) and isospin-constrained (columns 5--6 and 9) fits. The total uncertainties and correlations are calculated based on 
Eq.~\ref{eq:CombineRho}.} 
\label{tab:Syst} 
\begin{tabular}{l r r r r r c | r r r} \hline\hline
& \multicolumn{6}{c}{Fractional uncertainty (\%)} & \multicolumn{3}{|c}{Correlation  }\\
Source of uncertainty& $\RDz$	& $\RDstarz$		& $\RDp$	& $\RDstarp$		& \hspace{4mm}$\RD$		
& $\RDs$ & \hspace{2mm}$\Dz/\Dstarz$		& $\Dp/\Dstarp$	& \hspace{4mm}$D/\Dstar$\\ \hline
{\bf Additive uncertainties}&&&&&&\\
\hspace{0.1in} {\bf PDFs} &&&&&&\\
\hspace{0.2in} MC statistics			& 6.5		& 2.9		& 5.7		& 2.7		& 4.4		& 2.0 & $-0.70$	& $-0.34$	& $-0.56$ \\ 
\hspace{0.2in} \BDxltnu FFs		& 0.3		& 0.2		& 0.2		& 0.1		& 0.2		& 0.2 & $-0.52$	& $-0.13$	& $-0.35$\\ 
\hspace{0.2in} $D^{**}\to\ds(\piz/\pipm)$  & 0.7	& 0.5		& 0.7		& 0.5		& 0.7		& 0.5 & 0.22	& 0.40	& 0.53\\
\hspace{0.2in} ${\cal B}(\BDsslnu)$	& 1.0		& 0.4		& 1.0		& 0.4		& 0.8		& 0.3 & $-0.63$	& $-0.68$	& $-0.58$\\
\hspace{0.2in} ${\cal B}(\BDsstaunu)$& 1.2	& 2.0		& 2.1		& 1.6		& 1.8		& 1.7  & 1.00	& 1.00	& 1.00\\
\hspace{0.2in} $D^{**}\to\ds\pi\pi$ 	& 2.1		& 2.6		& 2.1		& 2.6		& 2.1		& 2.6 & 0.22	& 0.40	& 0.53\\
\hspace{0.1in} {\bf Cross-feed constraints}&&&&&&\\
\hspace{0.2in} MC statistics 	            & 2.6		& 0.9		& 2.1		& 0.9		& 2.4		&1.5 & 0.02	& $-0.02$	& $-0.16$\\ 
\hspace{0.2in} \fDss				& 6.2		& 2.6		& 5.3		& 1.8		& 5.0		& 2.0 & 0.22	& 0.40	& 0.53\\
\hspace{0.2in} Feed-up/feed-down& 1.9	& 0.5		& 1.6		& 0.2		& 1.3		& 0.4 & 0.29 	& 0.51 	& 0.47\\
\hspace{0.2in} Isospin constraints 	& --		& --		& --		& --		& 1.2		& 0.3 & --		& --		& $-0.60$\\
\hspace{0.1in} {\bf Fixed backgrounds} &&&&&&\\
\hspace{0.2in} MC statistics			& 4.3		& 2.3		& 4.3		& 1.8		& 3.1		& 1.5 & $-0.48$	& $-0.05$	& $-0.30$\\ 
\hspace{0.2in} Efficiency corrections	& 4.8		& 3.0		& 4.5		& 2.3		& 3.9		& 2.3 & $-0.53$	& 0.20	& $-0.28$\\[1mm]
{\bf Multiplicative uncertainties} 	&&&&&&\\ 
\hspace{0.2in} MC statistics		& 2.3		& 1.4		& 3.0		& 2.2		& 1.8		& 1.2 & 0.00	& 0.00	& 0.00\\ 
\hspace{0.2in} \BDxltnu FFs		& 1.6		& 0.4		& 1.6		& 0.3		& 1.6		& 0.4 & 0.00	& 0.00	& 0.00\\ 
\hspace{0.2in} Lepton PID		& 0.6		& 0.6		& 0.6		& 0.5		& 0.6		& 0.6	 & 1.00	& 1.00	& 1.00\\
\hspace{0.2in} \piz/\pipm from $D^*\to D \pi$	& 0.1		& 0.1		& 0.0		& 0.0		& 0.1		& 0.1 & 1.00	& 1.00	& 1.00\\ 
\hspace{0.2in} Detection/Reconstruction         & 0.7	      & 0.7		& 0.7		& 0.7		& 0.7		& 0.7	 & 1.00	& 1.00	& 1.00\\
\hspace{0.2in} ${\cal B}(\taum\to\ellm\bar{\nu}_\ell\nu_\tau)$& 0.2 & 0.2 & 0.2 & 0.2 & 0.2 & 0.2 & 1.00	& 1.00	& 1.00\\ 
&&&&&&&\\
{\bf Total syst. uncertainty}	&12.2	& 6.7		& 11.4	& 6.0		& 9.6		& 5.5 & $-0.21$	& 0.10	& 0.05\\ 
{\bf Total stat. uncertainty}		&19.2	& 9.8		& 18.0	& 11.0	& 13.1	& 7.1 & $-0.59$	& $-0.23$	& $-0.45$\\ 
&&&&&&&\\
{\bf Total uncertainty}		& 22.7	& 11.9	& 21.3	& 12.5	& 16.2	& 9.0 & $-0.48$	& $-0.15$	& $-0.27$\\ \hline\hline
\end{tabular}
\end{table*}

\subsection{Additive uncertainties: }
Additive uncertainties affect the results of the fit.  To asses their impact,
we vary the source of uncertainty 1000 times following a given distribution, 
and repeat the fit for each variation. We adopt as the uncertainty the 
standard deviation of the distribution of the resulting $R(\ds)$ values.
From this ensemble of fits, we also estimate the  
correlation between the uncertainties of \RD\ and \RDs.

\subsubsection{PDF Estimation}
\paragraph*{MC statistics:}
We employ a bootstrap algorithm~\cite{bootstrap} to estimate the uncertainty 
due to the limited size of the simulated event samples 
on which we base the 56 PDFs. 
We generate 1000 samples of simulated events by sampling the original MC sample with
replacement \cite{Poe:2001}. The PDFs are recalculated with each bootstrapped sample, and the fit
is repeated for each set of PDFs.
Figure \ref{fig:PDFs} shows the $1\sigma$ and $2\sigma$ bands for the projections
of five selected PDFs. The impact on the final result is 4.4\% for \RD\ and 2.0\% for \RDs.

\paragraph*{Form factors for \BDxltnu:} 
We estimate the impact on the signal and normalization PDFs due to the 
uncertainties on the FF parameters,  
$\rho^2_D$, $\Delta$, $\rho^2_{\Dstar}$, $R_0(1)$, $R_1(1)$, and $R_2(1)$, 
taking into account their uncertainties and correlations.
We recalculate the $\ds\tau\nu$ and $\ds\ell\nu$ PDFs with each set of 1000
Gaussian variations of the parameter values, and repeat the fit with each 
set of PDFs to determine the impact on \RDx .

\paragraph*{$D^{**}\to\ds(\piz/\pipm)$ fraction:} 
The simulation of \Dsslnu decays only includes
the two-body decays $\dss\to\ds\pi$ of the four $L=1$ charm meson states. 
The ratio of $\dss\to\ds\piz$ decays to $\dss\to\ds\pipm$ decays which is fixed by isospin relations 
has a significant impact on the PDFs, because $\dss\to\ds\piz$ decays result in a sharply peaked \mmiss\
distribution for the \dspizl samples. The measured uncertainty on the \piz detection efficiency is 3\%. 
We assume a 4\% uncertainty to the probability that a low momentum charged pion from $\dss\to\ds\pipm$ 
decays is misassigned to the \Btag decay. 
Combining these two uncertainties, we arrive at an uncertainty on the relative
proportion of the two-body decays of \dss\ of 5\%. We repeat the fit increasing and decreasing
this ratio by 5\%, and adopt the largest variation of the isospin-constrained fit results as the systematic uncertainty.

\paragraph*{\BDsslnu branching fractions:}
Since decays to the four \dss states are combined in the \BDxltnu\ samples, the PDFs depend on the
relative \BDsslnu\ branching fractions for the four $L=1$ states \cite{Amhis:2012bh}. 
The impact of the branching fraction uncertainties is assessed
by recalculating the \BDxltnu\ PDFs and adopting the variation of the fit results 
from the ensemble of PDFs as the uncertainty.

\paragraph*{\BDssltnu branching fractions:}
As noted above,
the sharp peak in the \mmiss distribution of the \dspizl samples 
constrains contributions from $B\to\ds\pi\ell\nu$ decays. Events with additional 
unreconstructed particles contribute to the tail of the \mmiss\ distribution and, 
thus, are more difficult to separate from other backgrounds and signal events.
This is the case for \BDsstaunu decays, which are combined with 
\BDsslnu decays in the \Dsslnu PDFs with the relative proportion $\RDss_{\rm PS}=0.18$. 
This value has been derived from the ratio of the available phase space.
The same estimate applied to \BDxlnu\ decays results in 
 $\RD_{\rm PS}=0.279$ and $\RDs_{\rm PS}=0.251$, 
values that are 58\% and 32\% smaller than the measured values.
Taking this comparison as guidance for the error on \RDss, we increase 
\RDss\ by 50\%, recalculate the \Dsslnu PDFs, and repeat the fit. As a result,
the values of \RD\ and \RDs\ decrease by 1.8\% and 1.7\%, respectively.
The impact is relatively small, because \BDsstaunu contributions are
small with respect to signal decays, which have much higher reconstruction 
efficiencies.

\paragraph*{Unmeasured $B\to D^{**} (\to D^{(*)} \pi \pi )\ell \nu_{\ell}$ decays:}
To assess the impact of other potential \BDsslnu contributions, 
we modify the standard fit by adding an additional component. Out of the four
contributions listed in Table \ref{tab:Dsss_sim},
the three-body decays of the \dss states with  $L=1$ give the best 
 agreement in the fits to the \dspizl samples.  For this decay chain, 
the \mmiss\ distribution has a long tail due to an additional undetected pion.
This  could account for some of the observed excess at $1<\mmiss<2\gev^2$ 
in Fig.~\ref{fig:Fit_Dss}.
We assign the observed change in \RDx\ as a systematic uncertainty.

\begin{table}
\caption{Additional \BDsslnu decays and the MC model implemented for their decays.
The fourth decay mode refers to three-body decay of the four $L=1$  \dss\ states.}
\begin{tabular}{ll} \hline\hline 
Decay                              						& Decay model  \\ \hline 
Non-resonant $B\to D^{(*)} \pi \ell \nu_{\ell}     $ 	& Goity-Roberts \cite{Goity:1994xn}\\ 
Non-resonant $B\to D^{(*)} \pi \pi \ell \nu_{\ell}$ 	& Phase Space  \\
$B\to D^{(*)} \eta \ell \nu_{\ell}                $ 			& Phase Space  \\
$B\to D^{**} (\to D^{(*)} \pi \pi )\ell \nu_{\ell}$ 		& ISGW2~\cite{Scora:1995ty}\\
\hline\hline
\end{tabular}
\label{tab:Dsss_sim}
\end{table}

\subsubsection{Cross-feed Constraints}

\paragraph*{MC statistics:}
Constraints on the efficiency ratios that link contributions from the same source 
are taken from MC simulation.
The impact of their statistical uncertainty is assessed by varying 
 the simulated event yields assuming Poisson errors.   

\paragraph*{The ratios \fDss:}
We assess the uncertainty on \fDss, the constraints linking the \Dsslnu yields in the \dsl 
and \dspizl samples, by estimating the relative efficiencies of the selection criteria that differ in
the two  samples. 
The main differences in the selection of these samples are due to differences in
the \dsl and \dspizl BDTs.

In the \dsl samples, we observed that differences between data and simulation cause a 
5\%-10\% underestimation of the continuum and \BB backgrounds after the BDT requirements 
are applied.
Since the \Dsslnu contributions have similar \eextra distributions, and these distributions are 
the key inputs to the BDTs, we applied the same 5\%-10\% corrections to these contributions.
We conservatively assign 100\% of 
this correction as the systematic uncertainty on the \Dsslnu efficiency in the \dsl samples.

Since \BDssltnu decays are difficult to isolate in samples other than the \dspizl control samples, 
we estimate the uncertainty on the \Dsslnu efficiency due to the \dspizl BDT selection by relying  
on the observed data-MC difference of the BDT selection efficiency for the $\ds\ell\nu$ sample.
We assign the full 8.5\% overestimate of the $\ds\ell\nu$ contribution as the 
systematic uncertainty on the \Dsslnu efficiency in the \dspizl samples.

The \fDss\ constraints also depend on the relative branching fractions of the four \BDsslnu decays that 
are combined in the \Dsslnu contributions. We  estimate their impact on \fDss from
the branching fraction variations observed in the evaluation of the PDF uncertainty.
The largest standard deviation for the four \fDss distributions is 1.8\%.

By adding the uncertainties on \fDss described above in quadrature, we obtain total uncertainties of
13.2\% for the $D$ samples, and 10.0\% for the \Dstar samples. 
Given that there are similarities between the BDT selections applied to the $D$ and \Dstar samples,
we adopt a 50\% correlation between their uncertainties.
With these uncertainties and correlations, we derive the total impact on the results, 
 5.0\% for \RD\ and 2.0\% for \RDs .

\paragraph*{Feed-down constraints:}
The feed-down constraints of the signal yields are corrected as part of the iteration of the fit. 
The uncertainties on these corrections are given by the statistical uncertainty on the ratios of the fitted
$\Dstar\ell\nu\Rightarrow\Dstar\ell$ and $\Dstar\ell\nu\Rightarrow D\ell$ yields.
They are 2.4\% and 4.4\% on the $\Dstarz\tau\nu$ and $\Dstarp\tau\nu$ 
feed-down constraints, respectively. 

\paragraph*{Feed-up constraints:}
We estimate the uncertainty on the $D\tau\nu$ and $D\ell\nu$ feed-up
constraints as 100\% of the corrections on the feed-down constraints.  This results in  6.8\% on the 
$\Dz(\ell/\tau)\nu$ feed-up and 9.9\% on the $\Dp(\ell/\tau)\nu$ feed-up.  These two effects combined 
lead to an uncertainty of 1.3\% on \RD\ and 0.4\% on \RDs .

\paragraph*{Isospin constraints:}
In the isospin-constrained fit, we employ five additional constraints to link the signal and normalization
yields of the samples corresponding to  \Bm and \Bz decays. 
Since we reweight these contributions with the $q^2\leq4\gev^2$
control sample, the uncertainty on the isospin constraints is given by the statistical uncertainty on 
the ratios of the $q^2\leq4\gev^2$ yields. This uncertainty is 3.4\% in the $D\ell$ samples and 
3.6\% in the $\Dstar\ell$ samples.  This translates into uncertainties of 1.2\% on \RD\ and 0.3\% on \RDs .

\subsubsection{Fixed Background Contributions}

\paragraph*{MC statistics:}
The yields of the continuum, \BB, and cross-feed backgrounds are fixed in the fit.
The uncertainty due to the limited size of the MC samples is estimated
generating Poisson variations of these yields, and repeating the fit with each set of values.
A significant part of this uncertainty is due to the continuum yields, 
since the size of simulated continuum sample is equivalent to only twice the data sample,

\paragraph*{Efficiency corrections:}
To account for the correlations among the various corrections applied to the continuum
and \BB backgrounds, we follow this multi-step procedure:
\begin{itemize}
\item We vary the continuum corrections within their statistical uncertainties of 3\%--9\% ,
given by the number of events in the off-peak data control samples. 
\item The branching fractions of the most abundant decays in the \BB background are varied within
their uncertainties \cite{Nakamura:2010zzi}.
\item The \BB\ correction is reestimated in the high \eextra\ control sample, and varied
within the statistical uncertainty of 1.9\%.
\item The BDT bias corrections are reestimated in the \mes sideband, and 
varied within their statistical uncertainties, 2.1\% in the $D\ell$ samples
and 3.6\% in the $\Dstar\ell$ samples.
\item The \BB\ background PDFs are recalculated.
\item The fit is repeated for each set of PDF and yield variations.
\end{itemize}

Table~\ref{tab:FixBkg} shows the size of the  continuum and \BB backgrounds and their uncertainties 
due to the limited size of the MC samples and the various corrections implemented by comparisons with 
control samples.

\begin{table} 
\caption{Continuum and other \BB background yields; the first uncertainty is
due to MC statistics, the second to efficiency corrections, and $\sigma$  refers to the
total uncertainty.} 
\label{tab:FixBkg} 
\renewcommand{\arraystretch}{1.1}
\begin{tabular}{l r @{ $\pm$ } r @{ $\pm$ } r r r @{ $\pm$ } r @{ $\pm$ } r r} \hline\hline
Sample \hspace{1mm}& \multicolumn{3}{c}{Continuum} & $\sigma$ (\%) 
& \multicolumn{3}{c}{\BB} & $\sigma$ (\%)\\ \hline
$\Dz\ell$		& 355	& 13	& 12	& 4.9 & \hspace{5mm}330	& 6	& 17	& 5.3\\ 
$\Dstarz\ell$	& 132	& 8	& 6	& 7.6 & 188	& 4	& 10	& 5.9 \\ 
$\Dp\ell$		& 157	& 9	& 6	& 6.9 & 191	& 5	& 9	& 5.5\\ 
$\Dstarp\ell$	& 12		& 3	& 1	& 23.6 & 72		& 3	& 4	& 6.9 \\ \hline\hline
\end{tabular}\renewcommand{\arraystretch}{1}
\end{table}

\subsection{Multiplicative Uncertainties}
\paragraph*{MC statistics:} 
The relative efficiency $\eps_{\rm sig}/\eps_{\rm norm}$ is estimated as the ratio of expected
yields, so the limited size of the MC samples contributes to its uncertainty. We estimate it assuming
Poisson errors on the MC yields.

\paragraph*{Form factors for \BDxltnu:} 
The $q^2>4\gev^2$ requirement introduces some dependence on the FF parameterization. This uncertainty
is assessed based on the effect of the FF variations calculated for the uncertainty on the PDFs.

\begin{figure}
\psfrag{a}[Bl]{\footnotesize{$\Dstarp\ell\nu$}}
\psfrag{b}[Bl]{\footnotesize{$\Dstarp\tau\nu$}}
\psfrag{pp}[Bc]{\footnotesize{$p_{\pip}$ (GeV)}}
\psfrag{p0}[Bc]{\footnotesize{$p_{\piz}$ (GeV)}}
\includegraphics[width=3.4in]{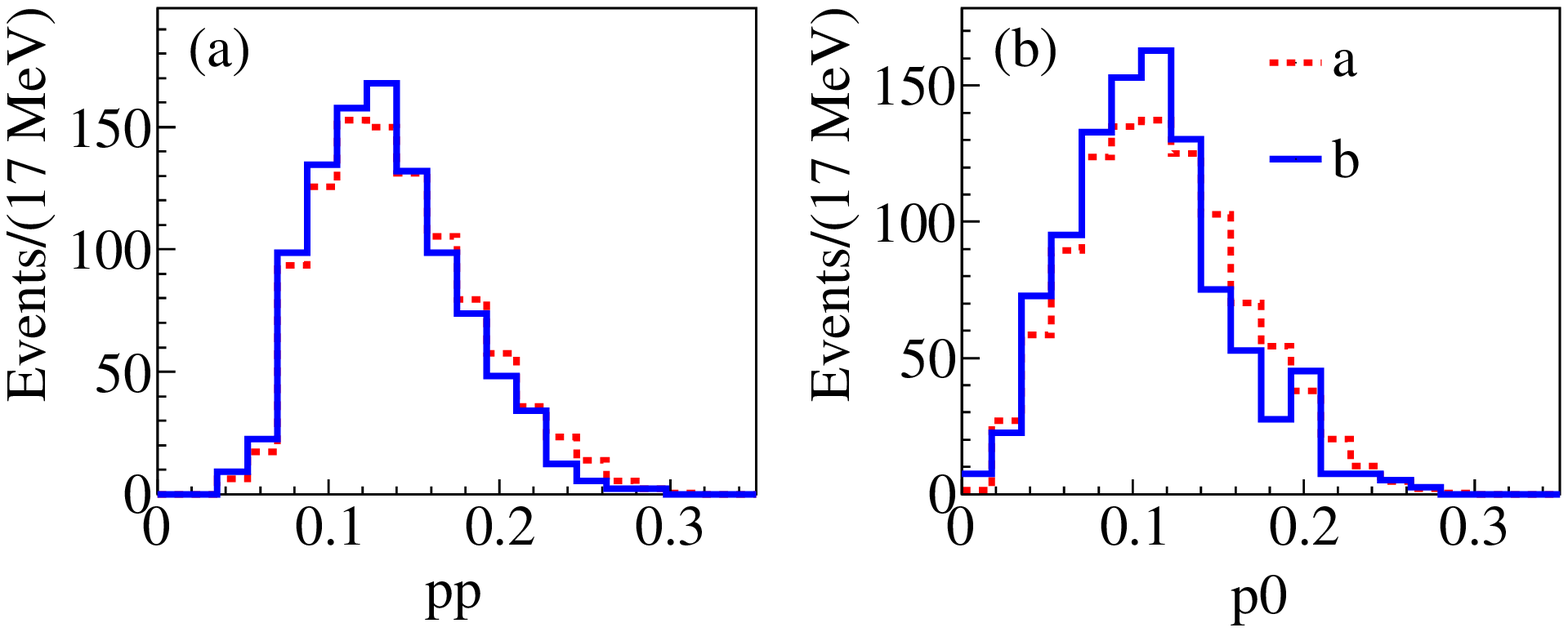}
\caption{(Color online). Pion momentum in the laboratory from $B\to\Dstarp\ell\nu$ and $B\to\Dstarp\tau\nu$ decays: 
(a) $\Dstarp\to\Dz\pip$,  and (b) $\Dstarp\to\Dp\piz$ decays. 
Histograms are normalized to 1000 entries.}
\label{fig:Syst_SoftPi}
\end{figure}

\paragraph*{\piz/\pipm from $D^*\to D \pi$:}
There is a significant momentum-dependent uncertainty on the reconstruction 
efficiency of soft pions originating from $\Dstar\to D\pi$ decays. However, the
momentum spectra of soft pions in signal and normalization decays are rather 
similar, see Fig.~\ref{fig:Syst_SoftPi}. As a result, 
the uncertainty on \RDx\ is less than 0.1\%.

\paragraph*{Detection and Reconstruction:} 
Given that signal and normalization decays are reconstructed by the same particles in
the final state, many of the uncertainties that impact their efficiencies cancel in
the ratios $\eps_{\rm sig}/\eps_{\rm norm}$.
Uncertainties due to final-state radiation, 
soft-pion reconstruction, and others related to the detector performance 
contribute less than 1\%. 
Similarly, the tagging efficiency for events with signal and normalization decays show
only very small differences.    

\paragraph*{$\taum\to\ellm\bar{\nu}_\ell\nu_\tau$ branching fraction:}
We use the world averages ${\cal B}(\taum\to e^-\nueb\nut)=(17.83\pm0.04)\%$ and
${\cal B}(\taum\to \mu^-\numb\nut)=(17.41\pm0.04)\%$ \cite{Nakamura:2010zzi}.

\subsection{Correlations}
Even though several of the uncertainties listed in Table~\ref{tab:Syst} have the same source, 
their impact on \RDx\ is largely uncorrelated, i.e., the correlation between uncertainties in different
rows of Table~\ref{tab:Syst} is negligible.
However, the correlation between the uncertainties on \RD\ and \RDs\ (different columns) is significant, and 
important for the comparison of these measurements with theoretical predictions.

For most of the additive systematic uncertainties, we estimate the correlation from the 
two-dimensional \RD--\RDs\ distribution resulting from the fit variations.
This is not possible for the $D^{**}\to\ds\piz/\pipm$ and $D^{**}\to\ds\pi\pi$ uncertainties.
These uncertainties affect the size of the \Dsslnu background in the 
\dsl samples in the same way that as \fDss does. Thus, we derive their correlations from
the \fDss correlations.
Since the signal and $\dss\tau\nu$ PDFs are very similar, we assign a 100\% correlation
on ${\cal B}(\BDsstaunu)$.

The multiplicative uncertainties on the efficiency due to the MC statistics are 
uncorrelated. The FFs for \BDlnu and \BDslnu decays are measured separately, so their uncertainties
are also not correlated. The uncertainty on ${\cal B}(\taum\to\ellm\bar{\nu}_\ell\nu_\tau)$ affects
all channels equally. 
We assume that the remaining small uncertainties
on the efficiencies due to detector effects are 100\% correlated as well.

The uncertainties and their correlations are listed in Table \ref{tab:Syst}.
We combine these correlations $\rho_i$  and the uncertainties by adding their covariance matrices,
\begin{equation}\label{eq:CombineRho}
\sum_{i} \left(\! \begin{array}{cc} \sigma_i^2 &\!\! \rho_i\sigma_i\sigma_i^* \\
\rho_i\sigma_i\sigma_i^* &  \sigma_i^{*2}  \end{array}\! \right) =
\left( \!\begin{array}{cc} \sigma_{\rm tot}^2 &\!\!\! \rho_{\rm tot}\sigma_{\rm tot}\sigma_{\rm tot}^* \\
\rho_{\rm tot}\sigma_{\rm tot}\sigma_{\rm tot}^* &  \sigma_{\rm tot}^{*2}  \end{array}\!\! \right).
\end{equation}
Here, $\sigma_i$ and $\sigma_i^*$ refer to the uncertainties on \RD\ and \RDs, respectively.

\section{Stability checks and Kinematic Distributions } 
\label{sec:Stability_checks}

\subsection{Stability tests}

We have checked the stability of the fit results for different data subsamples and different
levels of background suppression. 

\begin{figure}
\psfrag{Runs 123}[Bl]{\footnotesize{Period 1}}
\psfrag{Run 4}[Bl]{\footnotesize{Period 2}}
\psfrag{Run 5}[Bl]{\footnotesize{Period 3}}
\psfrag{Run 6}[Bl]{\footnotesize{Period 4}}
\psfrag{Electron}[Bl]{\footnotesize{Electron}}
\psfrag{Muon}[Bl]{\footnotesize{Muon}}
\psfrag{R\(D\)}[Bc]{\footnotesize{   \RD}}
\psfrag{R\(D*\)}[Bc]{\footnotesize{   \RDs}}
\psfrag{SM}[Bl]{\footnotesize{SM}}
\psfrag{All data}[Bl]{\footnotesize{All data}}
\psfrag{Bm}[Bl]{\footnotesize{\Bm}}
\psfrag{B0}[Bl]{\footnotesize{\Bz}}
\includegraphics[width=3.4in]{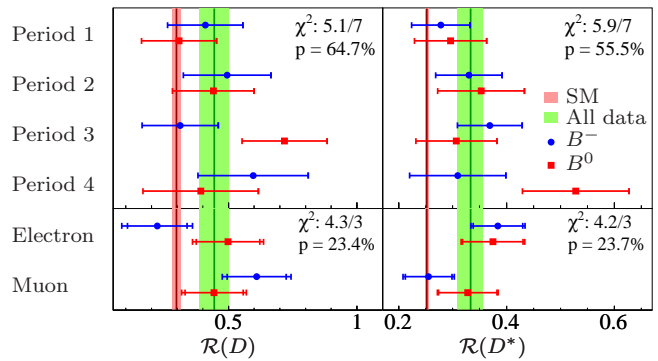}
\caption{(Color online). Measurements of \RD\ and \RDs\ for 
different data subsamples. Top: for four run periods with statistical uncertainties only.
Bottom: for electrons and muons with statistical and uncorrelated systematic uncertainties. 
The vertical bands labeled ``SM'' and ``All data" mark the SM predictions and the results of 
the fits to the whole data sample, respectively.}
\label{fig:Stability_Checks}
\end{figure}

To look for possible dependence of the results 
on the data taking periods, we divide the data sample into four periods 
corresponding to approximately equal luminosity, and fit each sample separately.
The results are presented in Fig.~\ref{fig:Stability_Checks}. 
The eight measurements each for \RD\ and \RDs, separately for \Bp\ and \Bz, are compared 
to the isospin-constrained fit results obtained from the complete data sample. 
Based on the values of $\chi^2$ for 7 degrees of freedom, we conclude that 
the results of these fits are statistically consistent with the fit to the whole data sample.

A similar test is performed for two samples identified by the final state lepton,
an electron or a muon.
This test includes the uncertainties on the background corrections that affect
the electron and muon samples differently. These uncertainties are statistically dominated and, thus, 
independent for both samples.
The results are presented in the bottom panels of Fig.~\ref{fig:Stability_Checks}. The $\chi^2$ tests 
confirm the stability of these measurements within the uncertainties.

\begin{figure}
\psfrag{R\(D\)}[Bl]{\footnotesize{\RD}}
\psfrag{R\(D*\)}[Bl]{\footnotesize{\RDs}}
\psfrag{Bm}[Bl]{\footnotesize{\Bm}}
\psfrag{B0}[Bl]{\footnotesize{\Bz}}
\includegraphics[width=3.4in]{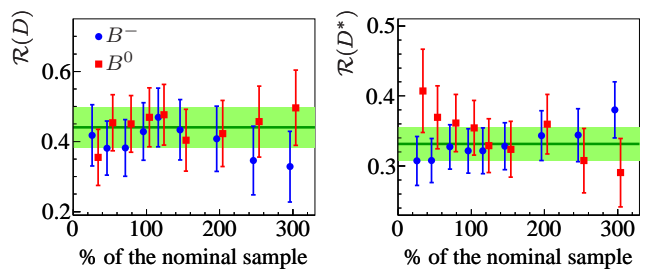}
\caption{(Color online). Measurements of \RD\ and \RDs\ for different BDT requirements, impacting the 
signal/background ratio.
The horizontal  bands mark the \RD\ and \RDs\ results for the isospin-constrained 
fit to the nominal (100\%) sample.
The data points represent the results of the fits for \Bp and \Bz\ mesons with their statistical uncertainties.}
\label{fig:Stability_BDTVar}
\end{figure}

To assess the sensitivity of the fit results on the purity of the data sample 
and the BDT selection,
we perform fits for samples selected with different BDT requirements.
We identify each sample by the relative number of events in the signal region 
($\mmiss>1\gev^2$) with respect to the nominal sample, which is labeled as the 100\% sample. 
The ratio of the number of fitted signal events $S$ to the number of background events $B$ varies from 
$S/B=1.27$ in the 30\% sample, to $S/B=0.27$ in the 300\% sample, while the backgrounds 
increase by a factor of 18. 
The BDT bias correction and the PDFs are recalculated for each sample.
Figure~\ref{fig:Stability_BDTVar} shows the results of fits to the different samples 
with tighter and looser BDT requirements. 
We take into account the large correlations between these nested samples 
and conclude that the results are stable for the very large variations 
of the BDT requirements.
 
\subsection{Gaussian Uncertainties} 
\label{sec:stability:gaussian}

\begin{figure}
\psfrag{a}[Bc]{\footnotesize{\BDtaunu yield}}
\psfrag{b}[Bc]{\footnotesize{\BDstaunu yield}}
\psfrag{n}[Br]{\footnotesize{$n^2_\sigma/2$}}
\includegraphics[width=3.4in]{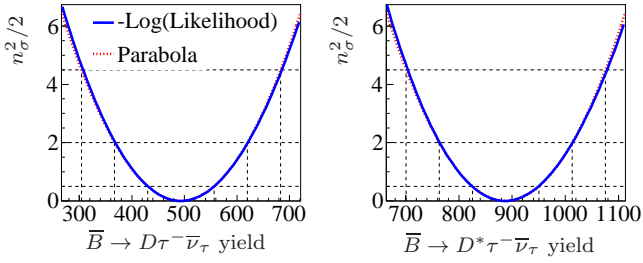}
\caption{(Color online). Likelihood scan for the two signal yields compared to a parabola. 
The dashed lines indicate the number of standard deviations ($n_\sigma$) away from the fit result.}
\label{fig:Stability_Likelihood}
\end{figure}

For a maximum likelihood fit with Gaussian uncertainties, the logarithm of the likelihood is 
described by the parabola $P(Y)=(Y-Y_{\rm fit})^2/2\sigma^2_{\rm fit}$,
where $Y_{\rm fit}$ is the fitted yield and  $\sigma_{\rm fit}$ is the uncertainty on $Y_{\rm fit}$. 
Figure \ref{fig:Stability_Likelihood} compares the likelihood scan of the signal yields for the isospin-constrained 
fit with the parabola that results from the fitted yields, presented in Table \ref{tab:FinalResults}.
There is a slight asymmetry in the likelihood function, but good agreement overall.
Thus, we conclude that the statistical uncertainty on \RD\ and \RDs\ may be considered Gaussian.

\begin{figure}
\psfrag{R\(D\)}[Bl]{\footnotesize{\RD}}
\psfrag{R\(D*\)}[Bl]{\footnotesize{\RDs}}
\includegraphics[width=3.4in]{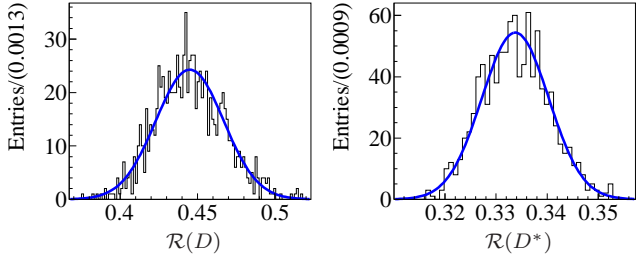}
\caption{(Color online). Histograms: \RDx\ distributions resulting from 1000 variations of \fDss.
Solid curves: Gaussian fits to the \RDx\ distributions.} 
\label{fig:Stability_Gaussian_fDss}
\end{figure}

Figure \ref{fig:Stability_Gaussian_fDss} shows the effect on
 \RD\ and \RDs\ from variations on \fDss, the largest source of systematic uncertainty.  
The distributions are well described by a Gaussian function.
This is also the case for the other major sources of systematic uncertainty.  

\subsection{Kinematic Distributions} 
\label{sec:stability:kinematic}
We further study the results of the fit by comparing the kinematic distributions of data events with 
the SM expectations.
Specifically, we focus on the signal-enriched region with $\mmiss>1.5\gev^2$
and scale each component in the simulation by the results of the fits.
To compare the data and MC distributions we calculate a $\chi^2$ per degree of freedom
which only includes the statistical uncertainty of bins with 8 or more events. 
The number of degrees of freedom is given by the number of bins minus the number of fitted signal yields.

\begin{figure*} 
\psfrag{E}[Bc]{\small{\eextra (GeV)}}
\psfrag{Dtau}[Bl]{\footnotesize{$D\tau\nu$}}
\psfrag{Dstau}[Bl]{\footnotesize{$\Dstar\tau\nu$}}
\psfrag{Dl}[Bl]{\footnotesize{$D\ell\nu$}}
\psfrag{Dsl}[Bl]{\footnotesize{$\Dstar\ell\nu$}}
\psfrag{Dssl}[Bl]{\footnotesize{$D^{**}(\ell/\tau)\nu$}}
\psfrag{Bkg}[Bl]{\footnotesize{Bkg.}}
\psfrag{a}[Bc]{$\Dz\ell$}
\psfrag{b}[Br]{$\Dstarz\ell$}
\psfrag{x}[Br]{$\Dp\ell$}
\psfrag{d}[Br]{$\Dstarp\ell$}
\psfrag{e}[Bl]{$D\ell$}
\psfrag{f}[Bl]{$\Dstar\ell$}
\includegraphics[width=6.5in]{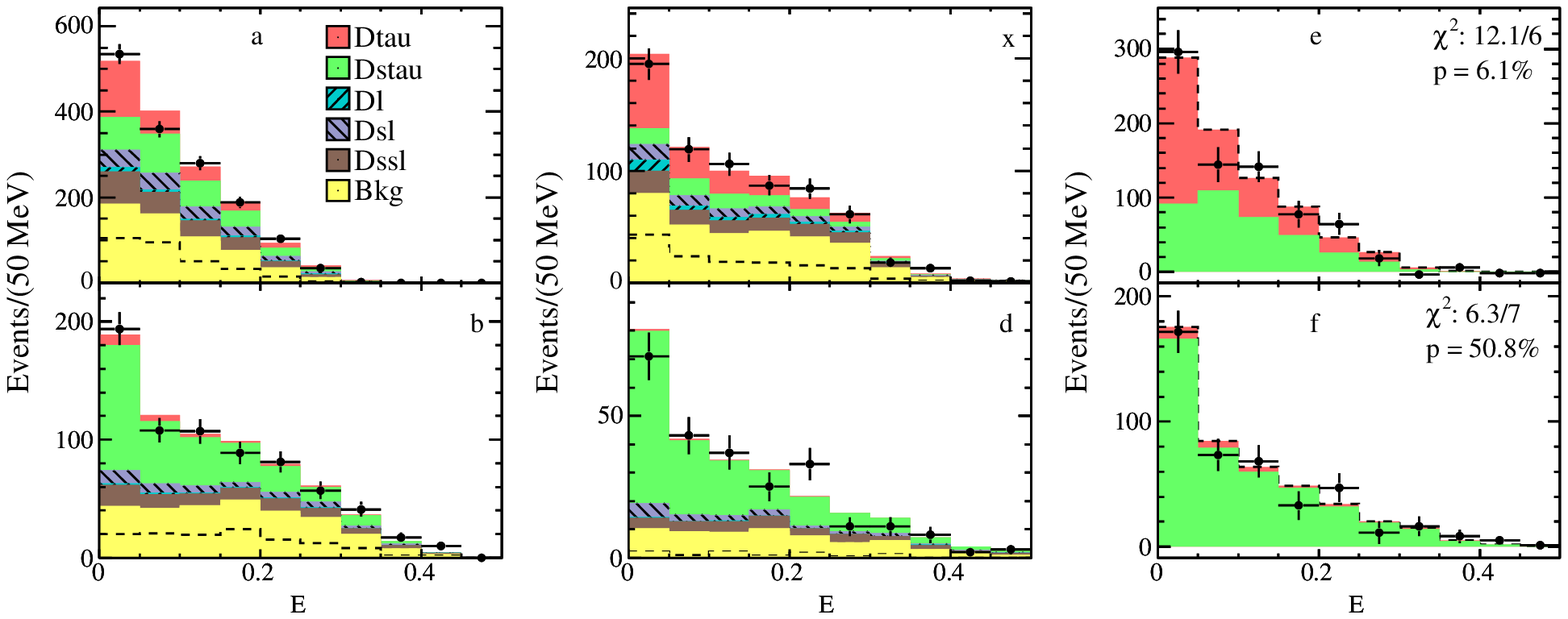}
\caption{(Color online). \eextra distributions for events with $\mmiss>1.5\gev^2$ scaled to the results of the 
isospin-unconstrained (first two columns) and isospin-contrained (last column) fits. 
The region above the dashed line of the background component corresponds to \BB background
and the region below corresponds to continuum. In the third column, the \Bz and \Bp samples are combined, 
and the normalization and background events are subtracted. }
\label{fig:Stability_EExtra}
\end{figure*}

\begin{figure*} 
\psfrag{M}[Bc]{\small{\mes (GeV)}}
\psfrag{P}[Bc]{\small{\pstarl (GeV)}}
\psfrag{a}[Bl]{$D\ell$}
\psfrag{b}[Bl]{$\Dstar\ell$}
\psfrag{x}[Bl]{$D\ell$}
\psfrag{d}[Bl]{$\Dstar\ell$}
\psfrag{e}[Bl]{$D\ell$}
\psfrag{f}[Bl]{$\Dstar\ell$}
\includegraphics[width=6.5in]{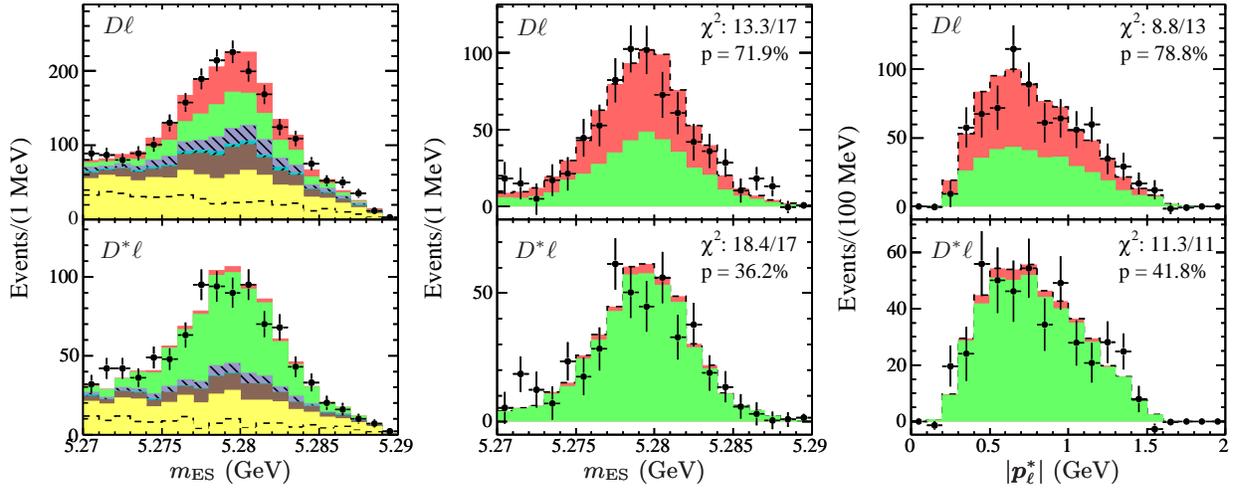}
\caption{(Color online). \mes distributions before (left) and after (center) subtraction of normalization of background events, 
and lepton momentum distributions after this subtraction (right) for events with $\mmiss>1.5\gev^2$ 
scaled to the results of the  isospin-constrained fit. 
The \Bz and \Bp samples are combined. See Fig.~\ref{fig:Stability_EExtra} for a legend.}
\label{fig:Stability_Kinematic}
\end{figure*}

Figure \ref{fig:Stability_EExtra} shows the \eextra distribution of events in the \dsl samples.
This variable is key in the BDT selection and overall background suppression.
There is a  clear enhancement of signal events at $\eextra=0$ in all four \dsl samples. The 
background contributions, which are significantly more uniform in
\eextra than those of signal, appear to be well reproduced.
We conclude that the simulation agrees well with the data distribution.

Figure \ref{fig:Stability_Kinematic} also shows clear signal enhancements in the 
\mes and \pstarl distributions of events in the $\mmiss>1.5\gev^2$ region.
The data and simulation agree well within the limited statistics.

\section{Results} 
\label{sec:Results}
\subsection{Comparison with SM expectations} 
Table \ref{tab:FinalResults} shows the results of the measurement of \RD\ and \RDs\
extracted from the fit without and with isospin constraints linking \Bp\ and \Bz\ decays.

\begin{table*} 
\caption{Results of the isospin-unconstrained (top four rows) and isospin-constrained fits 
(last two rows). The columns show the 
signal and normalization yields, the ratio of their efficiencies, \RDx, the signal branching fractions,  
and $\Sigma_{\rm stat}$ and $\Sigma_{\rm tot}$,
the statistical and total significances of the measured signal yields. 
Where two uncertainties are given, the first is statistical and the second is systematic. The second
and third uncertainties on the branching fractions
${\cal B}(\BDxtaunu)$ correspond to the systematic uncertainties due to \RDx\ and ${\cal B}(\BDxlnu)$,
respectively. The stated branching fractions for the isospin-constrained fit refer to \Bm\ decays.} 
\label{tab:FinalResults} \vspace{0.1in}
\begin{tabular}{ll r @{ $\pm$ } l r @{ $\pm$ } l r @{ $\pm$ } l r @{ $\pm$ } l @{ $\pm$ } l 
r @{ $\pm$ } l @{ $\pm$ } l @{ $\pm$ } l rr} \hline\hline
\multicolumn{2}{l}{Decay} &  \multicolumn{2}{c}{$N_\mathrm{sig}$} &  \multicolumn{2}{c}{$N_\mathrm{norm}$} 
&  \multicolumn{2}{c}{$\eps_{\rm sig}/\eps_{\rm norm} $} 
&  \multicolumn{3}{c}{\RDx} &  \multicolumn{4}{c}{${\cal B}(B\to\ds\tau\nu)\,(\%)$} 
&$\Sigma_{\text{stat}}$	& $\Sigma_{\text{tot}}$  \\ \hline
$\Bm\!\to$&$\Dz\taum\nutb$	&\hspace{0.03in}  314	& 60&\hspace{0.1in} 1995	& 55&\hspace{0.1in} 0.367	&0.011\hspace{0.07in}
				& 0.429	& 0.082	& 0.052\hspace{0.1in}	
				& 0.99	& 0.19\hspace{0.1in}& 0.12	& 0.04	& 5.5 &\hspace{0.15in}4.7 \\
$\Bm\!\to$&$\Dstarz\taum\nutb$& 639	& 62		& 8766	& 104	& 0.227	& 0.004	& 0.322	& 0.032	& 0.022	
				& 1.71	& 0.17	& 0.11	& 0.06	& 11.3 	&9.4 \\
$\Bzb\to$&$\Dp\taum\nutb$	& 177	& 31		& 986	& 35		& 0.384	& 0.014	& 0.469	& 0.084	& 0.053	
				& 1.01	& 0.18	& 0.11	& 0.04	& 6.1 	&5.2 \\ 
$\Bzb\to$&$\Dstarp\taum\nutb$& 245	& 27		& 3186	& 61		& 0.217	& 0.005	& 0.355	& 0.039	& 0.021	
				& 1.74	& 0.19	& 0.10	& 0.06	& 11.6 	&10.4 \\[1.5mm]
$\Bb\;\to$&$ D\taum\nutb$	& 489	& 63		& 2981	& 65		& 0.372	& 0.010	& 0.440	& 0.058	& 0.042	
				& 1.02	& 0.13	& 0.10	& 0.04	& 8.4		&6.8 \\
$\Bb\;\to$&$\Dstar\taum\nutb$	& 888	& 63		& 11953	& 122	& 0.224	& 0.004	& 0.332	& 0.024	& 0.018	
				& 1.76	& 0.13	& 0.10	& 0.06	& 16.4 	&13.2 \\ \hline\hline
\end{tabular}
\end{table*}

The \BDxtaunu branching fractions are calculated from the measured values of \RDx,
\begin{equation}
{\cal B}(\BDxtaunu) = \RDx\times {\cal B}(\BDxlnu).
\end{equation} 
For \Bm, we use the average branching fractions measured by 
\babar~\cite{PhysRevLett.104.011802, PhysRevD.79.012002,PhysRevD.77.032002}, 
\begin{linenomath} 
\begin{align}
{\cal B}(\Bm \to \Dz\ellm\nulb) & = (2.32\pm0.03\pm0.08)\%, \nonumber \\
{\cal B}(\Bm \to \Dstarz\ellm\nulb) & = (5.31\pm0.02\pm0.19)\%, \nonumber 
\end{align}
\end{linenomath} 
and for \Bzb, the corresponding branching fractions related by isospin.

We estimate the statistical significance of the measured signal branching fractions as 
$\Sigma_{\text{stat}}=\sqrt{2\Delta(\textrm{ln}\cal{L})}$, 
where $\Delta(\textrm{ln}\cal{L})$ is the increase in log-likelihood
for the nominal fit relative to the no-signal hypothesis.  The total significance $\Sigma_{\text{tot}}$ 
is determined as
\begin{equation} \label{eq:TotalSig}
\Sigma_{\text{tot}} = \Sigma_{\text{stat}} \frac{\sigma_{\text{stat}}}{\sqrt{\sigma_{\text{stat}}^2
+ \sigma_{\text{asys}}^2}}.
\end{equation}
In this expression, the statistical significance is scaled by the sum of the statistical 
uncertainty $\sigma_{\text{stat}}$ and the additive systematic uncertainty $\sigma_{\text{asys}}$.
The significance of the \BDtaunu signal is $6.8\sigma$, the first such measurement exceeding $5\sigma$.

\begin{figure}
\psfrag{R\(D\)}[Bl]{\small{\RD}}
\psfrag{R\(D*\)}[Bl]{\small{\RDs}}
\includegraphics[width=3.in]{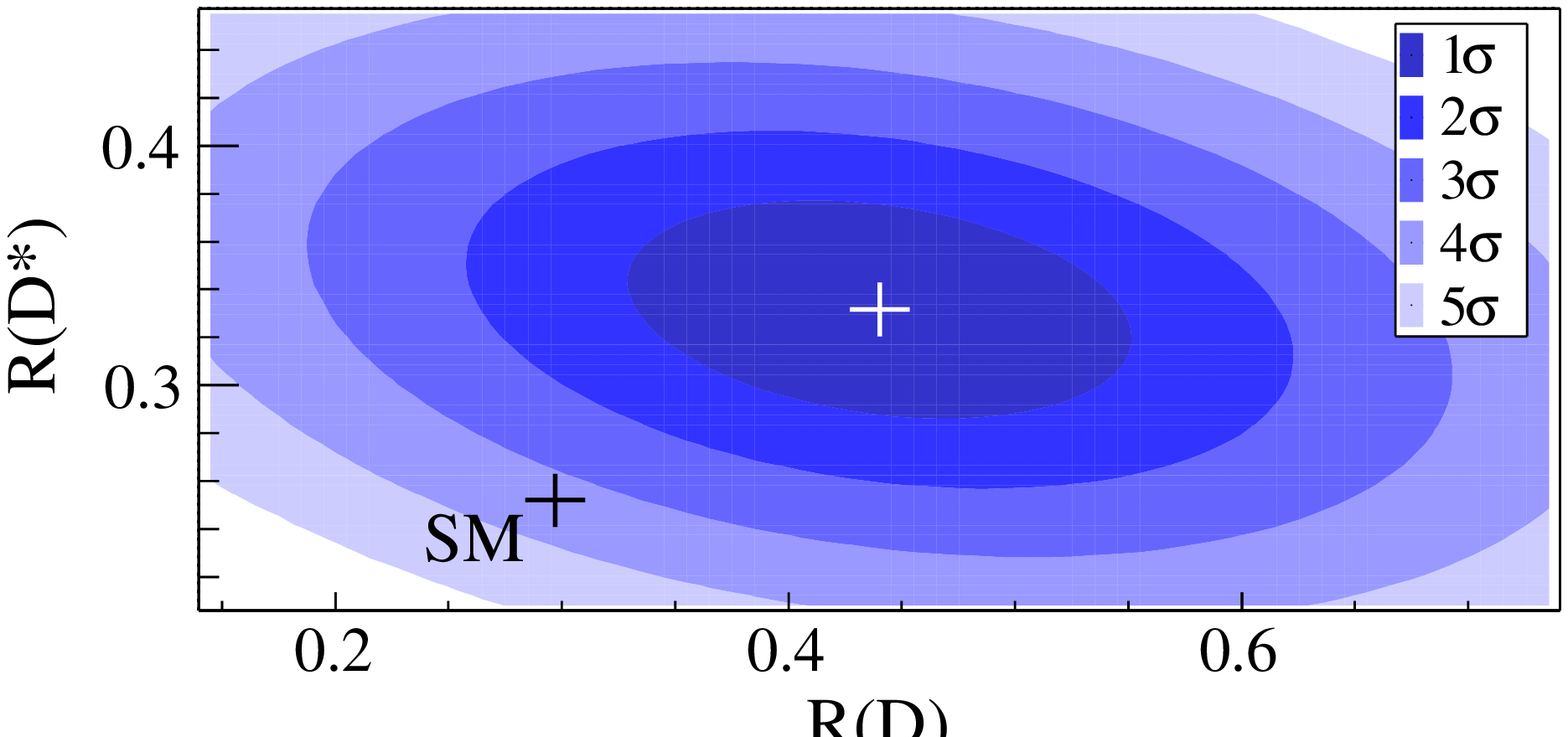} 
\caption{(Color online). Representation of  $\chi^2$ (Eq.~\ref{eq:Chi2}) in the \RD--\RDs\ plane.
The white cross corresponds to the measured \RDx, and the black cross to the SM predictions.
The shaded bands represent one standard deviation each.}
\label{fig:Higgs_Chi2}
\end{figure}

We compare the measured \RDx\ to the calculations based on the SM,
\begin{linenomath} 
\begin{align}
\RD_{\rm exp} &= 0.440\pm0.072 & \RDs_{\rm exp} &= 0.332\pm0.030,\nonumber\\
\RD_{\rm SM} &= 0.297 \pm 0.017 & \RDs_{\rm SM} &= 0.252 \pm 0.003,  \nonumber
\end{align}
\end{linenomath} 
and observe an excess over the SM predictions for \RD\ and \RDs\ of $2.0\sigma$ and
$2.7\sigma$, respectively. We combine these two measurements in the following way 
\begin{equation} \label{eq:Chi2}
\chi^2 = \left(\Delta, \Delta^*\right)
\left( \begin{array}{cc}
\sigma^2_\mathrm{exp}+\sigma^2_\mathrm{th} 	& \rho\, \sigma_\mathrm{exp}\;\sigma^*_\mathrm{exp}\\
\rho\, \sigma_\mathrm{exp}\;\sigma^*_\mathrm{exp}	& \sigma^{*2}_\mathrm{exp}+\sigma^{*2}_\mathrm{th}
\end{array} \right)^{-1}
\left( \begin{array}{c}
\Delta\\
\Delta^*
\end{array} \right),
\end{equation}
where $\Delta^{(*)}=\RDx_{\rm exp}-\RDx_{\rm th}$, and $\rho$ is the total correlation between the
two measurements, $\rho(\RD,\RDs)=-0.27$.
Since the total uncertainty is dominated by the experimental uncertainty, 
the expression in Eq.~\ref{eq:Chi2} is expected to be distributed as a $\chi^2$
distribution for two degrees of freedom.
Figure \ref{fig:Higgs_Chi2} shows this distribution
in the \RD--\RDs\ plane. The contours are ellipses
slightly rotated with respect to the \RD--\RDs\ axes, due to the non-zero correlation.

For the assumption that  $\RDx_{\rm th}=\RDx_{\rm SM}$, we obtain $\chi^2=14.6$, 
which corresponds to a probability of $6.9\times10^{-4}$. This means that the possibility 
that the measured \RD\ and \RDs\ both agree with the SM predictions is excluded at the 
$3.4\sigma$ level \footnote{In this paper, the significance of an observation with 
probability $p$ is expressed by the number of standard deviations 
$\sigma$ of a one-dimensional Gaussian function for this probability. The
shaded bands in Figs. \ref{fig:Higgs_Chi2}, \ref{fig:Exclusion2D_DDsTauNu_BaBar}, 
and \ref{fig:Higgs_Favored2D}
correspond to $p$ values of  0.683, 0.955, 0.997 and so on.}.
Recent calculations \cite{Nierste:2008qe,Tanaka:2010se,Bailey:2012jg,Becirevic:2012jf} have 
resulted in values of 
$\RD_{\rm SM}$\ that slightly exceed our estimate. For the largest of those values, the
significance of the observed excess decreases to $3.2\sigma$.

\subsection{Search for a charged Higgs}

To examine whether the excess in \RDx\ can be explained by contributions from a charged Higgs boson 
in the type II 2HDM, we study the dependence of the fit results on \tBmH.

\begin{figure}[b]
\psfrag{d}[br]{\footnotesize{Probability/$\!\gev^{2}$}}
\psfrag{x}[Br]{\footnotesize{Probability/\!\gev}}
\psfrag{m}[Bc]{\footnotesize{\mmiss (GeV$^2$)}}
\psfrag{p}[Bc]{\footnotesize{\pstarl (GeV)}}
\includegraphics[width=3.4in]{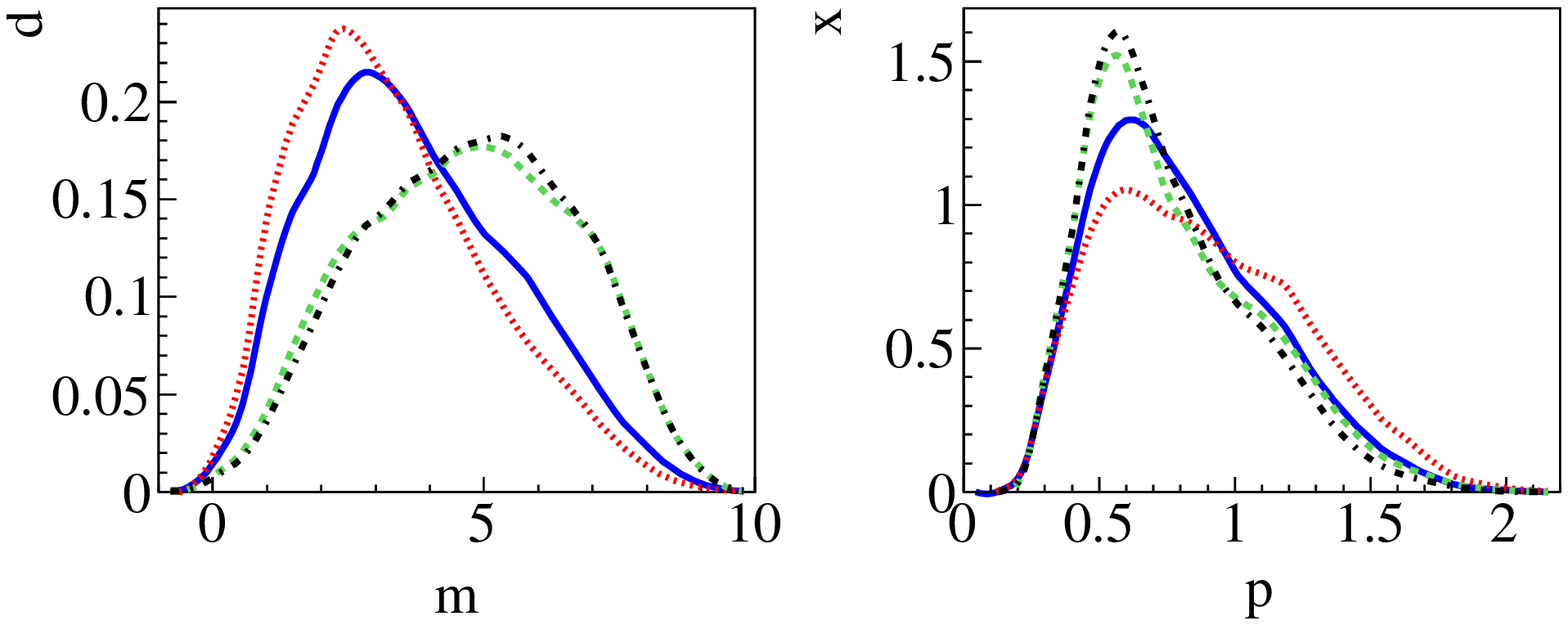} \\ \vspace{1mm}
\psfrag{a}[bl]{\footnotesize{ SM}}
\psfrag{b}[bl]{\footnotesize{ $\tBmH=0.3\gev^{-1}$}}
\psfrag{c}[bl]{\footnotesize{ $\tBmH=0.5\gev^{-1}$}}
\psfrag{d}[bl]{\footnotesize{ $\tBmH=1\gev^{-1}$}}
\includegraphics[width=3.4in]{images/Legend_tBmH.eps} 
\caption {(Color online). \mmiss and \pstarl projections of the $\Dz\tau\nu\Rightarrow\Dz\ell$ PDF for various
values of \tBmH.  }
\label{fig:Higgs_PDFs_1}
\end{figure}

\begin{figure}
\psfrag{E}[Bc]{\footnotesize{$\eps(\BDxtaunu)/\eps_{\rm SM}$ (\%)}}
\psfrag{Y}[Bc]{\footnotesize{\BDxtaunu yield}}
\psfrag{B}[Bc]{\footnotesize{\tBmH (GeV$^{-1}$)}}
\includegraphics[width=3.4in]{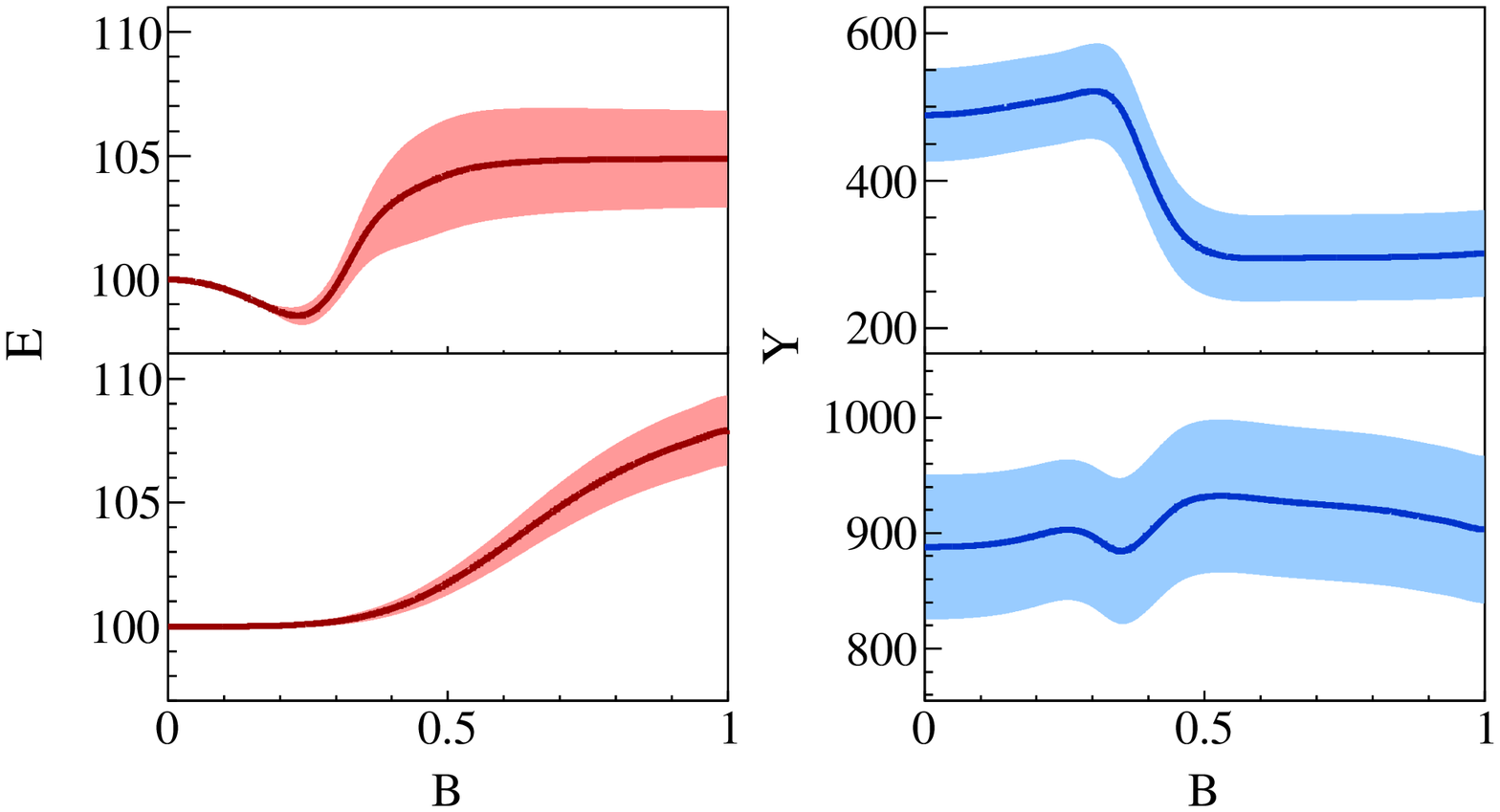}
\caption{(Color online). Left: Variation of the \BDtaunu (top) and \BDstaunu (bottom) efficiency in the 2HDM 
with respect to the SM efficiency. The band indicates the increase on statistical uncertainty 
with respect to the SM value.
Right: Variation of the fitted \BDtaunu (top) and \BDstaunu (bottom) yields as a function 
of \tBmH. The band indicates the statistical uncertainty of the fit.}
\label{fig:Higgs_Yields}
\end{figure}

For 20 values of \tBmH, equally spaced in the $[0.05, 1.00]\gev^{-1}$ range, we recalculate 
the eight signal PDFs, accounting for the charged Higgs contributions as described in
Sec.~\ref{sec:Theory}. Figure \ref{fig:Higgs_PDFs_1} shows the  \mmiss\ and \pstarl 
projections of the $\Dz\tau\nu\Rightarrow\Dz\ell$ 
PDF for four values of \tBmH.  The impact of charged Higgs contributions on the 
\mmiss\ distribution mirrors those in the $q^2$ distribution, see Fig.~\ref{fig:2HDMq2}, 
because of the relation
\begin{equation}
\mmiss = \left(p_{\epem}-p_{\Btag}-p_{\ds}-p_{\ell} \right)^2=\left(q-p_{\ell} \right)^2,\nonumber
\end{equation}
The changes in the \pstarl distribution are due to the change in the $\tau$ polarization.

We recalculate the value of the efficiency ratio $\eps_{\rm sig}/\eps_{\rm norm}$ as a function
of \tBmH (see Fig.~\ref{fig:Higgs_Yields}). The efficiency increases up to 8\% 
for large values of \tBmH, and, as we noted earlier, its uncertainty increases 
due to the larger dispersion of the  weights in the 2HDM reweighting. 

The variation of the fitted signal yields as a function of \tBmH is also shown in Fig.~\ref{fig:Higgs_Yields}. 
The sharp drop in the \BDtaunu\ yield at $\tBmH\approx0.4\gev^{-1}$ is due to the large shift 
in the \mmiss distribution which occurs when the Higgs contribution begins to dominate the total rate. 
This shift is also reflected in the \qt distribution and, as we will see in the next section, the data
do not support it.
The change of the \BDstaunu\ yield, mostly caused by the correlation with the \BDtaunu\ sample, 
is much smaller.

\begin{figure}[tb!]
\psfrag{R\(D\)}[bl]{\RD}
\psfrag{R\(D*\)}[bl]{\RDs}
\psfrag{t}[cc]{\tBmH (GeV$^{-1}$)}
\includegraphics[width=3in]{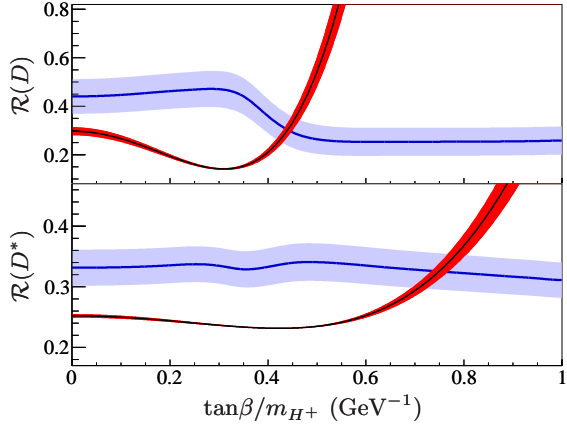}
\caption {(Color online). Comparison of the results of this analysis (light band, blue)
with predictions that include a charged Higgs boson of type II 2HDM (dark band, red).  
The widths of the two bands represent the uncertainties.
The SM corresponds to $\tanB/\mH=0$. }
\label{fig:PRL_Higgs}
\end{figure}

Figure \ref{fig:PRL_Higgs} compares the measured values of \RD\ and \RDs\ 
in the context of the type II 2HDM to the theoretical predictions as a function of  \tBmH. 
The  increase in the uncertainty on the signal PDFs and the efficiency ratio as a function 
of \tBmH are taken into account. Other sources of systematic uncertainty are kept constant 
in relative terms.

\begin{figure}
\psfrag{1000}[Bc]{\small{\hspace{4mm}1000}}
\psfrag{M}[Bc]{\small{\mH (GeV)}}
\psfrag{B}[Bc]{\small{\tanB}}
\includegraphics[width=3.45in]{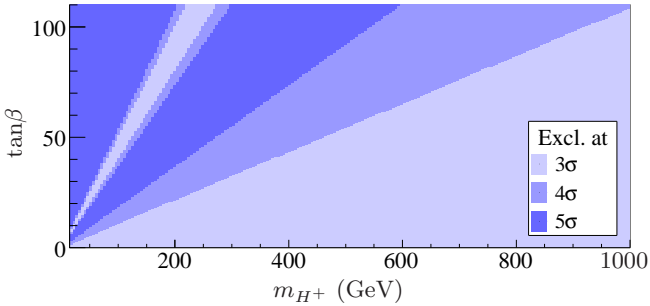} 
\caption{(Color online). Level of disagreement between this measurement of \RDx\ and the type II 2HDM
predictions for all values in the \tanB--\mH parameter space.}
\label{fig:Exclusion2D_DDsTauNu_BaBar}
\end{figure}

The measured values of \RD\ and \RDs\ match the predictions of this particular Higgs model
for  $\tanB/\mH=0.44\pm0.02\gev^{-1}$ and $\tanB/\mH=0.75\pm0.04\gev^{-1}$, respectively.
However, the combination of \RD\ and \RDs\ excludes the type II 2HDM charged Higgs boson 
at 99.8\% confidence level for any value of \tBmH, as illustrated in 
Fig.~\ref{fig:Exclusion2D_DDsTauNu_BaBar}.
This calculation is only valid for values of \mH greater than $15\gev$~\cite{Tanaka:1994ay,Tanaka:2010se}. 
The region for $\mH\leq 15\gev$ has already been excluded by $B\to X_s\gamma$ 
measurements~\cite{Misiak:2006zs}, and therefore, the type II 2HDM is excluded
in the full  \tanB--\mH parameter space.

The excess in both \RD\ and \RDs\ can be explained in more general charged Higgs 
models~\cite{Datta:2012qk,Fajfer:2012jt, Crivellin:2012ye, Becirevic:2012jf}.
The effective Hamiltonian for a type III 2HDM is
\begin{linenomath} 
\begin{align} \label{eq:Hamiltonian}
{\cal H}_{\rm eff} =& \frac{4G_F V_{cb}}{\sqrt{2}} \Bigl[(\overline{c}\gamma_\mu P_Lb)\, (\overline{\tau}
\gamma^\mu P_L\nu_\tau)  \nonumber \\
&+S_L (\overline{c}P_Lb)\, (\overline{\tau}P_L\nu_\tau)
+S_R (\overline{c}P_Rb)\, (\overline{\tau}P_L\nu_\tau)\Bigr],
\end{align}
\end{linenomath} 
where $S_{L}$ and $S_{R}$ are independent complex parameters, and $P_{L,R}\equiv (1\mp \gamma_5)/2$. 
This Hamiltonian describes the most general type of 2HDM for which $m_{H^+}^2\gg\qt$.

In this context, the ratios \RDx\ take the form
\begin{linenomath} 
\begin{align}
\RD &=\RD_{\rm SM}+A_{D}^{'} {\rm Re}(S_{R}+ S_{L})+B_{D}^{'} |S_{R}+ S_{L}|^2, \nonumber \\
\RDs &=\RDs_{\rm SM}+A_{\Dstar}^{'} {\rm Re}(S_{R}- S_{L})+B_{\Dstar}^{'} |S_{R}- S_{L}|^2. \nonumber 
\end{align}
\end{linenomath} 
The sign difference arises because \BDtaunu decays probe
scalar operators, while \BDstaunu decays are sensitive to pseudo-scalar operators.

\begin{figure}
\psfrag{rP}[Bc]{\small{$S_R+S_L$}}
\psfrag{rM}[Bc]{\small{$S_R-S_L$}}
\includegraphics[width=3.45in]{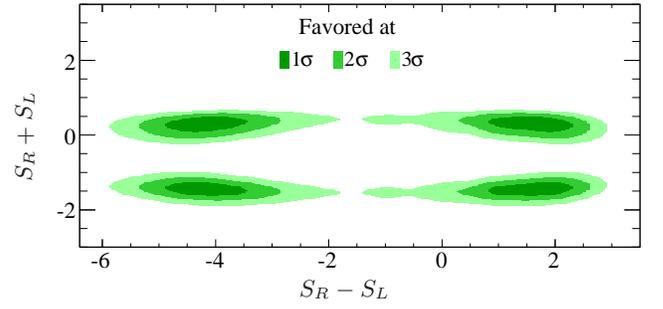} 
\caption{(Color online). Favored regions for real values of the type III 2HDM parameters $S_R$ and 
$S_L$ given by the measured values of \RDx.  The bottom two solutions are excluded by the 
measured \qt spectra.}
\label{fig:Higgs_Favored2D}
\end{figure}

The type II 2HDM corresponds to the subset of the type III 2HDM parameter space
for which $S_{R}= - m_bm_\tau {\rm tan}^2\beta/m_{H^+}^2$ and $S_{L}=0$. 

The \RDx\ measurements in the type II 2HDM context correspond to values of $S_R\pm S_L$ in the
range $[-7.4, 0]$. Given that the amplitude impacted by NP contributions takes the form
\begin{equation}
|H_s(S_R\pm S_L;\qt)| \propto |1+(S_R\pm S_L)\times F(\qt)|, 
\end{equation}
we can extend the type II results to the full 
type III parameter space by using the values of \RDx\ obtained 
with $H_s(S_R\pm S_L)$ for $H_s(-S_R\mp S_L)$. Given the small \tBmH dependence of \RDs\ 
(Fig.~\ref{fig:PRL_Higgs}), this is a good approximation for \BDstaunu decays.
For \BDtaunu decays, this is also true when the decay amplitude is dominated
either by SM or NP contributions, that is, for small or large values of $|S_R+S_L|$. 
The shift in the \mmiss and \qt spectra, which results in the 40\% drop on the value
of \RD\ shown in Fig.~\ref{fig:PRL_Higgs}, occurs in the intermediate region where 
SM and NP contributions are comparable. In this region, 
$H_s(S_R+ S_L) \neq H_s(-S_R- S_L)$, and, as a result, the large drop in
\RD\ is somewhat shifted. However, given that the asymptotic values of \RD\ are correctly
extrapolated, \RD\ is monotonous, and the measured value of \RDs\ is fairly constant,
the overall picture is well described by the 
$H_s(S_R\pm S_L) \approx H_s(-S_R\mp S_L)$ extrapolation.

Figure \ref{fig:Higgs_Favored2D} shows that for real values of $S_R$ and $S_L$, 
there are four regions in 
the type III parameter space that can explain the excess in both \RD\ and \RDs.
In addition, a range of complex values of the parameters are also compatible with
this measurement.

\begin{figure*}
\psfrag{Q}[Bc]{\small{$q^2$ (GeV$^2$)}}
\psfrag{a}[Br]{$D\ell$}
\psfrag{b}[Br]{$\Dstar\ell$}
\psfrag{x}[Br]{$D\ell$}
\psfrag{d}[Br]{$\Dstar\ell$}
\psfrag{e}[Br]{$D\ell$}
\psfrag{f}[Br]{$\Dstar\ell$}
\includegraphics[width=6.5in]{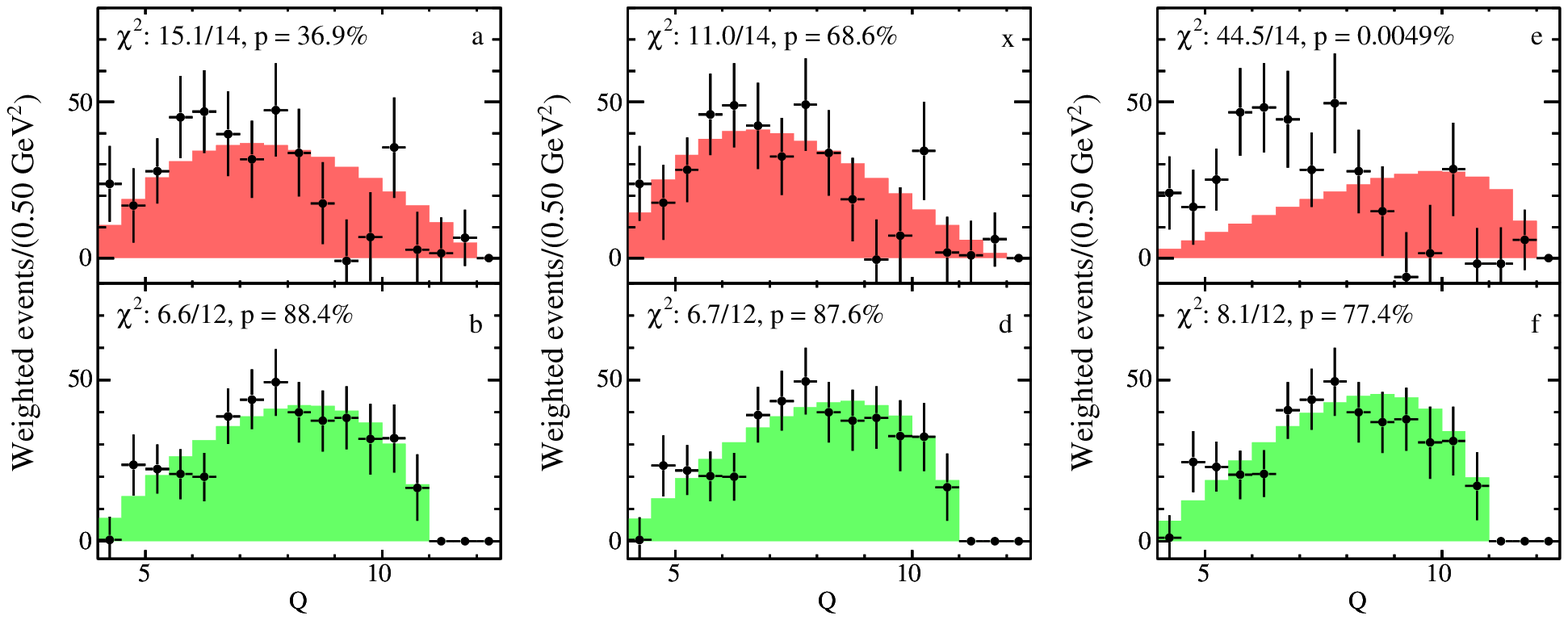}
\caption{(Color online) Efficiency corrected $q^2$ distributions for \BDtaunu (top) 
and \BDstaunu (bottom) events with $\mmiss>1.5\gev^2$ scaled to the results of the 
isospin-constrained fit. The points and the shaded histograms correspond to the measured 
and expected distributions, respectively.
Left: SM. Center: $\tBmH=0.30\gev^{-1}$. Right: 
$\tBmH=0.45\gev^{-1}$. The \Bz and \Bp samples are combined and the normalization 
and background events are subtracted. The distributions are normalized to the number of
detected events.
The uncertainty on the data points includes the statistical uncertainties of data and simulation.
The values of $\chi^2$ are based on this uncertainty.}
\label{fig:Higgs_Q2}
\end{figure*}

\subsection{Study of the $\boldsymbol{\qt}$ spectra}
As shown in Sec.~\ref{sec:theory:2hdm}, the \qt spectrum of \BDtaunu decays could be
significantly impacted by charged Higgs contributions.
Figure~\ref{fig:Higgs_Q2} compares the $q^2$ distribution of background subtracted
data, corrected for detector efficiency, with the expectations of three different scenarios.
Due to the subtraction of the large \BDstaunu feed-down in the $D\ell$ samples, the measured $q^2$ 
spectrum of \BDtaunu decays depends on the signal hypothesis. This dependence is very small, 
however, because the \qt 
spectrum of \BDstaunu decays is largely independent of \tBmH.

The measured \qt spectra agree with the SM expectations within the statistical uncertainties. 
For \BDtaunu decays, there might be a small shift to lower values, which is indicated by the 
increase in the $p$ value for $\tBmH=0.30\gev^{-1}$. 
As we showed in Sec.~\ref{sec:theory:2hdm}, the average \qt for  $\tBmH=0.30\gev^{-1}$ shifts
to lower values because the charged Higgs contribution to \BDtaunu decays, 
which always proceeds via an $S$-wave, interferes destructively with the SM $S$-wave. 
As a result, the decay proceeds via an almost pure $P$-wave and
is suppressed at large \qt by a factor of $p_D^2$, thus improving the agreement with data.
The negative interference suppresses the expected value of \RD\ as well, however,
so the region with small \tBmH is excluded by the measured \RD.

The two favored regions in Fig.~\ref{fig:Higgs_Favored2D} with $S_R+S_L\sim -1.5$ correspond 
to $\tBmH=0.45\gev^{-1}$ for \BDtaunu decays.
However, as we saw in Fig.~\ref{fig:2HDMq2}, the charged Higgs contributions dominate
\BDtaunu decays for values of $\tBmH>0.4\gev^{-1}$ and the \qt spectrum shifts significantly to 
larger values.
The data do not appear to support this expected shift to larger values of \qt.

\begin{table} 
\caption{Maximum $p$ value for the \qt distributions in Fig.~\ref{fig:Higgs_Q2} corresponding to the variations
due to the systematic uncertainties.} 
\label{tab:Chi2Syst} 
\begin{tabular}{l cc} \hline\hline
&\BDtaunu	& \BDstaunu  \\ \hline
SM					& 83.1\%	& 98.8\% \\
$\tBmH=0.30\gev^{-1}$ 	& 95.7\%	& 98.9\% \\
$\tBmH=0.45\gev^{-1}$\hspace{4mm} 	& 0.4\%	& 97.9\% \\ \hline\hline
\end{tabular}
\end{table}

To quantify the disagreement between the measured and expected \qt spectra, 
 we conservatively estimate the systematic uncertainties 
that impact the distributions shown in Fig.~\ref{fig:Higgs_Q2} (Appendix). 
Within these uncertainties, we find the variation that minimizes the $\chi^2$ value
of those distributions. Table \ref{tab:Chi2Syst} shows that, as expected, 
the conservative uncertainties give rise to large $p$ values in most cases. 
However, the $p$ value is only 0.4\% for \BDtaunu decays and $\tBmH=0.45\gev^{-1}$. 
Given that this value of \tBmH
corresponds to  $S_R+S_L\sim -1.5$, we exclude the two solutions at the bottom of 
Fig.~\ref{fig:Higgs_Favored2D} with a significance of at least $2.9\sigma$.

The other two solutions corresponding to $S_R+S_L\sim 0.4$ do not impact the \qt distributions of 
\BDtaunu to the same large degree, and, thus, we cannot exclude them with the current 
level of uncertainty.
However, these solutions also shift the \qt spectra to larger values due to the $S$-wave
contributions from the charged Higgs boson, so the agreement with the measured spectra
is worse than in the case of the SM. This is also true for any other solutions corresponding
to complex values of $S_R$ and $S_L$.

On the other hand, contributions to \BDtaunu decays proceeding via $P$-wave tend to shift the 
expected \qt spectra to lower values. 
Thus, NP processes with spin 1 could simultaneously explain the excess in 
\RDx\ \cite{Datta:2012qk,Becirevic:2012jf} and improve the agreement 
with the measured \qt distributions.

,
\section{Conclusions} 
\label{sec:Conclusions} 

In summary, we have measured the ratios $ \RDx = {\cal B}(\BDxtaunu)/{\cal B}(\BDxlnu)$ based
on the full \babar\ data sample, resulting in
\begin{linenomath} 
\begin{align}
\RD &= 0.440 \pm 0.058 \pm 0.042, \nonumber\\ 
\RDs &= 0.332 \pm 0.024 \pm 0.018, \nonumber
\end{align}
\end{linenomath} 
where the first uncertainty is statistical and the second is systematic.
These results supersede the previous \babar\ measurements  \cite{Aubert:2007dsa}.
Improvements of the event selection have increased
the reconstruction efficiency of signal events by more than a factor of 3, and the overall 
statistical uncertainty has been reduced by more than a factor of 2.

Table \ref{tab:Prev_RDx} shows the results of previous \BDxtaunu analyses.
In 2007 and 2010, the Belle collaboration measured the absolute \BDxtaunu branching fractions
which we translate to \RDx\ with ${\cal B}(\Bm\to\Dz\ellm\nulb)=(2.26\pm0.11)\%$ 
\cite{Nakamura:2010zzi} 
and ${\cal B}(\Bz\to\Dstarp\ellm\nulb)=(4.59\pm0.26)\%$ \cite{Dungel:2010uk}. 
For the translation of \RDs, we choose Belle's measurement of the branching fraction, instead of the
world average, because of the current large spread of measured values.
For Belle 2009, we average the results for \Bz and \Bm decays.

\begin{figure}
\psfrag{R\(D\)}[Bl]{\small{\RD}}
\psfrag{R\(D*\)}[Bl]{\small{\RDs}}
\psfrag{Belle 2007}[Bl]{\footnotesize{\textcolor[rgb]{0.45, 0.32, 0.27}{Belle 2007}}}
\psfrag{BaBar 2008}[Bl]{\footnotesize{\textcolor{blue}{\babar~2008}}}
\psfrag{Belle 2009}[Bl]{\footnotesize{\textcolor[rgb]{0.45, 0.32, 0.27}{Belle 2009}}}
\psfrag{Belle 2010}[Bl]{\footnotesize{\textcolor[rgb]{0.45, 0.32, 0.27}{Belle 2010}}}
\psfrag{BaBar 2012}[Bl]{\footnotesize{\textcolor{blue}{\babar~2012}}}
\includegraphics[width=3.3in]{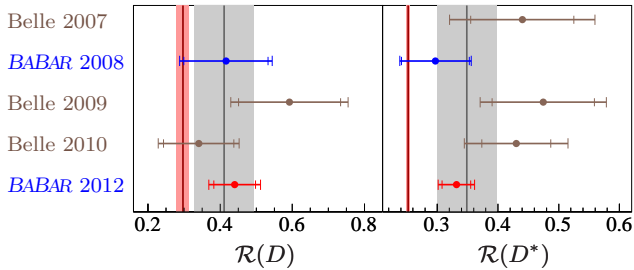} 
\caption{(Color online). Comparison of the previous measurements of \RDx\ with statistical and total uncertainties
(Table \ref{tab:Prev_RDx}) with this 
measurement (\babar\ 2012).
The vertical bands represent the average of the previous measurements (light shading) \
and  SM predictions (dark shading), separately for \RD\ and \RDs.  The widths of the bands
represents the uncertainties. }
\label{fig:Conclusions_RDx}
\end{figure}

The values measured in this analysis are compatible with those measured by the Belle Collaboration,
as illustrated in Fig.~\ref{fig:Conclusions_RDx}.

\begin{table}
\caption{Previous measurements of \RDx.} 
\label{tab:Prev_RDx} 
\begin{tabular}{l  r @{ $\pm$ } l @{ $\pm$ } l  r @{ $\pm$ } l @{ $\pm$ } l}\hline\hline
Measurement \hspace{5mm} &  \multicolumn{3}{c}{\RD} & \multicolumn{3}{c}{\RDs} \\ \hline
Belle 2007 \cite{Matyja:2007kt}			& \multicolumn{3}{c}{---} 		& 0.44	& 0.08	& 0.08 \\
\babar~2008 \cite{Aubert:2007dsa}		& 0.42	& 0.12 & 0.05 &\hspace{2mm} 0.30	& 0.06 	& 0.02 \\
Belle 2009 \cite{Adachi:2009qg}		& 0.59	& 0.14	& 0.08 	& 0.47	& 0.08	& 0.06 \\
Belle 2010 \cite{Bozek:2010xy}		& 0.34	& 0.10 	&0.06 	& 0.43	& 0.06	& 0.06 \\
\hline\hline
\end{tabular}
\end{table}

The results presented here exceed the SM predictions of 
$\RD_{\rm SM} = 0.297 \pm 0.017$ and $\RDs_{\rm SM} = 0.252 \pm 0.003$
by $2.0\sigma$ and $2.7\sigma$, respectively.
The combined significance of this disagreement, including the negative correlation
between \RD\ and \RDs, is $3.4\sigma$.
Together with the measurements by the Belle Collaboration, which also exceed the SM expectations, 
this could be an indication of 
NP processes affecting \BDxtaunu decays.

These results are not compatible with a charged Higgs boson in the type II 2HDM, 
and, together with $B\to X_s\gamma$ measurements, exclude this model in the full  \tanB--\mH parameter space.
More general charged Higgs models, or NP contributions with nonzero spin, are compatible with the
measurements presented here.

An analysis of the efficiency corrected \qt spectra of \BDtaunu and \BDstaunu decays shows 
good agreement with the SM expectations, within the estimated uncertainties.
The combination of the measured values of \RDx\ and the \qt spectra exclude a 
significant portion of the
type III 2HDM parameter space. 
Charged Higgs contributions with small scalar terms, $|S_R+S_L|<1.4$, 
are compatible with the measured \RDx\ and \qt distributions, but NP contributions with spin 1
are favored by data.

\begin{acknowledgments}
The concept for this analysis is to a large degree based on earlier \babar\
work and we acknowledge the guidance provided by M.~Mazur. The authors
consulted with theorists A.~Datta, S.~Westhoff, S.~Fajfer, J.~Kamenik, and I.~Ni\v{s}and\v{z}i\'{c} 
on the calculations of the  charged Higgs contributions to the decay rates.
We are grateful for the 
extraordinary contributions of our \pep2\ colleagues in
achieving the excellent luminosity and machine conditions
that have made this work possible.
The success of this project also relied critically on the 
expertise and dedication of the computing organizations that 
support \babar.
The collaborating institutions wish to thank 
SLAC for its support and the kind hospitality extended to them. 
This work is supported by the
US Department of Energy
and National Science Foundation, the
Natural Sciences and Engineering Research Council (Canada),
the Commissariat \`a l'Energie Atomique and
Institut National de Physique Nucl\'eaire et de Physique des Particules
(France), the
Bundesministerium f\"ur Bildung und Forschung and
Deutsche Forschungsgemeinschaft
(Germany), the
Istituto Nazionale di Fisica Nucleare (Italy),
the Foundation for Fundamental Research on Matter (Netherlands),
the Research Council of Norway, the
Ministry of Education and Science of the Russian Federation, 
Ministerio de Econom\'{\i}a y Competitividad (Spain), and the
Science and Technology Facilities Council (United Kingdom).
Individuals have received support from 
the Marie-Curie IEF program (European Union) and the A. P. Sloan Foundation (USA). 
\end{acknowledgments}
\begin{figure*} 
\psfrag{x}[bc]{\small{$q^2$ (GeV$^2$)}}
\psfrag{a}[Bl]{{\small $D\ell$}}
\psfrag{b}[Bl]{{\small $D\ell$}}
\psfrag{c}[Bl]{{\small $D\ell$}}
\psfrag{d}[Bl]{{\small $\Dstar\ell$}}
\psfrag{e}[Bl]{{\small $\Dstar\ell$}}
\psfrag{f}[Bl]{{\small $\Dstar\ell$}}
\includegraphics[width=7in]{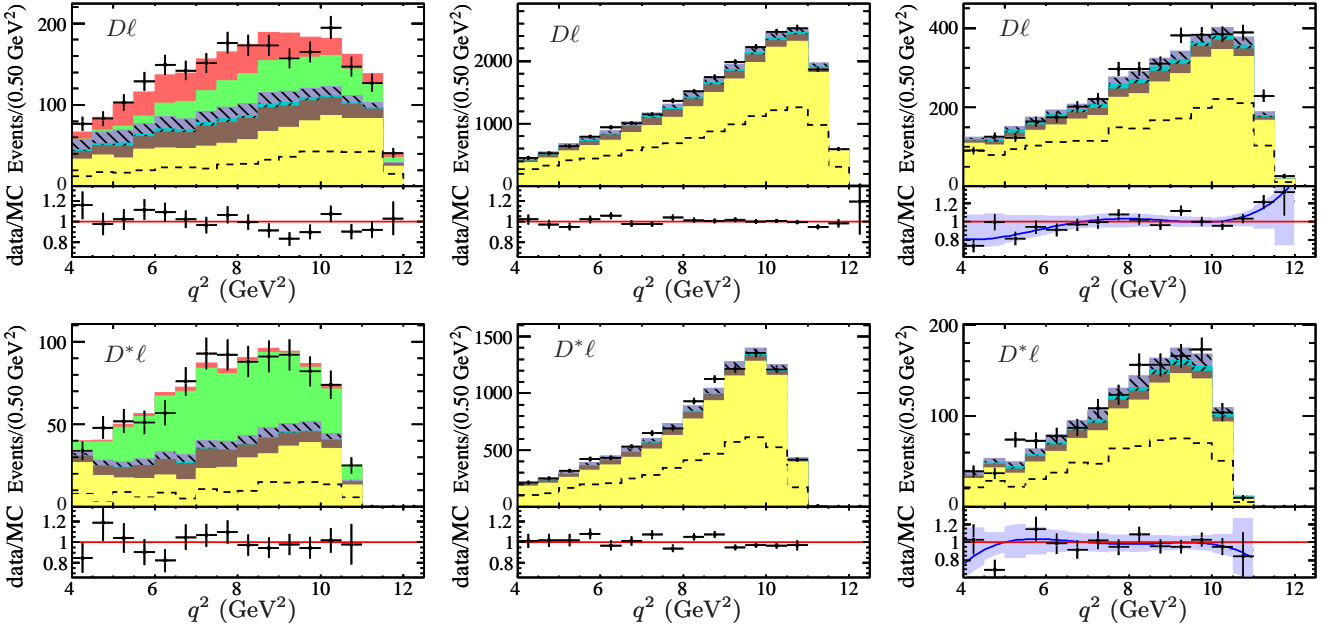}
\caption{(Color online). Assessment of the uncertainties on the $q^2$ distributions of background events 
with $\mmiss>1.5\gev^2$.
Left: results of the isospin-constrained fit for the SM.
Center: sample with $0.5<\eextra<1.2\gev$
and $5.27<\mes<5.29\gev$. Right: sample satisfying the 
BDT requirements in the $5.20<\mes<5.26\gev$ region. 
The data/MC plots show a fourth order polynomial fit and the total systematic
uncertainty considered.
The simulation in the control samples is normalized to the number of events in data.
 See Fig.~\ref{fig:Stability_EExtra} for a legend.}
\label{fig:Higgs_Q2Bkg}
\end{figure*}

\begin{figure*}
\psfrag{Dss}[Bl]{\footnotesize{$\dss\ell\nu$}}
\psfrag{Dsstau}[Bl]{\footnotesize{$\dss\tau\nu$}}
\psfrag{Dpipi}[Bl]{\footnotesize{$\dss(\to\ds\pi\pi)\ell\nu$}}
\psfrag{Norm}[Bl]{\footnotesize{Normalization}}
\psfrag{Sig}[Bl]{\footnotesize{Signal}}
\psfrag{Sigm2}[Bl]{\footnotesize{Signal $\mmiss>1.5\gev^2$}}
\psfrag{Q}[Bc]{$q^2$ (GeV$^2$)}
\psfrag{a}[Br]{$D\ell$}
\psfrag{b}[Br]{$\Dstar\ell$}
\psfrag{x}[Br]{$D\ell$}
\psfrag{d}[Br]{$\Dstar\ell$}
\psfrag{e}[Br]{$D\ell$}
\psfrag{f}[Br]{$\Dstar\ell$}
\includegraphics[width=7in]{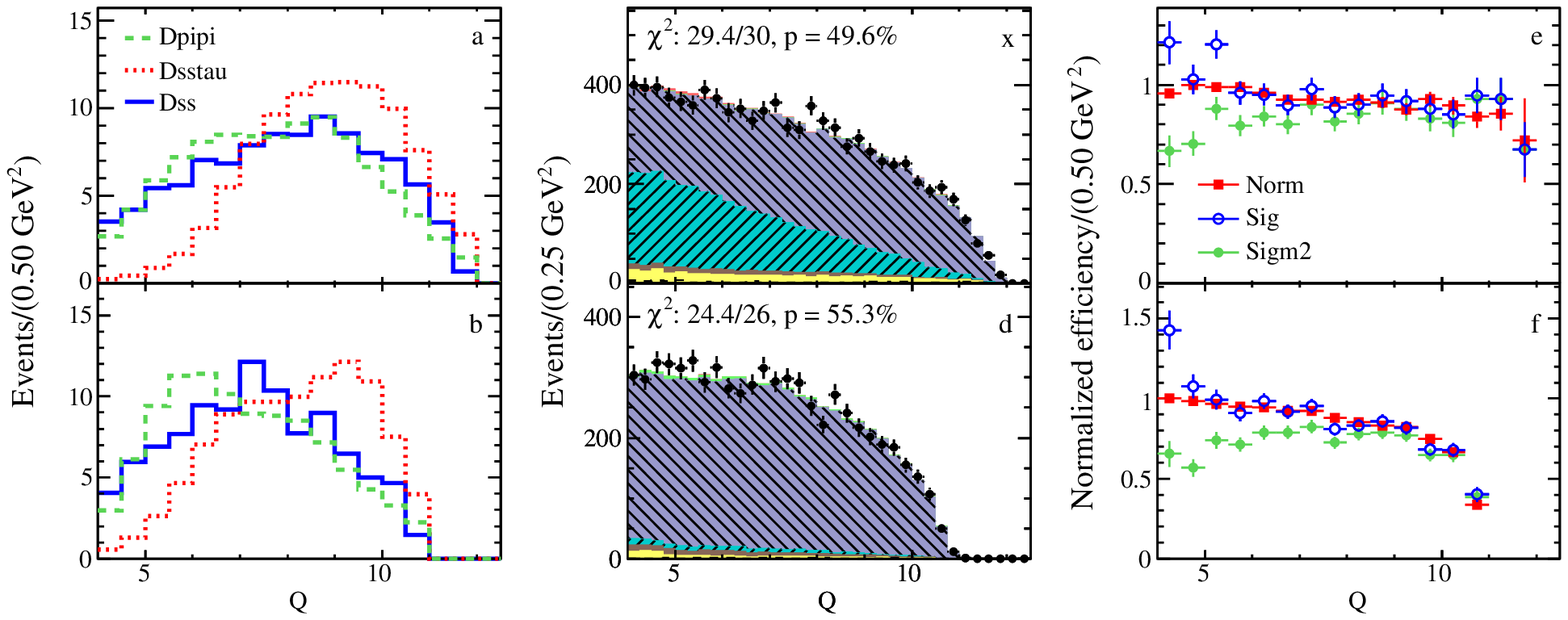}
\caption{(Color online). Left: \qt distributions for the different \BDssltnu contributions, all normalized to 100 events.
Center: $q^2$ distributions for events with $\mmiss<1.5\gev^2$ scaled to the results of the 
isospin-constrained fit for the SM. 
See Fig.~\ref{fig:Stability_EExtra} for a legend.
Right: \qt dependence of the efficiency. 
The scale for the efficiency of the normalization decays is chosen so that the maximum value is 1. 
The efficiency data for the signal are adjusted so that they overlap with the data for normalization 
decays in the central part of the \qt range.  
The signal efficiencies with and without the \mmiss selection have the same scale. }
\label{fig:Stability_Q2_NormEffDss}
\end{figure*}

\section*{APPENDIX: SYSTEMATIC UNCERTAINTIES ON THE $\boldsymbol{\qt}$ SPECTRA} 
\label{sec:Appendix}

To assess the systematic uncertainty on the measured \qt distributions of
\BDxtaunu decays, we examine their sensitivity to the estimated contributions
from background and normalization events.
The \qt distributions of signal and the various backgrounds are presented in
Fig.~\ref{fig:Higgs_Q2Bkg} (left). There is good agreement between the data and 
the background contributions as  derived 
from the isospin-constrained fit. To further examine the shape of the fixed contributions from \BB 
and continuum background, we show two comparisons with data control samples:
one for medium values of \eextra in the \mes peak regions without the BDT requirements 
imposed, and the other for the \mes sidebands with the BDT requirements.  While the first 
sample shows excellent agreement over the full \qt range, the smaller  second  sample shows 
some deviations at low and high \qt.  
We approximate the deviation of the data from the simulation by a fourth order polynomial,
and we adopt this difference plus the statistical uncertainty of each bin
as the overall uncertainty of the  
\BB and continuum backgrounds. We conservatively consider it uniformly distributed
between the limits of the band shown in Fig.~\ref{fig:Higgs_Q2Bkg} and uncorrelated 
between different bins.

The systematic uncertainty on the shape of the \qt distribution of \BDssltnu decays is estimated by varying
the relative abundance of the contributions shown in Fig.~\ref{fig:Stability_Q2_NormEffDss}. 
We allow a variation of \RDss, the ratio of 
\BDsstaunu decays to \BDsslnu decays, between $-20\%$ and $+50\%$. We also allow a contribution 
of up to 30\% of \BDsslnu decays with the \dss decaying into $\ds\pipi$. 
In addition, we assume a $\pm15\%$ variation of the total \BDssltnu yield.

The \qt spectrum of normalization decays, both well reconstructed and cross-feed \BDxlnu decays,
is well described by the simulation, see Fig.~\ref{fig:Stability_Q2_NormEffDss}. 
Given that the normalization decays are well understood theoretically, we adopt the statistical
uncertainty of the simulated distributions as the overall uncertainty of this contribution.
Except for $\qt<5\gev^2$, where the rate of signal decays is highly suppressed,
the efficiency and detector effects are very similar for signal and normalization. Thus, we also derive 
the overall uncertainty from the statistical uncertainty of the
simulated signal \qt distributions.

Since it is not feasible to repeat the \mmiss--\pstarl fit for each variation of the background contributions,
we adopt the following procedure to account for the impact of these  changes
on the $\chi^2$: for each of the three \qt distributions in Fig.~\ref{fig:Higgs_Q2} and each variation
of the background components, we determine the 
\BDtaunu and \BDstaunu yields by a fit that minimizes the $\chi^2$ of those distributions.

\bibliography{paper}

\begin{thebibliography}{48}
\bibitem [{con()}]{convention}%
  \BibitemOpen
  \href@noop {} {}\bibinfo {note} {Throughout this letter, $\ell$ refers only
  to the light leptons $e$ and $\mu$, $D^{(*)}$ refers to a $D$ or a $D^*$
  meson, and charge-conjugate decay modes are implied}\BibitemShut {NoStop}%
\end{thebibliography}

\end{document}